\documentclass[a4paper]{article}

\usepackage{amsfonts}
\usepackage{amssymb,todonotes}
\definecolor{darkblue}{rgb}{0.1,0.1,0.45}
\usepackage[colorlinks=true, pdfstartview=FitV, linkcolor=darkblue, citecolor=darkblue, urlcolor=darkblue]{hyperref}
\usepackage{amsmath}
\usepackage{amsthm}
\usepackage{enumerate}
\usepackage{graphicx}
\usepackage{amsmath}
\usepackage{amssymb}
\def \QED{\hfill $\blacksquare$\\[2pt]}
\usepackage{amsfonts}
\usepackage{epsfig}
\usepackage{graphicx}
\def\le{\left}
\def \ri{\right}

\usepackage{framed}

\newtheorem*{prop*}{Proposition}
\newtheorem{teor}{Theorem}

\newtheorem{RHP}{Riemann-Hilbert problem}
\newtheorem{Assump}{Assumption}
\newtheorem{col*}{Corollary}

\newcommand{\e}{\mathrm{e}}
\newcommand{\ii}{\mathrm{i}}
\renewcommand{\i}{\mathrm{i}}

\renewcommand{\d}{\mathrm{d}}
\renewcommand{\Im}{\operatorname{Im}}
\renewcommand{\(}{\left(}
\renewcommand{\)}{\right)}
\newcommand{\w}{\mathrm{w}}
\newcommand{\dsfrac}{\displaystyle\frac}
\newcommand{\ds}{\displaystyle}
\newcommand{\ol}{\overline}

\providecommand{\B}{\mathbf}

\renewcommand{\Re}{\operatorname{Re}}

\newtheorem{lem}{Lemma}[section]
\newtheorem{rem}{Remark}[section]
\newtheorem{com}{Remark}[section]

\newcounter{subsect}[sect]

\newtheorem{cor}{Corollary}
\newtheorem{prop}{Proposition}[subsect]

\title{Laguerre polynomials and transitional asymptotics of the modified Korteweg-de Vries equation for step-like initial data}

\begin{document}
\begin{center}
\begin{Large}
\textbf{
Laguerre polynomials and transitional asymptotics of the modified Korteweg-de Vries equation for step-like initial data} 
\end{Large}
\end{center}
\begin{center}
M. Bertola$^{\dagger}$ \footnote{Marco.Bertola@\{concordia.ca, sissa.it\}},
A. Minakov$^{\dagger\ddagger}$  \footnote{ominakov@sissa.it}

\bigskip
\begin{minipage}{0.7\textwidth}
\begin{small}
\begin{enumerate}
\item [${\dagger}$] {\it  Department of Mathematics and
Statistics, Concordia University\\ 1455 de Maisonneuve W., Montr\'eal, Qu\'ebec,
Canada H3G 1M8}  
\item[${\ddagger}$]{\it SISSA/ISAS, via Bonomea 265, Trieste, Italy }
\end{enumerate}
\end{small}
\end{minipage}
\vspace{0.5cm}
\end{center}

\begin{abstract}
We consider the compressive wave for the modified Korteweg--de Vries equation with background constants $c>0$ for $x\to-\infty$ and $0$ for $x\to+\infty.$ We study the asymptotics of solutions in the transition zone
$4c^2t-\varepsilon t<x<4c^2t-\beta t^{\sigma}\ln t$ for $\varepsilon>0,$ $\sigma\in(0,1),$ $\beta>0.$ In this region we have a bulk of nonvanishing oscillations, the number of which grows as $\frac{\varepsilon t}{\ln t}.$
Also we show how to
obtain Khruslov--Kotlyarov's asymptotics in the domain $4c^2t-\rho\ln t<x<4c^2t$ with the help of parametrices constructed out of Laguerre polynomials in the corresponding Riemann-Hilbert problem.

\end{abstract}

\tableofcontents

\section{Introduction and results}

We consider the   Cauchy problem for the modified Korteweg--de Vries equation with step--like initial datum in the long-time limit, namely
\begin{equation}\label{mkdv}
q_t(x,t)+6q^2(x,t)q_x(x,t)+q_{xxx}(x,t)=0
\end{equation}
\begin{equation}\label{ic}
q(x,0)=q_0(x)\to\begin{cases}0\qquad {\rm as}\quad
x\to+\infty,\\c\qquad {\rm as }\quad x\to-\infty,\qquad
c>0,\end{cases}
\end{equation}
The problem received a lot of attention in the span of more than four decades;
the first asymptotic results were obtained for the  Korteweg--de Vries equation. Physicists started to develop a qualitative description since  the pioneer work of A.~Gurevich and L.~Pitaevsky \cite{GP} (1973). Later in this direction R.~Bikbaev,
V.~Novokshenov  and others  actively worked on the topic (sf. \cite{BikN1}-\cite{Bikb6}). In particular, the Cauchy problem (\ref{mkdv}) -- (\ref{ic}) with more general type of initial data was considered by  R.~Bikbaev \cite{Bikb4} in 1992.

\begin{figure}[ht]
\begin{center}
\epsfig{width=200mm,figure=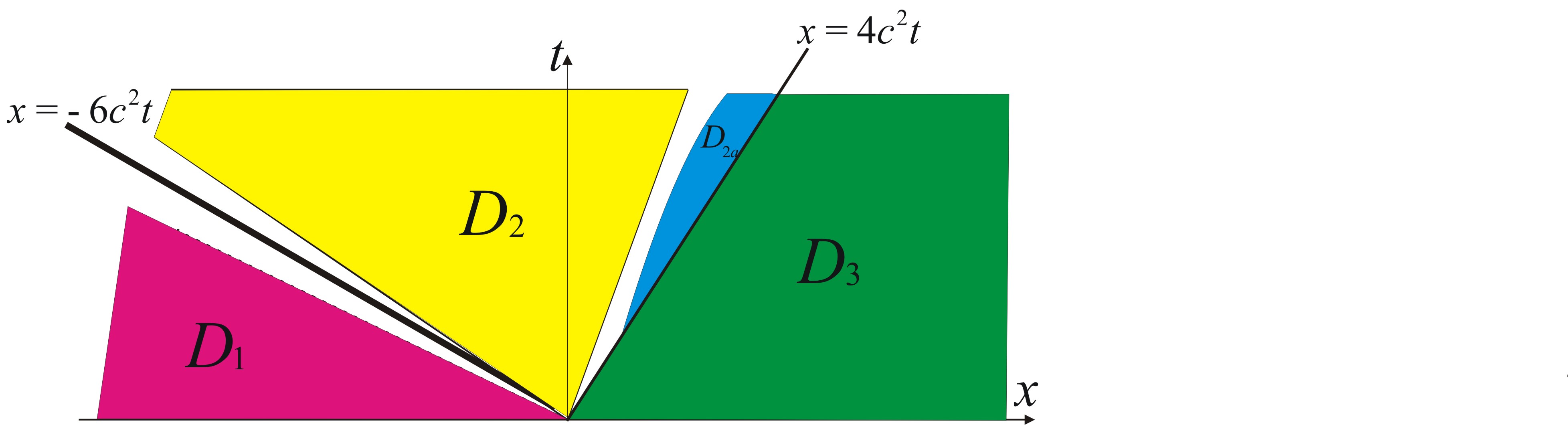}
\end{center}
\caption{Domains in $x,t>0$-half-plane with qualitatively different asymptotic behaviour of $q(x,t):$
$D_1 - $ asymptotics tends to the constant $c$, $D_3$ - possible solitons, otherwise the solution $q(x,t)$ of
the Cauchy problem  tends to 0, $D_2$ - modulated elliptic asymptotics, $D_{2a}$ - asymptotic solitons.}
\label{Domains_in_xt_plane}
\end{figure}

The above research utilized the semi--heuristic Whitham method.
The physical intuition  suggested that the $(x,t)$-plane gets  divided into three domains (see Figure \ref{Domains_in_xt_plane}),
\begin{figure}[ht]
\begin{center}
\includegraphics[width=100mm]{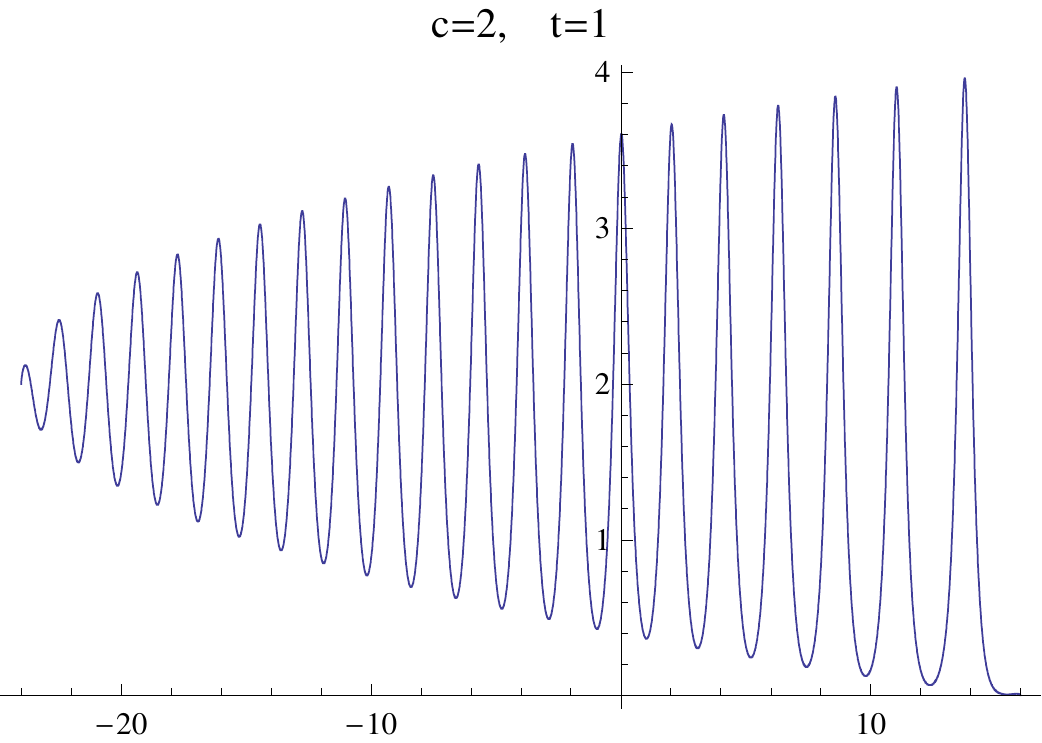}
\end{center}
\caption{Graphics of $q_{el}$ in the domain $-6c^2t<x<4c^2t$ for initial data (\ref{pure_step_ic}).
}
\label{Graphics_q_el_1}
\end{figure}
\begin{figure}[ht]
\begin{center}
\includegraphics[width=100mm]{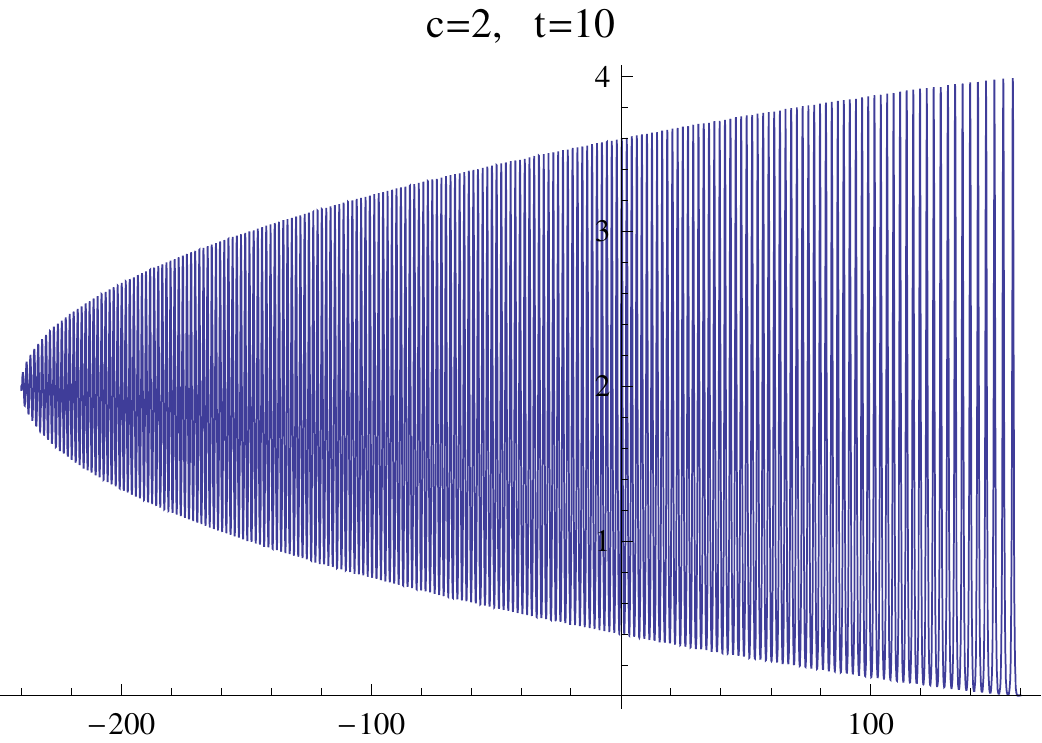}
\end{center}
\caption{Graphics of $q_{el}$ in the domain $-6c^2t<x<4c^2t$ for initial data (\ref{pure_step_ic}).}
\label{Graphics_q_el_2}
\end{figure}
where the solution exhibits completely different asymp- \newpage \noindent totic behaviours; in  the left and in the right domains ($D_1$ and $D_3$) the solution tends to constants (one of them equals zero in our case) while in the middle domain it behaves as a modulated elliptic wave.

Rigorous mathematical papers were almost absent with the exception of papers concerning the  region of asymptotic solitons. It was proved that near the wave front there exists a strip--like domain 
where the
so-called {\it asymptotic solitons} arise. They are attributable to  the presence of an endpoint  of the  simple continuous spectrum of the corresponding Lax operator, as opposed to usual solitons which are  generated by points of the  discrete spectrum.
For the KdV equation this observation was pointed out  in the pioneer work by E.~Khruslov \cite{Kh1}, \cite{Kh2}
(1975, 1976),  which appeared even before the first studies of the scattering and inverse scattering problem for Sturm-Liouville operators with step-like potential. 
This was done even before a full proof of existence of the time dynamics for spectral functions in the case of step-like data
(see the book by Marchenko \cite{Mar} 1977, Deift-Trubowitz \cite{DT} 1979, Koen-Kappeler \cite{CK} 1985). The work by Khruslov was inspired by prior results of  Gurevich, Pitaevskii \cite{GP} 
and for the MKdV we refer to E.~Khruslov and V.~Kotlyarov \cite{KK} (1989). A comprehensive  review of the results in this direction can be found in \cite{KK2}, \cite{KK3} and references therein. Thus, besides the three specified  regions,   there is an additional transition region near the leading edge where a train of  asymptotic solitons runs.

Similar phenomena occur in the semiclassical asymptotics of integrable equations and large size random matrices,  see \cite{Claeys Grava 2010}, \cite{Bertola Buckingma Lee Pierce}, \cite{Bertola Tovbis}.

The so--called {\it Riemann--Hilbert method} (a.k.a. {\it inverse scattering}) and the corresponding steepest descent method \cite{DZ93} have been actively developed now for more than 20 years. Recently these methods were successfully applied to the study of solutions of step type in other regions of the $x,t$ half-plane, not only in the soliton domains (\cite{BK07} --\cite{BV},  \cite{EGKT}, \cite{KM} -- \cite{MK10}).

 Throughout the paper, we make some additional assumptions on the initial data, namely Assumptions \ref{Assump_no_ds}, \ref{Assump_2}, \ref{Assump_3} described in section \ref{sect_prel}.
The Cauchy problem (\ref{mkdv})-(\ref{ic}) 
 with a pure step initial data \begin{equation}\label{pure_step_ic}
                    \widetilde q_0(x)=\begin{cases}
                            0,\quad\ \ \quad x>0,\\c>0,\quad x<0,
                           \end{cases}
                   \end{equation}

\noindent  for the modified Korteweg -- de Vries equation \eqref{mkdv} was recently studied by V.~Kotlyarov and A.~Minakov \cite{KM} via the Riemann--Hilbert approach. (More general initial data with nonzero backgrounds, defined by two different and nonzero constants as $x\to\pm\infty$,
were studied  in \cite{M1}, \cite{M2}, \cite{KM2}, see also \cite{Leach_Needham}, \cite{Marchant}). In particular, in the domain $-6c^2 t<x<4c^2t $ the asymptotics of the  Cauchy problem solution is described by a modulated elliptic wave.

\begin{teor} {\rm\cite{KM}}\label{teorKM} In the region $-6c^2 t<x<4c^2t $
the solution of the initial value problem (\ref{mkdv}), (\ref{ic}), \eqref{pure_step_ic}  takes
form of a modulated elliptic wave
\[q(x,t)=q_{ell}(x,t)+\mathrm{o}\(1\),\quad t\rightarrow\infty,\] where
\begin{equation}{\label{qmod}}q_{ell}(x,t)=\sqrt{c^2-d^2(\xi)}
\dsfrac{\Theta\(\pi\i+\i t
B(d(\xi))+\i\Delta(d(\xi))|\tau(d(\xi))\)}{\Theta\(\i t
B(d(\xi))+\i\Delta(d(\xi))|\tau(d(\xi))\)}, \quad
\xi=\dsfrac{x}{12t}.
\end{equation}
Here $$\Theta(z,\tau)=\sum\limits_{m\in\mathbb{Z}}\e^{\frac12 \tau m^2+z m}$$ is the theta function determined  by its $b$-period $\tau=\tau(d)$, the functions
 $B(d)$, $\tau(d)$, $\Delta(d)$ are explicitly defined via (\ref{Bg})-(\ref{Delta})
\begin{equation}{\label{Bg}}B(d)=24\ds\int\limits_{\i d}^{\i
c}\dsfrac{(k^2+\mu^2(d))(k^2+d^2)\d k}{\w_+(k,d)}, \qquad \textrm{where}\quad \w(k,d)=\sqrt{(k^2+c^2)(k^2+d^2)},
\end{equation}
\begin{equation}{\label{tau}}\tau(d)=-\pi\i\ds\int\limits_{\i d}^{\i c}\dsfrac{\d k}{\w_+(k,d)}
\(\ds\int\limits_0^{\i d}\dsfrac{\d k}{\w(k,d)}\)^{-1},
\end{equation}
\begin{equation}{\label{Delta}}\Delta(d)=
 \int\limits_{\i d}^{\i c}\dsfrac{\log\left(a_+(k)a_-(k)\right)\d k}{\w_+(k,d)}
\left(
 \i\int\limits_{0}^{\i d}\dsfrac{\d k}{\w(k,d)}\right)^{-1} ,
\end{equation}
and the functions $d(\xi),$ $\mu(d(\xi))$
 are defined implicitly through formulas
 (\ref{dmu1}), (\ref{dmu2})
 \begin{equation}
 {\label{dmu1}}
\int\limits_0^1(\mu^2-\lambda^2d^2)\sqrt{\dsfrac{1-\lambda^2}{c^2-\lambda^2d^2}}\d\lambda=0,
\end{equation}
\begin{equation}{\label{dmu2}}
\dsfrac{c^2}{2}+\xi=\mu^2+\dsfrac{d^2}{2}.
\end{equation}

\end{teor}

{ The method used in \cite{KM} can be  readily applied not only for the pure initial datum \eqref{pure_step_ic}, but for more general initial datum of the form 
\eqref{ic} satisfying the  Assumptions \ref{Assump_no_ds}, \ref{Assump_2}, \ref{Assump_3} from section \ref{sect_prel}.}
\noindent Our first result improves on Thm. \ref{teorKM} by specifying the error term, namely
 we show that


\begin{teor}\label{thrm_Elliptic_est}

Let $q(x,t)$ be the solution of initial value problem \eqref{mkdv}, \eqref{ic}, with initial datum satisfying Assumptions \ref{Assump_no_ds}, \ref{Assump_2}, \ref{Assump_3} of section \ref{sect_prel}. Then we have three typical behaviours in the following regions:
\begin{enumerate}
\item  {\rm Elliptic region} (far from  the leading edge and asymptotic solitons).
Let $\varepsilon>0$ be  sufficiently small. Then
$$q(x,t)=q_{el}(x,t)+\mathcal{O}(t^{-1})\qquad\textrm{ for }\quad   \frac{-c^2}{2}+\varepsilon<\frac{x}{12t}<\frac{c^2}{3}-\varepsilon,
$$
\item   The {\rm beginning of transition}. Let $\sigma\in(0,1)$ and $\frac{1}{K}<\beta<K,$ where $K>0$. Then
$$q(x,t)=q_{el}(x,t)+\mathcal{O}\({t^{-\sigma}}\)\qquad\textrm{ for }\quad   \frac{c^2}{3}-\varepsilon<\frac{x}{12t}<\frac{c^2}{3}-\frac{\beta t^{\sigma}\ln t}{12t}.
$$
\item In the {\rm middle of transition.} Let $0<\sigma_1<\sigma_2<1$ and $\frac{1}{K}<\beta_1<\beta_2<K,$ where $K>0$. Then
$$q(x,t)=q_{el}(x,t)+\mathcal{O}\({t^{-\sigma_1}}\)\qquad\textrm{ for }\quad   \frac{c^2}{3}-\frac{\beta_2 t^{\sigma_2}\ln t}{12t}<\frac{x}{12t}<\frac{c^2}{3}-\frac{\beta_1 t^{\sigma_1}\ln t}{12t}.
$$
\end{enumerate}
\end{teor}
\vskip0.5cm\noindent Since we  mainly consider the regime in which $$\xi\equiv\frac{x}{12t}\to\frac{c^2}{3},$$ it is useful to introduce parameters associated with $\xi, d,$ which are small in this regime. Namely, since  in our regime we have
$$d\to c,\quad \xi\to\frac{c^2}{3},$$
we  use the following notations 

\begin{equation}
\label{etav}
\eta=1-\frac{d}{c},\quad v=1-\frac{3\xi}{c^2}.
\end{equation}
It can be shown (see \cite{KM2015}) that the quantities $B$ (\ref{Bg}), $v$ (\ref{etav}) 
have the following expansions:
\begin{equation}\label{B}
 \frac{B}{\pi} = 8c^3\eta\(1+\sum\limits_{j=1}^{M-1} \eta^jP_j(\eta)+\mathcal{O}(\eta^{M}\ln^{M}\eta)\),
\qquad M=2,3,\dots
\end{equation}
where $P_j$ are polynomials of $\ln\eta$ of degree $j,$ the first two of which are
$$P_1(\eta) = -\(2+\frac12\ln\frac{\eta}{8}\),\quad P_2(\eta) = \frac{1}{16}\(13-42\ln2+36\ln^22+2\ln\eta(7-12\ln2+2\ln\eta)\)$$
and \begin{equation}\label{vineta}
 v = \eta\ln\frac{8\e}{\eta}+\sum\limits_{j=2}^{M} \eta^jQ_j(\eta)+\mathcal{O}(\eta^{M}\ln^{M}\eta),
 \qquad M=2,3,\dots
\end{equation}
where $Q_j$ are polynomials of $\ln\eta$ of degree $j$. The first polynomials $Q_j$ are
$$Q_2(\eta) = \frac12\(-2-9\ln2+9\ln^22+\(3+\ln\frac{\eta}{64}\)\ln\eta\),$$
$$Q_3(\eta) = \frac{1}{16}\(9-6\ln2+18\ln^22(-5+6\ln2)\)+\frac{\ln\eta}{8}(1+30\ln2-54\ln^22)+\frac{\ln^2\eta}{8}(-5+18\ln2)-\frac{\ln^3\eta}{4}.$$

\noindent Further, as a consequence of Thm \ref{thrm_Elliptic_est}, we can decompose the elliptic wave into a train of asymptotic solitons (see the end of Section \ref{sect_ell}),
\begin{cor}\label{cor_1_q_el} Let $M$ be 1,2, or 3. Suppose that  $(x, t)$ lies  on a curve
\begin{equation}
\label{curve}
x=4c^2t - {\beta t^{\sigma} \ln t},\ \ \beta>0,\ \ \sigma\in(0,\frac{M}{M+1}).
\end{equation}
 Let $\widetilde \eta$ be the unique solution of
$$v=\widetilde\eta\ln\frac{8\e}{\widetilde\eta}+\sum\limits_{j=2}^{M}\widetilde\eta^jQ_j(\widetilde\eta)$$ that behaves as $\widetilde\eta\asymp\frac{v}{\ln\frac{1}{v}}$ as $v\to0$ ($Q_j$ are as in (\ref{vineta})). Let
\begin{equation}\label{ztilde}\widetilde z = 8c^3t\widetilde\eta\(1+\sum\limits_{j=1}^{M-1}\widetilde\eta^jP_j(\widetilde\eta)\)-\frac12,
\end{equation}
 where $P_j$ are as in (\ref{B}) and let $n$ be the greatest integer not exceeding $\frac{\widetilde z}{2},$ i.e.
$n=\lfloor\widetilde z/2\rfloor$, $2n\leq\widetilde z<2n+2.$
Then, as $t\to\infty$
 \begin{equation}\label{eq_cor_1}q(x,t) =
\frac{2c}{\cosh\left[2c(x-4c^2t)+(2n+\frac32)\ln t+\alpha^{(M)}_n\right]}+r^{(M)}(x,t),\end{equation}
where the errors $r^{(M)}$ are $$r^{(M=1)}=\mathcal{O}(t^{2\sigma-1}\ln^2 t)+\mathcal{O}(t^{-\sigma}),$$
$$r^{(M=2)}=\mathcal{O}(t^{3\sigma-2}\ln^3 t)+\mathcal{O}(t^{\sigma-1}\ln t)+\mathcal{O}(t^{-\sigma}),$$
$$r^{(M=3)}=\mathcal{O}(t^{4\sigma-3}\ln^4 t)+\mathcal{O}(t^{\sigma-1}\ln t)+\mathcal{O}(t^{-\sigma}),$$
and the phases $\alpha_n^{(M)}$ are
$$\alpha_n^{(M=1)}=(2n+\frac32)\ln\frac{32c^3}{n}+2n-\ln\frac{4}{(h^*)^2},$$
$$\alpha_n^{(M=2)}=(2n+\frac32)\ln\frac{32c^3}{n}+2n-\ln\frac{4}{(h^*)^2}+\frac{n^2}{4c^3t}\ln\frac{32c^3t}{\e^2 n},$$
$$\alpha_n^{(M=3)}=(2n+\frac32)\ln\frac{32c^3}{n}+2n-\ln\frac{4}{(h^*)^2}+\frac{n^2}{4c^3t}\ln\frac{32c^3t}{\e^2 n}+$$
$$+\frac{n^3}{64c^6t^2}\(-31+\frac{25}{2}\ln\frac{32c^3\e t}{n}-\(\ln\frac{32c^3\e t}{n}\)^2
\).$$
\end{cor}

\begin{com} It is possible to find expressions for  $\alpha^{(M)}_n$ and the error $r^{(M)}$ for any given integer $M$ but the expressions become quickly unwieldy. 
\end{com}
\begin{com}
 We see that on the curve \eqref{curve}, we move through the bulk of solitons, and not constrained to a particular soliton. Hence the quantity $n,$ which describes
the number of soliton on which we are located, varies with time. We can give a rough approximation of how it changes. Namely, the quantity $\widetilde z$ \eqref{ztilde} behaves like
$$\frac{\widetilde z}{2}=\frac{4c^3tv}{\ln\frac{1}{v}}
\le(1+\mathcal{O}\le(\frac{1}{\ln v}\ri)\ri)-\frac14+\mathcal{O}\le(\frac{1}{\ln v}\ri),
$$
hence on the curve $$v=\frac{\beta t^{\sigma}\ln t}{4c^2t}$$ we have $$n\sim\frac{\widetilde z}{2}=\frac{\beta c t^{\sigma}}{1-\sigma}\(1+\mathcal{O}\(\frac{\ln\ln t}{\ln t}\)\).$$
\end{com}

\begin{com}
 In section \ref{sect_mesoscopic} we give an alternative proof of Corollary \ref{cor_1_q_el} (and more generally, for all integer $M$)
by constructing parametrices in terms of Laguerre polynomials for the corresponding Riemann-Hilbert problem.
\end{com}

The Corollary \ref{cor_1_q_el} does not deal with the situation when $\sigma\to 0$ or $\sigma=0,$ that corresponds to the cases when the number of solitons grows but not very fast, or when it it bounded;
for finite number of solitons Khruslov and Kotlyarov obtained the following formula:

\begin{teor}
 [\cite{KK}] \label{teorKK}
Fix an integer $N\geq1$.
Then for
$$x>4c^2t-\dsfrac{(N+\frac12)\ln t}{c}$$
the solution of the initial value problem (\ref{mkdv})-(\ref{ic}) 
with initial datum satisfying
\begin{equation}\label{first_moment_init_data}
 \int\limits_{-\infty}^0(1+|x|)|q_0(x)|\d x+\int\limits_0^{+\infty}(1+|x|)|q_0(x)|<\infty 
\end{equation}
admits the asymptotic representation:
\begin{eqnarray}\label{qsumofasymptoticsolitons}
q(x,t)&=&q_{as}(x,t)+\mathrm{O}\left(t^{-\frac{1}{2}+\varepsilon}\right),\\
q_{as}(x,t)&=&\sum\limits_{n=0}^{N-1}\dsfrac{2c}{\cosh\left(2c(x-4c^2t)+\left(2n+\dsfrac{3}{2}\right)
\ln t+\tilde\alpha_n\right)} \label{qass},
\end{eqnarray}
\begin{equation} {\label{tildealpha}}
\tilde\alpha_n=-\ln\left[\dsfrac{|h_0|}{4^{2n+1}(2c)^{6n+3}}
\Gamma(n+1)\Gamma\(n+\dsfrac{3}{2}\)\right]=-\ln\left[\dsfrac{2\Gamma(n+1)\Gamma\(n+\dsfrac{3}{2}\)}
{(h^*)^2\pi \(32c^3\)^{2n+\frac32}}\right], \end{equation}
where $|h_0|$ is related to $h^*$ (\ref{asingularity}) in the following way:
\begin{equation}\label{|h0|}
|h_0|=\dsfrac{1}{{h^*}^2\pi(2c)^{3/2}}.
\end{equation}
\end{teor}

\noindent We prove a refined version of this theorem with improved error estimates.
\begin{teor} \label{teorKK_refined}
Let $N\geq1$ be a fixed integer. Then in the domain $4c^2t\geq x\geq 4c^2t-\frac{N+\frac14}{c}\ln t$ the solution of the initial value  problem (\ref{mkdv}), (\ref{ic})
 with initial datum satisfying Assumptions \ref{Assump_no_ds}, \ref{Assump_2}, \ref{Assump_3} of section \ref{sect_prel} has the following asymptotics
$$q(x,t)=\sum\limits_{n=0}^{N-1}\frac{2c}{\cosh\(2c(x-4c^2t)+(2n+\frac32)\ln t+\alpha_n(\frac{x}{t})\)}+\mathcal{O}(t^{-1})=$$ $$=
\sum\limits_{n=0}^{N-1}\frac{2c}{\cosh\(2c(x-4c^2t)+(2n+\frac32)\ln t+\widetilde\alpha_n\)}+\mathcal{O}(\frac{\ln t}{t}),$$
where $v=1-\frac{x}{4c^2t},$ and phase $\widetilde\alpha_n$ is in (\ref{tildealpha}),
\begin{equation}\label{alpha_n}\alpha_n = \alpha_n\(\frac{x}{t}\) =\ln\frac{\pi\ (h^*)^2}{2\Gamma(n+1)\Gamma(n+\frac32)}+\left(2n+\frac32\right)\ln(16c^3(2+v)).
\end{equation}
\end{teor}
%

\noindent The method in \cite{KK} is based on the corresponding Gelfand-Levitan-Marchenko equation. In Sec. \ref{sect_as_sol} we show how to obtain this asymptotics using the
parametrices in terms of Laguerre polynomials in the corresponding Riemann-Hilbert problem. 
\vskip 13pt
\noindent
Theorems \ref{thrm_Elliptic_est}, \ref{teorKK} and Corollary \ref{cor_1_q_el} do not cover the whole  transition zone $$4c^2t-\varepsilon t\leq x\leq4c^2t,$$ where $\varepsilon>0$. Indeed  there remains an unexplored region
$$4c^2t-\beta t^{\sigma}\ln t\leq x\leq 4c^2t-\rho\ln t,$$
with fixed $\beta>0$, $0<\sigma<1$, $\rho>0.$ We study the  asymptotics in this region in section \ref{sect_mesoscopic}.

Comparing phases in the argument of $\cosh$ in (\ref{eq_cor_1}) and (\ref{tildealpha}), we see that those arguments are equal up to terms of  order $n^{-1},$
$$\alpha_{n}^{(M=1)}-\widetilde\alpha_n=\mathcal{O}\(\frac{1}{n}\).$$ 
 This suggests  that the formula (\ref{qsumofasymptoticsolitons}) is valid in a wider region; to this end we have the following theorem
\begin{teor} 
\label{thrm_as_sol_mes} 
Let $\sigma\in[0,\frac12)$,  $\beta\in(\frac{1}{K},K),$ $K>0,$ and suppose that  $(x,t)$ lies on the curve $x=4c^2t-\beta t^{\sigma} \ln t.$ Let $\widetilde \eta,$ $\widetilde z$ and $n$
be as in Corollary \ref{cor_1_q_el} with $M=1.$ Then, as $t\to\infty$
\begin{equation}\label{KKMesoscopic}q(x,t)=\frac{2c}{\cosh\(2c(x-4c^2t)+(2n+\frac32)\ln t+\widetilde\alpha_n \)} +\mathcal{O}(t^{2\sigma-1}\ln^2 t)	,
\end{equation}
where $\widetilde\alpha_n$ is given  in (\ref{tildealpha}), and the $\mathcal{O}$ estimate is uniform when $\sigma,$ $\beta$ change in compact domains of their domain, i.e. for $0\leq\sigma\leq\sigma_0,$ $\frac{1}{K_0}\leq\beta\leq K_0.$
\end{teor}

\begin{com}
 Theorem \ref{thrm_as_sol_mes} covers the regime in which the point $(x,t)$ travels along solitons with slowly growing number $\asymp t^{\sigma}, $ $\sigma<<1.$ 
\end{com}
\begin{com}
 We leave open the question about direct verification of consistency of results of Theorem \ref{thrm_Elliptic_est}, Corollary \ref{cor_1_q_el} and results which can be obtained with the help of method of chapter \ref{sect_mesoscopic}, 
formula (\ref{q_mesoscopic}), for the regime $$x=4c^2t-\beta t^{\sigma}\ln t$$ when $\sigma$ is in the range
 $$\sigma\in\left[\frac{M}{M+1},\frac{M+1}{M+2}\right),\quad \textrm{ for }\quad M\geq1.$$ While this might be checked for several first $M=1,2,3,$ to verify this for a general $M$
requires deeper understanding of relation between $z$ (\ref{z_tBDelta}) and $\gamma$ (\ref{gamma_eq_mes}).
\end{com}

\section{Preliminaries}\label{sect_prel}

 In this section we collect some well-known results. The
MKdV equation \eqref{mkdv} admits a Lax pair representation in the form \cite{Wadati}, \cite{Shabat}
\begin{eqnarray}\label{x-eq}
&&\Phi_x(x,t;k)+\i k\sigma_3\Phi(x,t;k)=Q(x,t)\Phi(x,t;k),\\\label{t-eq}
&&\Phi_t(x,t;k)+4\i k^3\sigma_3\Phi(x,t;k)=\widehat Q(x,t;k)\Phi(x,t;k),
\end{eqnarray}
where \begin{equation}\label{Q_Qhat}
       Q(x,t)=\begin{pmatrix}
          0 & q(x,t) \\ -q(x,t) & 0
         \end{pmatrix},\quad \widehat Q(x,t;k)=4k^2Q-2\i k(Q^2+Q_x)\sigma_3+2Q^3-Q_{xx},\quad \sigma_3=\begin{pmatrix}1&0\\0&-1\end{pmatrix}.
      \end{equation}
If we substitute $q(x,t)=c$ (constant)  in \eqref{Q_Qhat}, then the Lax pair equations \eqref{x-eq}, \eqref{t-eq} admit  explicit solutions  \cite{KM}
\begin{equation}\nonumber
 E_0(x,t;k)=\e^{-(i k x+4ik^3t)\sigma_3},\quad\textrm {and }\qquad E_c(x,t;k)=K(k)\e^{-(\i x X(k)+2\i t (2k^2-c^2)X(k))\sigma_3}, 
\end{equation}
for $c=0$ and $c\neq 0$, respectively. In the formula above we have set $X(k)=\sqrt{k^2+c^2}$ and $$K(k)=\begin{pmatrix}a_{\gamma}(k)&b_{\gamma}(k)\\b_{\gamma}(k)&a_{\gamma}(k)\end{pmatrix},\quad a_{\gamma}(k)=\frac12(\gamma(k)+\gamma^{-1}(k)),\ b_{\gamma}(k)=\frac12(\gamma(k)-\gamma^{-1}(k)),\ \gamma(k)=\sqrt[4]{\frac{k-\i c}{k+\i c}}.$$

\begin{prop} \label{prop_Jakovleva_1}\textrm{(\cite{Jakovleva})}
Let $q_0(x)$ satisfies \eqref{first_moment_init_data} and \begin{equation}\label{q_0_limits_boundedness}\sup\limits_{x\in\mathbb{R}}|q_0(x)|<\infty,\quad \lim\limits_{x\to+\infty}q_0(x)=0,\quad \lim\limits_{x\to-\infty}q_0(x)=c>0.
    \end{equation}
Denote $$\sigma_{l}(x)=\int\limits_{-\infty}^x|q_0(\widetilde x)-c|\d\widetilde x,\quad \sigma_{r}(x)=\int\limits^{+\infty}_x|q_0(\widetilde x)|\d\widetilde x,
\quad\sigma_{l,1}(x)=\int\limits_{-\infty}^x\sigma_l(\widetilde x)\d\widetilde x, \quad\sigma_{r,1}(x)=\int\limits_{-\infty}^x\sigma_r(\widetilde x)\d\widetilde x$$
(these quantities exist under the above condition \eqref{first_moment_init_data}), 
then there exist {\rm Jost solutions} of the equation \eqref{x-eq} that possess the following integral representations
\begin{equation}
\label{integral_representation_Jost_solution}
\Psi_l(x,k)=E_c(x,k)+\int\limits_{-\infty}^{x}L_l(x,y)E_c(y,k)\d y,\quad \Psi_r(x,k)=E_0(x,k)+\int\limits^{+\infty}_{x}L_r(x,y)E_0(y,k)\d y
\end{equation}
with kernels
$$L_l(x,y)=\begin{pmatrix}
            L_{l,1}(x,y) & -L_{l,2}(x,y)\\ L_{l,2}(x,y) & L_{l,1}(x,y)
           \end{pmatrix},\quad 
L_r(x,y)=\begin{pmatrix}
            L_{r,1}(x,y) & -L_{r,2}(x,y)\\ L_{r,2}(x,y) & L_{r,1}(x,y)
           \end{pmatrix},
$$
which satisfy the following estimates:
\begin{eqnarray}\begin{cases}
|L_{l,1}(x,y)|\leq\frac12\sigma_{l}(\frac{x+y}{2})M_{2,l} M_{1,l},\\
|L_{l,2}(x,y)|\leq\frac{|q_0(\frac{x+y}{2})-c|}{2}+\sigma_{l}(\frac{x+y}{2})\frac{\(\sigma_{l}(x)-\sigma_{l}(\frac{x+y}{2})\)}{2}M_{2,l}M_{1,l}
,\\
|L_{r,1}(x,y)|\leq\frac12\sigma_{r}(\frac{x+y}{2})M_{2,r}M_{1,r},\\
|L_{r,2}(x,y)|\leq\frac{|q_0(\frac{x+y}{2})|}{2}+\sigma_{r}(\frac{x+y}{2})\frac{\(\sigma_{r}(x)-\sigma_{r}(\frac{x+y}{2})\)}{2}M_{2,r}M_{1,r},
\label{estimates_Jost _kernel}
\end{cases}\end{eqnarray}
where we denoted for brevity
\begin{equation}\label{M1M2M3}
\begin{cases}
 M_{1,l}=\exp\left[\max\limits_{z\in(-\infty,x)}|q_0(z)+c|\cdot\(\sigma_{l,1}(x)-\sigma_{l,1}(\frac{x+y}{2})\)\right],\qquad 
M_{2,l}=\max\limits_{z\in(-\infty,x)}|q_0(z)+c|,
\\ M_{1,r}=\exp\left[\max\limits_{z\in(x,+\infty)}|q_0(z)|\cdot\(\sigma_{r,1}(x)-\sigma_{r,1}(\frac{x+y}{2})\)\right],\qquad 
M_{2,r}=\max\limits_{z\in(x,+\infty)}|q_0(z)|.
\end{cases}
\end{equation}

\end{prop}
{\it Proof.} For convenience of the reader we will sketch the proof for $L_{l}.$ Substituting the first part of \eqref{integral_representation_Jost_solution} into \eqref{x-eq}, we come to the system 
$$\begin{cases}
 L_{l1,x}(x,y)+L_{l1,y}(x,y)=(q_0(x)+c)L_{l2}(x,y),\quad L_{l2,x}(x,y)-L_{l2,y}(x,y)=-(q_0(x)-c)L_{l1}(x,y),\quad y<x,
\\
L_{l,2}(x,x)=\frac{-(q_0(x)-c)}{2},\quad \lim\limits_{y\to-\infty}L_{l,2}(x,y)=0.
\end{cases}$$
Making the change of variables 
$$u=\frac{x+y}{2},\quad v=\frac{x-y}{2},\qquad H_{1}(u,v)\equiv L_{l,1}(x,y),\quad H_{2}(u,v)\equiv L_{l,2}(x,y),$$
we come to the integral equations
\begin{equation}\label{int_eq_1}H_{1}(u,v)=-\int\limits_{-\infty}^u(q_0(\widetilde u+v)+c)H_2(\widetilde u,v)\d \widetilde u,\qquad 
H_{2}(u,v)=\frac{-(q_0(u)-c)}{2}-\int\limits_{0}^v(q_0(u+\widetilde v)-c)H_1(u, \widetilde v)\d \widetilde v,\end{equation}
or 
\begin{equation}\label{int_eq_2}H_1(u,v)=\frac12\int\limits_{-\infty}^u(q_0(\widetilde u+v)+c)(q_0(\widetilde u)-c)\d\widetilde u+\int\limits_{-\infty}^u(q_0(\widetilde u+v)+c)\int\limits_{0}^v(q_0(\widetilde u+\widetilde v)-c)
H_1(\widetilde u,\widetilde v)\d\widetilde v\d \widetilde u,\end{equation}
\begin{equation}\label{int_eq_3}H_2(u,v)=\frac{-(q_0(u)-c)}{2}+\int\limits_{0}^v(q_0(u+\widetilde v)-c)\int\limits_{-\infty}^u(q_0(\widetilde u+\widetilde v)+c)
H_2(\widetilde u,\widetilde v)\d\widetilde u\d\widetilde v.\end{equation}
Applying the method of successive approximations (the integral operator becomes a contraction for $u$ sufficiently large) to the above integral equations, we come to the statement of the proposition.
\hfill $\blacksquare$\\[1pt] 
\begin{prop}\label{prop_Jakovleva_2}(\cite{Jakovleva})
 Let initial function $q_0(x)$ satisfies \eqref{first_moment_init_data} and also 
\begin{equation}\label{initial_funct_deriv_bounded}\sup\limits_{x\in\mathbb{R}}|q_0'(x)|<\infty,\quad \lim\limits_{x\to\pm\infty}q_0'(x)=0,\quad \int\limits_{-\infty}^{+\infty}|q_0'(x)|\d x<\infty.
\end{equation}
Denote $$\sigma_{2,l}(x)=\int\limits_{-\infty}^x|q_0'(\widetilde x)|\d \widetilde x, \qquad \sigma_{2,r}(x)=\int\limits^{+\infty}_x|q_0'(\widetilde x)|\d \widetilde x.$$
Then the partial derivatives of $L_{l},$ $L_{r}$ satisfy the following estimates:
$$|L_{1l,y}|\leq \frac{\sigma_l(\frac{x+y}{2})}{4}\left[M_{3,l}+M_{1,l}M_{2,l}M_{3,l}\(\sigma_{l,1}(x)-\sigma_{l,1}(\frac{x+y}{2})\)+\(2\sigma_{l}(x)-\sigma_{l}(\frac{x+y}{2})\)M_{2,l}^2M_{1,l}\right]
+$$
$$+\frac{|q_0(\frac{x+y}{2})-c|}{4}|q_0(x)+c|,$$
$$|L_{2l,y}|\leq\frac{|q_0'(\frac{x+y}{2})|}{4}+\frac{\sigma_{l}(\frac{x+y}{2})\(\sigma_{l,2}(x)-\sigma_{l,2}(\frac{x+y}{2})\)}{4}M_{2,l}M_{1,l}\(1+\(\sigma_{l}(x)-\sigma_{l}(\frac{x+y}{2})\)M_{2,l}\)+$$
$$+\frac{|q_0(\frac{x+y}{2})-c|(\sigma_{l}(x)-\sigma_{l}(\frac{x+y}{2}))}{4}M_{2,l}+\frac{\sigma_{l}(\frac{x+y}{2})|q_0(x)-c|}{4}M_{2,l}M_{1,l}.$$
Similar estimates are valid for $L_{r},$ we just need to replace index $l$ with $r$, $c$ with $0$, and $-\infty$ with $+\infty$ in the above expressions.
Here $M_1,M_2$ are defined in \eqref{M1M2M3}, and 
$$M_{3,l}=\max\limits_{z\leq x}|q_0'(z)|,\quad M_{3,r}=\max\limits_{z\geq x}|q_0'(z)|$$
\end{prop}
{\it Proof.}  The statements of the lemma follows directly from analysis of integral equations \eqref{int_eq_1}, \eqref{int_eq_2}, \eqref{int_eq_3} and Proposition \ref{prop_Jakovleva_1}.
\hfill $\blacksquare$\\[1pt] 
\begin{rem}
 Conditions \eqref{q_0_limits_boundedness} follow from \eqref{initial_funct_deriv_bounded}.
\end{rem}

\begin{cor}\label{cor_analyticity_Jost}\begin{enumerate}[(a)]
            \item 
Provided that the conditions \eqref{first_moment_init_data} are satisfied, the first column $\Psi_{l,1}$ is analytic in $\Im\sqrt{k^2+c^2}>0$, the second column $\Psi_{l,2}$ is analytic in $\Im\sqrt{k^2+c^2}<0,$ and the first column of the right Jost solution $\Psi_{r,1}$
is analytic in $\Im k<0,$ and the second column $\Psi_{r,2}$ is analytic in $\Im k>0.$
\item Suppose that in addition to \eqref{first_moment_init_data} the following condition is satisfied:
 \begin{equation}
 \label{exp_decreasing}\int\limits_{-\infty}^0|q_0(x)-c|\e^{2|x|\sqrt{l_0^2-c^2}}\d
x+\int\limits_{0}^{+\infty}|q_0(x)|\e^{2xl_0}\d
x<\infty,\qquad \textrm{ for some }\quad  l_0>c>0.
\end{equation} 
Then the Jost solutions $\Psi_{l},$ $\Psi_{r}$ are analytic in a $l_0-c$ neighborhood of the contour $k\in\Sigma=\mathbb{R}\cup[\i c,-\i c].$
\end{enumerate}
\end{cor}
{\it Proof.} The first part of the corollary follows from the fact that under \eqref{first_moment_init_data} the kernels in \eqref{integral_representation_Jost_solution} are summable:
$$ \int\limits_{-\infty}^x|L_l(x,y)|\d y<\infty,\quad \quad \int\limits^{+\infty}_x|L_r(x,y)|\d y<\infty.$$
The second statement of the corollary follows directly from the estimates \eqref{estimates_Jost _kernel} and the fact that under \eqref{exp_decreasing} 
$$\int\limits_{-\infty}^0 \e^{2|y| \sqrt{l_0^2-c^2}}\sigma_l(y)\d y<\infty,
\qquad\int\limits^{+\infty}_0 \e^{2y l_0}\sigma_r(y)\d y<\infty.$$
The first of the last estimates gives us that $\Psi_{l}$ is analytic in $0<|\Im\sqrt{k^2+c^2}|<\sqrt{l_0^2-c^2},$ which includes $l_0-c$ neighborhood of $\Sigma,$
and the second estimate gives us analyticity of $\Psi_{r}$ in $|\Im k|<l_0,$ which also includes $l_0-c$ neighborhood of $\Sigma.$
\hfill $\blacksquare$\\[1pt]

\noindent
Given the Jost solutions of \eqref{x-eq} for $t=0,$ we define the transition matrix $T(k)$ by 
\begin{equation}\nonumber
 \Psi_{l}(x,k)=\Psi_{r}(x,k)T(k).
\end{equation}
{\color{black} The {\it Jost solution}  have the (defining) property that $\Psi_l$ is asymptotic to $E_c$ as $x\to-\infty$ and $\Psi_r$ is asymptotic to $E_0$ as $x\to+\infty$ (whence the subscripts, ``l(eft)''  and ``r(ight)'').}
Due to symmetries of the $x$-equation \eqref{x-eq} the transition matrix has the following structure:
\begin{equation}\label{T_a_b}T(k)=\begin{pmatrix}
        a(k)& -\ol{b(\ol{k})} \\ b(k) & \ol{a(\ol{k})}
       \end{pmatrix},\qquad \textrm{ where } a(k)=\det\(\Psi_{l1},\Psi_{r2}\),\ b(k)=\det\(\Psi_{r1},\Psi_{l1}\).
\end{equation}
\noindent
The functions $a^{-1}(k)$ and $r(k):=\frac{b(k)}{a(k)}$ are called the transmission and reflection coefficients, respectively.
Under the condition  \eqref{first_moment_init_data},
they have the following properties:

\begin{lem}\label{lem_prop_a_r}
\begin{enumerate}
\item Under conditions \eqref{first_moment_init_data}, \eqref{initial_funct_deriv_bounded}, the function $a(k)$ is analytic in $\left\{k:\  0<\Im k\right\} \setminus[0,\i c]$, and it can be extended continuously up to the boundary
with the exception of the point $\i c,$ where $a(k)$ may have at
most a root singularity of the order $(k-\i c)^{-1/4};$ further, $$a(k)=1+\mathcal{O}(k^{-1}) \quad \textrm{ as }\quad k\to\infty;$$
function $r(k)$ is defined in $k\in\mathbb{R}$ except for points where $a(k)=0$ and $$r(k)=\mathcal{O}(k^{-1}).$$

\item $a(k)$, $r(k)$ satisfy the symmetry condition
\begin{equation}\label{asymmetry}\ol{a(-\ol{k})}=a(k),\quad \ol{r(-\ol{k})}=r(k);\end{equation}

\item if we assume also \eqref{exp_decreasing}, then $r(k)$ is meromorphic in a $l-c_0$ neighborhood of $\Sigma=\mathbb{R}\cup[\i c,-\i c]$ with poles at zeros of $a(k).$ In the specified domain 
$r(k)$ has the following asymptotics:
$$r(k)=\mathcal{O}(k^{-1}) \quad \textrm{ as }\quad k\to\infty.$$


\end{enumerate}
\end{lem}
{\it Proof.}  The first two items have been proved in \cite{Jakovleva}. The third statement follows from Corollary \ref{cor_analyticity_Jost}, (b) and estimates in Proposition \ref{prop_Jakovleva_2},
which are used after integrating by parts in \eqref{integral_representation_Jost_solution}.
\hfill $\blacksquare$\\[1pt] 

\begin{Assump}\label{Assump_no_ds}
For simplicity, we suppose the absence of usual solitons generated by the discrete spectrum:
$$\forall
k\in\mathbb{C}_+\backslash[\i c,0] \ \ \ a(k)\neq0.$$
\end{Assump}


\noindent We have the following
\begin{lem}\label{lem_a_b} The spectral coefficients $a(k), b(k)$ do not vanish on the segment $(-\i c,\i
c)$.
\end{lem}
{\it Proof.} 
Using the relations (they follow from the symmetry and determinantal properties of the corresponding Jost solutions, see \cite{KM}, chapter II, p.4-5)
\begin{equation}
a(k)\ol{a(\ol{k})}+b(k)\ol{b(\ol{k})}=1,\quad
\ol{a(-\ol{k})}=a(k), \quad \ol{b(-\ol{k})}=b(k), \quad a_-(k)=-\i
\ol{b_+(\ol{k})},
\label{extrasym}
\end{equation}
we get that $a_{\pm}(k)\neq 0, b_{\pm}(k)\neq 0$ on $(\i c,-\i c).$
\hfill $\blacksquare$\\[1pt] 

\begin{Assump}\label{Assump_2}
From the definition of $a(k)$ \eqref{T_a_b} and the integral representations for Jost solutions \eqref{integral_representation_Jost_solution} it follows that generically expansion of $a(k)$ in a vicinity of the point $k=\i c$ 
starts from the term $(k-\i c)^{-1/4}$ and then continues with $(k-\i c)^{1/4},$ $(k-\i c)^{3/4},..$. However, it might happen that the coefficient in front of the term $(k-\i c)^{-1/4}$ vanishes for some particular initial data.
Here we suppose the generic situation, i.e. 
\begin{equation}
\label{asingularity}a(k)=\dsfrac{h^*}{2}\sqrt[4]{\dsfrac{2\i c}{k-\i
c}}\(1+\mathrm{O}\(\sqrt{\dsfrac{k-\i c}{\i}}\)\),\ k\to\i
c,\qquad \qquad h^*\in\mathbb{R}\backslash
\left\{0\right\}.
\end{equation}
\noindent The fact that $h^*\in\mathbb{R}$ follows from the symmetry 
(\ref{asymmetry}).
\end{Assump}


\begin{Assump}\label{Assump_3} We will also assume that the reflection coefficient admit an analytic continuation to a $\delta-$neighborhood of $\mathbb{R}\cup[\i c,0]$ with some $\delta>0.$
\end{Assump}
This assumption is satisfied for example in each of the following cases:
\begin{enumerate} 
 \item the initial function is smooth \eqref{initial_funct_deriv_bounded} and tends to its background limits 
 exponentially fast \eqref{exp_decreasing}. Lemma \ref{lem_prop_a_r} than ensures analyticity of $r(k)$ in a neighbourhood of contour $\Sigma$ under assumption \ref{Assump_no_ds}
of absence of discrete spectrum. 
Here $l_0$ satisfies the inequality $l_0>c+\delta>0.$
\item the initial data is of the form {\color{blue} \eqref{pure_step_ic}.}
In this case the inverse of the transmission coefficient
$a(k)=\widetilde{a}(k)$ and the reflection coefficient $r(k)=\widetilde{r}(k)$ can be computed explicitly and have the
following form:
$$\widetilde{a}(k)=\dsfrac{1}{2}\(\sqrt[4]{\dsfrac{k-\i c}{k+\i
c}}+\sqrt[4]{\dsfrac{k+\i c}{k-\i c}}\),\quad \widetilde{r}(k)=\frac{\(\sqrt[4]{\frac{k-\i c}{k+\i
c}}-\sqrt[4]{\frac{k+\i c}{k-\i c}}\)}{\(\sqrt[4]{\frac{k-\i c}{k+\i
c}}+\sqrt[4]{\frac{k+\i c}{k-\i c}}\)},$$ where the cut is taken
along the segment $[-\i c,\i c]$ and the branch of the root is
taken such that $\widetilde a(k)$ tends to $1$ as $k\rightarrow\infty.$
The constant $h^{*}$ in \eqref{asingularity} is $h^*=1.$
\end{enumerate}



\begin{com} 
We should notice that the "pure" step initial function (\ref{pure_step_ic}) is not in the class 
\begin{equation}\nonumber\int\limits_{-\infty}^0|x||q(x,t)-c|\d
x+\int\limits_{0}^{+\infty}x|q(x,t)|\d x<\infty,
\end{equation} since
$q(x,t)=c+\mathcal{O}(|x|^{-\frac34})$ as $x\to-\infty$ for $t>0$; nevertheless,  due to the analyticity of $\widetilde{a}(k)$, $\widetilde{r}(k),$ the asymptotic analysis
in this case can be done in the same manner as in the case of smooth
and fast decreasing  initial functions.
\end{com}

The solution of the initial value problem (\ref{mkdv}), (\ref{ic}) can be reconstructed from the solution of the following Riemann-Hilbert problem (\cite{KM}):
\begin{RHP}\label{RH_problem_1}\begin{enumerate} find a $2\times 2$ matrix-valued function $M(\xi,t;k)$ such that
 \item analyticity: $M(\xi,t;k)$ is analytic in $\mathbb{C}\setminus\Sigma,$ and continuous up to the boundary. Here the oriented contour $\Sigma$ is $\Sigma=\mathbb{R}\cup[\i c,-\i c].$
 \item jump: $M_-(\xi,t;k)=M_+(\xi,t;k)J(\xi,t;k),$ where
 $$J(\xi,t;k)=\begin{pmatrix}1 & -\overline{r(k)}\e^{-2it\theta(k,\xi)}\\-r(k)\e^{2\i t\theta(k,\xi)} & 1+|r(k)|^2\end{pmatrix},\quad k\in\mathbb{R}\setminus\left\{0\right\},$$
$$J(\xi,t;k)=\begin{pmatrix}1&0\\f(k)\e^{2\i t\theta(k,\xi)} & 1\end{pmatrix},\quad k\in(\i c,0),$$
$$J(\xi,t;k)=\begin{pmatrix}1&-\overline{f(\ol{k})}\e^{-2\i t\theta(k,\xi)} \\0& 1\end{pmatrix},\quad k\in(0,-\i c),$$
where $r(k)$ is the reflection coefficient associated with spectral problem associated with MKdV, and
 \begin{equation}
 \label{f_funct} f(k)=r_-(k)-r_+(k)=\frac{\i}{a_-(k)a_+(k)},\quad k\in(\i c,-\i c)
 \end{equation} 
 where $a^{-1}(k)$ is the transmission coefficient.
\item asymptotics: $M(x,t;k)\to I$ as $k\to\infty.$
\end{enumerate}
\end{RHP}

The jump matrices in the RH problem \eqref{RH_problem_1} satisfy the  Schwartz symmetry $J^{-1}(k)=\(\ol{J^{T}(\ol{k})}\)$ for $k\in\Sigma\setminus\mathbb{R}.$ Hence, it follows from the vanishing lemma \cite{Zhou}
that the solution of the RH problem \eqref{RH_problem_1} exists. Further, from analyticity of the reflection coefficient it follows that we can deform the contour $\Sigma$ into such a contour that
the corresponding singular integral equation, which is equivalent to the RHP \ref{RH_problem_1}, admits $x$ and $t$ differentiation. Then, in the spirit of the well-known result of Zakharov -- Shabat \cite{Zakharov-Shabat}, one can prove that 
the solution of the initial value problem (\ref{mkdv}), (\ref{ic}) can be reconstructed by the following formula (see \cite{M_disser}, chapter 2 for details):
$$q(x,t)=\lim\limits_{k\to\infty}\(2\i k M(x,t;k)\)_{21}=\lim\limits_{k\to\infty}\(2\i k M(x,t;k)\)_{12}.$$

\section{Asymptotics in the domain $D_2$}\label{sect_ell}
\begin{figure}[t]
\begin{center}
\includegraphics[width=100mm]{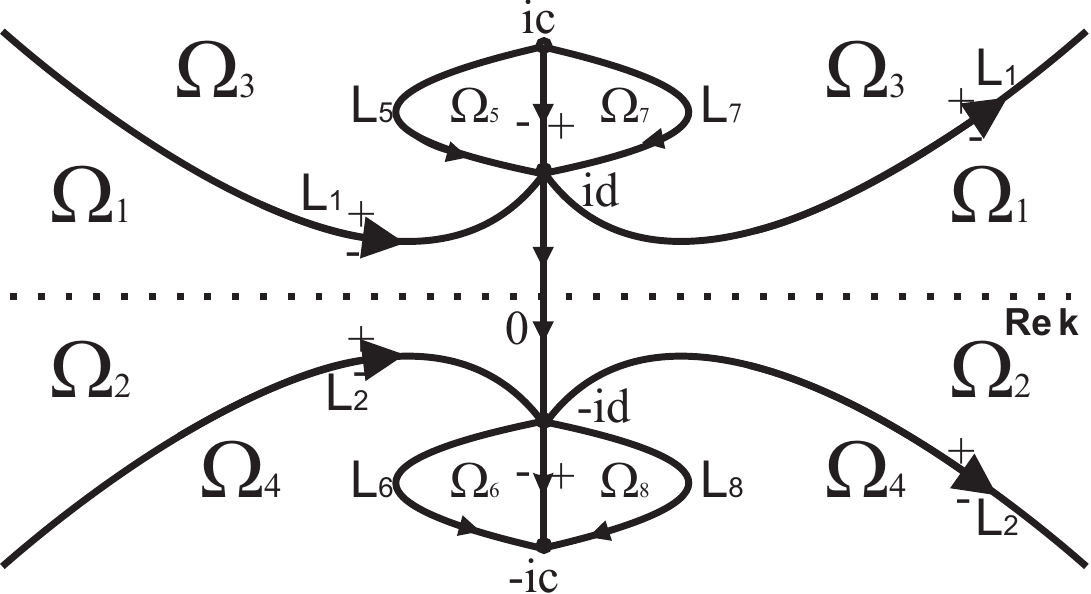}
\end{center}
\caption{Contour $\Sigma^{(3)}$ for the RH problem for
$M^{(3)}(\xi,t,k)$} \label{Contourelliptic}
\end{figure}

%
%
{
The region $D_2$ corresponds to  
$\frac{-c^2}{2}+\varepsilon<\xi<\frac{c^2}{3}-\frac{\beta
t^{\sigma}\ln t}{t}$, $\sigma\in(0,1)$, $\varepsilon>0.$}
The main result in this section is Theorem \ref{thrm_Elliptic_est}.
It was shown in \cite{KM} that the RH problem \ref{RH_problem_1}  is equivalent to the following one: 
\begin{RHP}\label{RH_problem_3}
Find  a  matrix-valued function
$M^{(3)}(\xi,t,k)$  such that 
\begin{enumerate}
\item  {\rm analyticity}: it is analytic in
$k\in\mathbb{C}\backslash\Sigma^{(3)}$ (contour $\Sigma^{(3)}$ is shown in Figure \ref{Contourelliptic}), 
\item {\rm asymptotics}; we have $M^{(3)} (x,t;k) \to I$ as $k\to \infty$ and near the points $\pm \i c$ we have  that every entry is bounded by $\mathcal O( |z\mp \i c|^{-\frac 1 4})$ (respectively).
\item {\rm jump}: the boundary values satisfy the following jump conditions on the
contour $\Sigma^{(3)}$ 
\begin{equation}
\label{RHM3beg}
M^{(3)}_-(\xi,t,k)=M^{(3)}_+(\xi,t,k)J^{(3)}(\xi,t,k), \qquad
M^{(3)}(\xi,t,k)\rightarrow I,\quad k\rightarrow\infty.
\end{equation}
The jump matrix is given by the formula
\begin{align*}
J^{(3)}(\xi,t,k)&=\left(\begin{array}{ccc}1&0\\
{-r(k){F^{-2}(k, \xi)}e^{2itg(k,\xi)}}&1
\end{array}\right),& k\in L_1 \\
&=\left(\begin{array}{ccc}1& {- \ol{r( \ol{k} )} F^{2}(k, \xi)}
{e^{-2itg(k,\xi)}}\\0&1
\end{array}\right),& k\in L_2
\end{align*}
\begin{align*}
J^{(3)}(\xi,t,k)& =\left(\begin{array}{ccc}1&
{\widehat{f}^{-1}(k)}{F^2(k, \xi)
e^{-2itg(k,\xi)}}\\0&1\end{array}\right), &k\in
L_7,\\
&=\left(\begin{array}{ccc}1& -{\widehat{f}^{-1}(k)}{ F^2(k,\xi)
e^{-2itg(k,\xi)}}\\0&1\end{array}\right),& k\in L_5
\end{align*}
\begin{align}\nonumber
J^{(3)}(\xi,t,k)
\nonumber&=\left(\begin{array}{ccc}1&0\\{-\ol{\widehat{f}^{-1}(\ol{k})}F^{-2}(k,\xi)}{e^{2itg(k,\xi)}}&1\end{array}\right),\ k\in L_8,\\
\nonumber&=\left(\begin{array}{ccc}1&0\\\ol{\widehat{f}^{-1}(\ol{k})}F^{-2}(k,\xi){e^{2itg(k,\xi)}} &1\end{array}\right),\ k\in L_6\\
\nonumber&=\begin{pmatrix}e^{itB_g(\xi)+\ii\Delta(\xi)}&0\\0&
e^{-itB_g(\xi)-\ii\Delta(\xi)}\end{pmatrix},\
k\in(-\ii d,\ii d)\\
&=\left(\begin{array}{ccc}0&\i\\\i&0\end{array}\right),\ k\in(\ii
c,\ii d)\bigcup(-\ii d,-\ii c).\label{RHM3end}
\end{align}
\end{enumerate}
\end{RHP}

The quantities $B$, $\Delta$, $\tau$ in the formulation of the RHP \ref{RH_problem_3}  are defined by (\ref{Bg})-(\ref{Delta}): the parameter $d$ is determined by (\ref{dmu1})-(\ref{dmu2}). The function $F$ is determined by a  scalar
boundary value 
problem:
\begin{enumerate}
 \item analyticity: $F$ is analytic in $\mathbb{C}\setminus[i c,-i c];$
 \item 
  jump 
   conditions: $$F_-F_+=-\i f,\ k\in[\i c,\i d],\quad
 F_-F_+=\frac{-\i}{\ol{f(\ol{k})}},\ k\in[-\i d,-\i c],
 \quad \frac{F_+}{F_-}=\e^{\i\Delta},\ k\in[\i d,-\i d];$$
 \item asymptotics: $F\to 1$ as $k\to\infty.$
\end{enumerate}
The function $\widehat f$ is the continuation of the function $f$ (\ref{f_funct}) from the segment $[\i c,-\i c],$
 using the last of the relations \eqref{extrasym}
 $$
 \widehat f(k)=\frac{-1}{a(k)\ol{b(\ol{k})}}, \ k\not\in [\i c, -\i c]\qquad
\widehat f_+=-\widehat f_-=f,\ k\in[\i c,-\i c].
$$
The function $g$ appearing in \eqref{RHM3end} is analytic in $k\in\mathbb{C}\setminus[\i c,-\i c]$ and is given by 
$$
g(k,\xi)=12\int\limits_{\i c}^{k}\frac{(s^2+\mu^2(\xi))\sqrt{s^2+d^2(\xi)}\ \d s}{\sqrt{s^2+c^2}}.
$$
 where the path of integration is arbitrarily chosen (notice that the integral on a loop surrounding the segment $[\i c,-\i c]$ is zero because the integrand is an even function, with the cuts of the integrand being $[\i d,\i c]\cup[-\i d, -\i c]$).

\begin{figure}[t]
\begin{center}
\includegraphics[width=100mm]{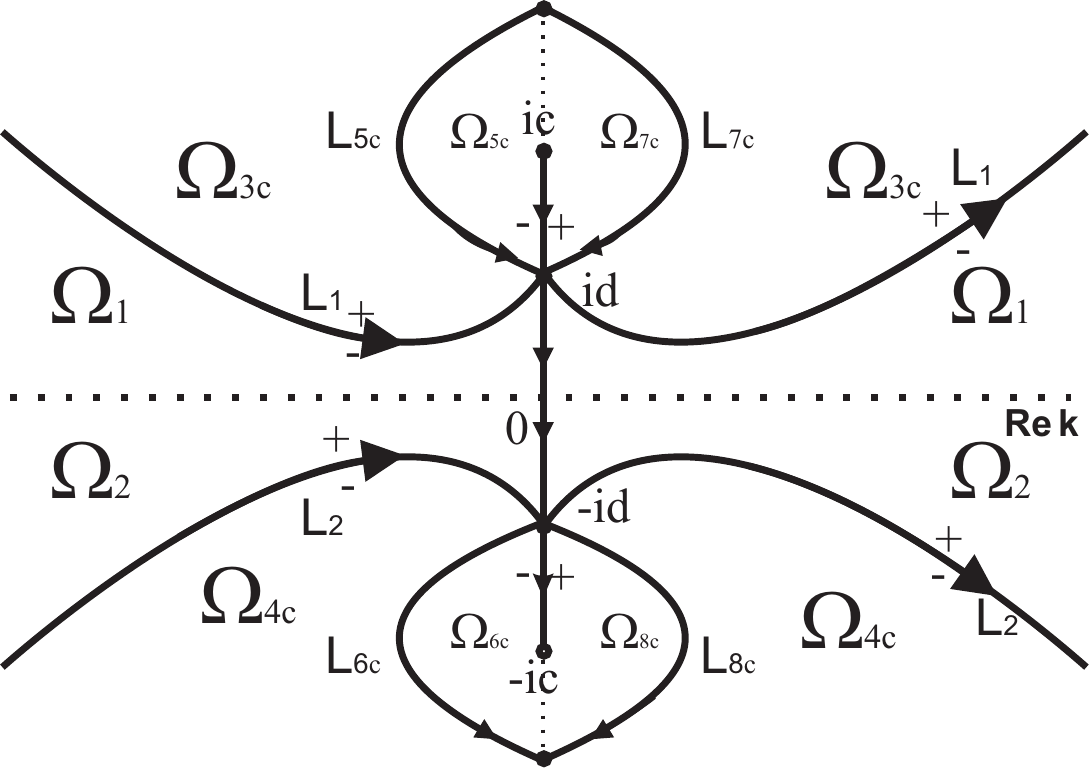}
\end{center}
\caption{Contour $\Sigma^{(3)}$ for the RH problem for
$M^{(3c)}(\xi,t,k)$} \label{Contourelliptic_c}
\end{figure}

First we "move" the lenses $L_7, L_5$, $L_8,L_6$ in such a way,
that they envelope the points $\pm ic,$ respectively (see Figure \ref{Contourelliptic_c}). Further, we fold
contours $L_1, L_2$, to the intervals $(\i d,\i (d-\widetilde
r))$, $(-\i (d-\widetilde r),-\i d)$, respectively. (Here
$\widetilde r=(c^2-d^2) r$, and $r>0$ is a fixed sufficiently small constant.)

This is achieved by the transformation
\[M^{(3c)}(\xi,t,k)=M^{(3)}(\xi,t,k)G^{(3c)}(\xi,t,k),\]
where
\[G^{(3c)}(\xi,t,k)=\left(\begin{array}{ccc}1&
{\widehat{f}^{-1}(k)}{F^2(k, \xi)
e^{-2itg(k,\xi)}}\\0&1\end{array}\right),\quad
k\in\(\Omega_{7c}\backslash\Omega_7\)\cup\(\Omega_{5c}\backslash\Omega_5\),\]

\[G^{(3c)}(\xi,t,k)=\left(\begin{array}{ccc}1&0\\{-\ol{\widehat{f}^{-1}(\ol{k})}F^{-2}(k,\xi)}{e^{2itg(k,\xi)}}&1\end{array}\right),
\quad
k\in\(\Omega_{8c}\backslash\Omega_8\)\cup\(\Omega_{6c}\backslash\Omega_6\).\]
\[G^{(3c)}=\left(\begin{array}{ccc}1&0\\
{r(k){F^{-2}(k, \xi)}e^{2itg(k,\xi)}}&1
\end{array}\right), k\in \Omega_{1}\setminus\Omega_{1c},\]
\[G^{(3c)}=\left(\begin{array}{ccc}1& {- \ol{r( \ol{k} )} F^{2}(k, \xi)}
{e^{-2itg(k,\xi)}}\\0&1
\end{array}\right), k\in \Omega_{2}\setminus\Omega_{2c},\]

This transformation modifies  the matrices in the jump relations on the intervals $(\i
d,\i (d-\widetilde r)),$ $(-\i (d-\widetilde r),-\i d),$ namely
\[J^{(3c)}=\begin{pmatrix}\e^{\i tB+\i\Delta}&0\\\frac{f}{F_+F_-}\e^{\i t(g_-+g_+)}&\e^{-\i tB-\i\Delta}\end{pmatrix},
k\in(\i d,\i (d-\widetilde r)),\]
\[J^{(3c)}=\begin{pmatrix}\e^{\i tB+\i\Delta}&-\overline{f(\overline{k})}F_+F_-\e^{-\i t(g_-+g_+)}\\0&\e^{-\i tB-\i\Delta}\end{pmatrix},
k\in(-\i (d-\widetilde r),-\i d),\] Under this transformation
lines $L_7$, $L_5$ "move" up, so they do not intersect the segment
$\(\i c,-\i c\)$, and the lines $L_8$, $L_6$ "moves" down. We have
no additional jump across $(\i c, \i (c+\delta))\cup\(-\i
(c+\delta),-\i c\).$ The matrix  $M^{(3c)}(\xi,t,k)$ still retains the same  singularities
near the points $\pm \i c$ as $M^{(3)}$.

The phase function $g(k,\xi)$ has the
following behaviour near the points $\i d(\xi)$, $-\i d(\xi): $
\[\hskip-0cm g_{\pm}(k,\xi)=\mp\dsfrac{B}{2}+\i
8(d^2-\mu^2)\sqrt{2d}(c^2-d^2)\(\dsfrac{k-\i
d}{-\i(c^2-d^2)}\)^{\frac32}\times\]
\[\times\(1-\frac{3(9c^2d^2-5d^4-3d^2\mu^2-c^2\mu^2)}{20d(d^2-\mu^2)}\(\frac{k-\i
d}{-\i(c^2-d^2)}\)+\mathcal{O}\(\frac{k-\i
d}{-i(c^2-d^2)}\)^2\),\]
\[\hskip12cm k\rightarrow\i d(\xi),
\]

\[\hskip-0cm g_{\pm}(k,\xi)=\mp\dsfrac{B}{2}-\i
8(d^2-\mu^2)\sqrt{2d}(c^2-d^2)\(\dsfrac{k+\i
d}{\i(c^2-d^2)}\)^{\frac32}\times\]
\[\times\(1-\frac{3(9c^2d^2-5d^4-3d^2\mu^2-c^2\mu^2)}{20d(d^2-\mu^2)}\(\frac{k+\i
d}{\i(c^2-d^2)}\)+\mathcal{O}\(\frac{k+\i d}{i(c^2-d^2)}\)^2\),\]
\[\hskip12cm
k\rightarrow-\i d(\xi).
\]
The sign +(-) is taken 
according to the boundary values as indicated in Fig. \ref{Contourelliptic}.
 In the region $\xi\to\frac{c^2}{3}$ we have $d\to c,\ \mu\to\frac{c}{\sqrt{3}}.$
This suggests to introduce new local coordinates
$$
g=:\mp\frac{B}{2}+\i (c^2-d^2)z^{3/2},\quad tg=\mp\frac{tB}{2}+\i
\widehat\tau z^{3/2}=:\mp\frac{tB}{2}+\i\frac23\zeta^{3/2}, \qquad |k-\i
d|<r(c^2-d^2),$$
$$
g=:\mp\frac{B}{2}-\i (c^2-d^2)z_d^{3/2},\quad
tg=\mp\frac{tB}{2}-\i \widehat\tau z_d^{3/2}=:
\mp\frac{tB}{2}-\i\frac23\zeta_d^{3/2}, \qquad |k+\i
d|<r(c^2-d^2),$$ where we have introduced the new fast variable
\begin{equation}
\label{deftau}
\widehat\tau:=t(c^2-d^2).
\end{equation}
We also define the functions
$$\phi(k):=\begin{cases}
\frac{F(k)\e^{\frac{\i tB}{2}}}{\sqrt{-\i\widehat{f}(k)}},\quad
|k-\i d|<r(c^2-d^2),\ \Re k>0,
\\
\frac{F(k)\e^{\frac{-\i tB}{2}}}{\sqrt{\i\widehat{f}(k)}},\quad
|k-\i d|<r(c^2-d^2),\ \Re k<0,
\end{cases}
$$
$$
\phi_d(k):=\begin{cases}
F(k)\e^{\frac{\i
tB}{2}}\sqrt{\i\overline{\widehat{f}(\overline{k})}},\quad |k+\i
d|<r(c^2-d^2),\ \Re k>0,
\\
F(k)\e^{\frac{-\i
tB}{2}}\sqrt{-\i\overline{\widehat{f}(\overline{k})}},\quad |k+\i
d|<r(c^2-d^2),\ \Re k<0,
\end{cases}.
$$
Inside the disks  $|k\mp \i d|<r(c^2-d^2)$ the jumps
$J^{(3c)}$ can be written in the form
$$\phi^{\sigma_3}\begin{pmatrix}
1&-\i \e^{\frac23\zeta^{3/2}}\\0&1
\end{pmatrix}
\phi^{-\sigma_3},k\in L_7\cup L_5, \qquad
\phi_+^{\sigma_3}\begin{pmatrix} 1&0\\\i
\e^{-\frac23\zeta^{3/2}}0&1
\end{pmatrix}
\phi_-^{-\sigma_3},k\in (\i d,\i(d-\widetilde r)),$$
$$\phi_+^{\sigma_3}\begin{pmatrix}0&\i \\\i& 0\end{pmatrix}\phi_-^{-\sigma_3},\ k\in(\i d,\i (d+\widetilde r))$$
and
$$\phi_d^{\sigma_3}\begin{pmatrix}
1&0\\-\i \e^{\frac23\zeta_d^{3/2}}&1
\end{pmatrix}
\phi_d^{-\sigma_3},k\in L_8\cup L_6, \qquad
\phi_{d,+}^{\sigma_3}\begin{pmatrix} 1&\i
\e^{-\frac23\zeta_d^{3/2}}\\0&1
\end{pmatrix}
\phi_{d,-}^{-\sigma_3},k\in (-\i(d-\widetilde r),-\i d),$$
$$\phi_{d,+}^{\sigma_3}\begin{pmatrix}0&\i \\\i& 0\end{pmatrix}\phi_{d,-}^{-\sigma_3},\ k\in(-\i (d+\widetilde r),-\i d).$$
This leads to parametrices in terms of Airy functions; namely, we
define
 $$\Psi(\zeta) =
\Psi_{Ai}(\zeta)\e^{-\pi\i\sigma_3/4},\qquad
\Psi_d(\zeta_d)=\Psi(\zeta_d)\begin{pmatrix}0&1\\-1&0\end{pmatrix}$$
where $\Psi_{Ai}$ is found in the Appendix (\eqref{Psi_Ai}, \eqref{v01m1}). Then
$$\Psi(\zeta) =
\zeta^{-\sigma_3/4}\frac{1}{\sqrt{2}}\begin{pmatrix}1&1\\1&-1\end{pmatrix}\mathcal{E}(\zeta)\e^{\frac23\zeta^{3/2}\sigma_3},
\  \Psi_d(\zeta_d) =
\zeta_d^{-\sigma_3/4}\frac{1}{\sqrt{2}}\begin{pmatrix}-1&1\\1&1\end{pmatrix}\mathcal{E}_d(\zeta_d)\e^{-\frac23\zeta_d^{3/2}\sigma_3},
$$
$$\mathcal{E}(\zeta)=I+\begin{pmatrix}\frac{-1}{48} & \frac{-1}{8}
\\
\frac{1}{8} &
\frac{1}{48}\end{pmatrix}\zeta^{-\frac32}+\mathcal{O}(\zeta^{-3}),
\quad \mathcal{E}_d(\zeta_d)=I+\begin{pmatrix}\frac{1}{48} &
\frac{-1}{8}
\\
\frac{1}{8} &
\frac{-1}{48}\end{pmatrix}\zeta_d^{-\frac32}+\mathcal{O}(\zeta_d^{-3}).
$$
Now we are ready to define the approximate solution to the RH
problem for $M^{(3c)}.$ Define
$$M_{\infty}=\begin{cases}
M_{ell}(k),\quad |k\mp\i d|>r(c^2-d^2),
\\
B(k)\(\frac{3\widehat\tau}{2}\)^{\sigma_3/6}\Psi(\zeta)\phi^{-\sigma_3}(k)\e^{-\frac23\zeta^{3/2}\sigma_3},\quad
|k-\i d|<r(c^2-d^2),
\\
B_d(k)\(\frac{3\widehat\tau}{2}\)^{\sigma_3/6}\Psi_d(\zeta_d)\phi_d^{-\sigma_3}(k)\e^{\frac23\zeta_d^{3/2}\sigma_3},\quad
|k+\i d|<r(c^2-d^2).
\end{cases}$$
Here the analytic matrices $B(k), B_d(k)$ are to be determined in such a
way that they minimize the jump on the boundaries of the disks
for the error matrix $E=MM_{ell}^{-1}.$ Hence we take
$$B(k)=M_{ell}(k)\phi^{\sigma_3}(k)\frac{1}{\sqrt{2}}\begin{pmatrix}1&1\\1&-1\end{pmatrix} z^{\sigma_3/4},
\quad
B_d(k)=M_{ell}(k)\phi_d^{\sigma_3}(k)\frac{1}{\sqrt{2}}\begin{pmatrix}-1&1\\1&1\end{pmatrix} z_d^{\sigma_3/4},$$
and it is straightforward to check that $B(k), B_d(k)$ are indeed
have identity jumps inside the disks, and singularity at the
points $k=\pm\i d$ (due to $M_{ell}$) has the order no more than $(k\mp\i d)^{-1/2},$
i.e.  it is a removable singularity. Hence, these functions are
indeed analytic inside the disks. The error matrix has a jump matrix $J_E$ ($E_-=E_+J_E$) as follows
\begin{equation}
\label{eqB}
J_E=M_{ell}\phi^{\sigma_3}\mathcal{E}\phi^{-\sigma_3}M_{ell}^{-1}=B\(\frac{3\widehat\tau}{2}\)^{\hskip-2mm \frac{\sigma_3}{6}}\zeta^{\frac{-\sigma_3}{4}}\frac{1}{\sqrt{2}}\begin{pmatrix}1&1\\1&-1\end{pmatrix}\mathcal{E}
\frac{1}{\sqrt{2}}\begin{pmatrix}1&1\\1&-1\end{pmatrix}\zeta^{\frac{\sigma_3}{4}}\(\frac{3\widehat\tau}{2}\)^{\hskip-2mm\frac{-\sigma_3}{6}}B^{-1},
\end{equation}
$$\hfill\ |k-\i d|=\widetilde r,$$
\begin{equation}
\label{eqBd}
J_E=M_{ell}\phi_d^{\sigma_3}\mathcal{E}_d\phi_d^{-\sigma_3}M_{ell}^{-1}=B_d\(\frac{3\widehat\tau}{2}\)^{\hskip-2mm\frac{\sigma_3}{6}}\zeta_d^{\frac{-\sigma_3}{4}}\frac{1}{\sqrt{2}}
\begin{pmatrix}-1&1\\1&1\end{pmatrix}\mathcal{E}_d
\frac{1}{\sqrt{2}}\begin{pmatrix}-1&1\\1&1\end{pmatrix}\zeta_d^{\frac{\hskip-2mm\sigma_3}{4}}\(\frac{3\widehat\tau}{2}\)^{\hskip-2mm\frac{-\sigma_3}{6}}B_d^{-1},
\end{equation}
$$\hfill\hfill
\ |k+\i d|=\widetilde r.$$
The terms in the middle of the conjugations by $B, B_d$ in \eqref{eqB}, \eqref{eqBd} admit the following estimate
$$\(\frac{3\widehat\tau}{2}\)^{\hskip-2mm \frac{\sigma_3}{6}}\zeta^{\frac{-\sigma_3}{4}}\frac{1}{\sqrt{2}}\begin{pmatrix}1&1\\1&-1\end{pmatrix}\mathcal{E}
\frac{1}{\sqrt{2}}\begin{pmatrix}1&1\\1&-1\end{pmatrix}\zeta^{\frac{\sigma_3}{4}}\(\frac{3\widehat\tau}{2}\)^{\hskip-2mm\frac{-\sigma_3}{6}}
=\frac{2}{3\widehat\tau}\begin{pmatrix}0&\frac{5}{48z^2}\\\frac{-7}{48z}&0\end{pmatrix}+\mathcal{O}(\widehat\tau^{-2})
$$
$$
\(\frac{3\widehat\tau}{2}\)^{\hskip-2mm\frac{\sigma_3}{6}}\zeta_d^{\frac{-\sigma_3}{4}}\frac{1}{\sqrt{2}}
\begin{pmatrix}-1&1\\1&1\end{pmatrix}\mathcal{E}_d
\frac{1}{\sqrt{2}}\begin{pmatrix}-1&1\\1&1\end{pmatrix}\zeta_d^{\frac{\hskip-2mm\sigma_3}{4}}\(\frac{3\widehat\tau}{2}\)^{\hskip-2mm\frac{-\sigma_3}{6}}
=\frac{2}{3\widehat\tau}\begin{pmatrix}0&\frac{5}{48z_d^2}\\\frac{-7}{48z_d}&0\end{pmatrix}+\mathcal{O}(\widehat\tau^{-2}).$$

Since the analytic matrices $B, B_d$ are bounded for $\widehat\tau\to\infty$ on the circles $|k\mp\i d|=r(c^2-d^2),$ we get that
uniformly for $k\in\Sigma^{(3c)}\cup\left\{k:|k\mp\i c|=r(c^2-d^2)\right\}$ we have
$$J_E(k)=I+\mathcal{O}\(\widehat\tau^{-1}\), \textrm{ and hence } \quad E(k)=I+\mathcal{O}\(\frac{1}{\widehat\tau k}\)
\textrm { for } k\to\infty.$$
Hence, the solution to the MKdV equation $q(x,t)$ is estimated by the function $q_{el}=\lim\limits_{k\to\infty}(2\i k M_{el}(x,t;k))$ with the accuracy $\widehat\tau^{-1}:$
$$q(x,t)=q_{el}(x,t)+\mathcal{O}(\widehat\tau^{-1}).$$
To achieve the statement of Theorem \ref{thrm_Elliptic_est} it remains to estimate $\widehat\tau^{-1}$ in terms of $t$,  which requires to unravel its definition from \eqref{deftau} and \eqref{dmu1}, \eqref{dmu2}.
Denote $\eta=1-\frac{d(\xi)}{c}, \quad v=1-\frac{3\xi}{c^2}$: then for $d, v\to0$ we have $\eta=\frac{v}{\ln\frac1v}(1+\mathrm{o}(1))$ (\cite{KM2015}).
Hence, on the curve $\xi=\frac{c^2}{3}-\frac{\beta t^{\sigma}\ln t}{t}, \sigma\in(0,1),\ \beta>0,$ we have $\widehat\tau=\frac{2c^2\beta\ \ t^{\sigma}}{1-\sigma}(1+_\mathcal{O}(1)).$
Therefore, for fixed small $\varepsilon>0$ we have that
$$\begin{cases}q(x,t)=q_{el}(x,t)+\mathcal{O}(t^{-1})\qquad\textrm{ for }\quad   \frac{-c^2}{2}+\varepsilon<\frac{x}{12t}<\frac{c^2}{3}-\varepsilon,
 \\
q(x,t)=q_{el}(x,t)+\mathcal{O}\(\frac{1-\sigma}{\beta t^{\sigma}}\)\qquad\textrm{ for }\quad   \frac{c^2}{3}-\varepsilon<\frac{x}{12t}<\frac{c^2}{3}-\frac{\beta t^{\sigma}\ln t}{t},\quad \sigma\in(0,1).
\end{cases}$$
This completes the proof of Theorem \ref{thrm_Elliptic_est}. \hfill $\blacksquare$
\vskip1cm
\noindent
We now prove Corollary \ref{cor_1_q_el}.  

\noindent {\bf Proof.} [Proof of Corollary \ref{cor_1_q_el}.]
Denote 
\begin{equation}\label{z_tBDelta}z:=\frac{tB+\Delta}{\pi},\quad \tau^*=\frac{4\pi^2}{\tau},\quad \eta:=1-\frac{d}{c},\quad v:=1-\frac{3\xi}{c^2}.\end{equation}
Then (see \cite{KM2015}, p.16-17)
\begin{equation}\label{qelfromKM}
 q_{el}(x,t)=\frac{2c}{\cosh\(\frac{\tau^*}{4}(z-2n-1)\)}+\mathcal{O}\(\e^{\frac{\tau^*}{4}}\)\quad \textrm{ for } 2n\leq z\leq2n+2.
\end{equation}
Furthermore, (\cite{KM2015} (B.43), (B.54)),
$$\eta = \e^{\frac{\tau^*}{4}}+\mathcal{O}\(\e^{\frac{\tau^*}{2}}\),\qquad v=\eta\ln\frac{8\e}{\eta}+\mathcal{O}(\eta^2\ln^2\eta).$$
Consequently,  on a curve $$\xi=\frac{c^2}{3}-\frac{\beta t^{\sigma}\ln t}{12 t}$$ we have 
\begin{equation}
v=\frac{\beta t^{\sigma}\ln t}{4c^2t}\quad \textrm { and hence }\quad \eta\asymp t^{\sigma-1},\quad
\e^{\frac{\tau^*}{4}}=\mathcal{O}(t^{\sigma-1}),
\label{Oest}
\end{equation}
and we can substitute the $\mathcal{O}$ estimate \eqref{Oest} in (\ref{qelfromKM}).
We also  have the expansions (\cite{KM2015})
\begin{equation}\label{taustar}
 \frac{\tau^*}{4} = \ln\frac{\eta}{8}+\frac{\eta}{2}+\frac{3\eta^2}{16}+\mathcal{O}(\eta^3),\quad \frac{4}{\tau^*}=\frac{1}{\ln\frac{\eta}{8}}+\mathcal{O}\(\frac{\eta}{\ln^2\eta}\),
\end{equation}
\begin{equation}\label{Deltapl12}
 \frac{\Delta}{\pi}+\frac12 = \frac{\ln\frac{4}{(h^*)^2}}{\ln\frac{8}{\eta}}+\mathcal{O}(\eta),
\end{equation}
\begin{equation}\nonumber
 \frac{B}{\pi} = 8c^3\eta\(1+\sum\limits_{j=1}^{M-1} \eta^jP_j(\eta)+\mathcal{O}(\eta^{M}\ln^{M}\eta)\),
\ \  M \geq 2,
\end{equation}

\begin{equation}\nonumber
  v = \eta\ln\frac{8\e}{\eta}+\sum\limits_{j=2}^{M} \eta^jQ_j(\eta)+\mathcal{O}(\eta^{M+1}\ln^{M+1}\eta),\ \ M\geq 2.
\end{equation}
The polynomials $P_j, Q_j$ are described after  \eqref{B}, \eqref{vineta}.
Now, expressing directly the argument of $\cosh$ in (\ref{qelfromKM}) in terms of $x,t$ (i.e. in terms of $v,t$) does not yet lead us to the statement of the corollary.
An indication of this is that the $n$-dependence of the phases in (\ref{eq_cor_1}) is much more involved than in (\ref{qelfromKM}). But the real issue is that, due to the inversion of the  Lambert equation, $\eta$ is invertible in terms of $v$ as a series in  $\frac{\ln\ln v}{\ln v},$ i.e. it has a very slow convergent rate. Hence, the leading term of $\tau^*$ (see \ref{taustar}), i.e. $\ln\eta$, being multiplied by the leading term from $tB$ (see (\ref{B})),
i.e. $t\eta\asymp t^{\sigma}$, gives us an expression of the type $$t^{\sigma}\sum\limits_{j=0}^{\infty}b_j\(\frac{\ln\ln v}{\ln v}\)^j,$$ and since $\ln v\asymp\ln t,$ we can not truncate this series. To overcome this issue, let us notice that
$$
\frac{2c}{\cosh\(\frac{\tau^*}{4}(z-2n-1)\)}=:\frac{2c}{\cosh a}\in[K t^{-\delta},2c]
$$
 for some $K>0,\delta>0$ if and only if $$a:=\frac{\tau^*}{4}(z-2n-1)\in[-\delta\ln t-C,\delta\ln t+C]$$
for some $C>0.$ So, we  treat the argument of $\cosh$ only in a domain around the peak, where $\frac{2c}{\cosh a}$ is at least of the order $\asymp t^{-\delta},$ since in the remaining part it is $\mathcal{O}(t^{-\delta}).$
We have $$a=\frac{\tau^*}{4}\(\frac{tB}{\pi}+\frac{\Delta}{\pi}+\frac12-2n-\frac32\),\quad \textrm{ or }\quad 2n+\frac32+\frac{4a}{\tau^*}=\frac{tB}{\pi}+\frac{\Delta}{\pi}+\frac12,$$
and substituting here (\ref{Deltapl12}), (\ref{B}), (\ref{taustar}), and expressing further the leading term of $tB$, i.e. $8c^3t\eta$, we obtain
\begin{equation}\label{aseq1}
 8c^3t\eta=2n+\frac32-\frac{\ln\frac{4}{(h^*)^2}+a}{\ln\frac{8}{\eta}}-8c^3t\eta\sum\limits_{j=1}^{M-1}\eta^jP_j(\eta)+\mathcal{O}\(\eta\)+\mathcal{O}\(t\eta^{M+1}\ln^M\eta\).
\end{equation}
Now let us find an expression for $-2c(x-4c^2t)=8c^3vt.$ Substituting (\ref{vineta}), and expressing then subsequently $8c^3t\eta$, $\eta$ and $\ln\frac{8\e}{\eta}$ by the l.h.s. of (\ref{aseq1}), after some computation we obtain
\begin{eqnarray*}
8c^3vt&=&\le(2n+\frac32
\ri)\ln\frac{32c^3\e t}{n}-\ln\frac{4}{(h^*)^2}-a-\frac{\ln\frac{4}{(h^*)^2}+a}{\ln\frac{8}{\eta}}+\mathcal{O}(\eta\ln\eta)+
\\
&&+8c^3t\left[\sum\limits_{j=2}^{M}\eta^jQ_j(\eta)
-\eta\ln\frac{8\e}{\eta}\sum\limits_{j=1}^{M-1}\eta^jP_j(\eta)+\mathcal{O}(\eta^{M+1}\ln^{M+1}\eta)\right]-
\\
&&-2n\le(1+\frac{3}{4n}\ri)\ln\left[1+\frac{3}{4n}-\frac{\ln\frac{4}{(h^*)^2}+a}{2n\ln\frac{8}{\eta}}-\frac{4c^3t\eta}{n}\sum\limits_{j=1}^{M-1}\eta^jP_j(\eta)+
\mathcal{O}\(\frac{t\eta^{M+1}\ln^M\eta}{n}\)+\mathcal{O}(t^{-1})\right].
\end{eqnarray*}
Taking, in this formula $M=1,2,3,$ and expressing $a$ as the third summand in the r.h.s., we obtain subsequently
$$a = -8c^3vt+\le(2n+\frac32\ri)\ln\frac{32c^3t}{n}+2n-\ln\frac{4}{(h^*)^2}+\mathcal{O}(t^{2\sigma-1}\ln^2t)+\mathcal{O}(t^{-\sigma}),$$
$$a = -8c^3vt+\le(2n+\frac32\ri)\ln\frac{32c^3t}{n}+2n-\ln\frac{4}{(h^*)^2}+\frac{n^2}{4c^3t}\(-2+\ln\frac{32c^3t}{n}\)+\mathcal{O}(t^{3\sigma-2}\ln^3t)+\mathcal{O}(t^{\sigma-1}\ln^2t)+\mathcal{O}(t^{-\sigma}).$$
$$a = -8c^3vt+\le(2n+\frac32\ri)\ln\frac{32c^3t}{n}+2n-\ln\frac{4}{(h^*)^2}
+\frac{n^2\ln\frac{8}{\e^2\eta}}{4c^3t}+\frac{n^3}{8c^6t^2}\(Q_3-\ln\frac{8}{\eta}P_2-\ln\frac{8}{\e^2\eta}P_1+\frac12P_1^2\)+$$
$$
+\mathcal{O}(t^{4\sigma-3}\ln^4t)+\mathcal{O}(t^{\sigma-1}\ln^2t)+\mathcal{O}(t^{-\sigma}),$$
and, after further computations and substitutions of $\eta$ using (\ref{aseq1}),
\begin{eqnarray}
a &=& -8c^3vt+(2n+\frac32)\ln\frac{32c^3t}{n}+2n-\ln\frac{4}{(h^*)^2}
+\cr
&&+\frac{n^2\ln\frac{32c^3t}{\e^2n}}{4c^3t}+\frac{n^3}{128c^6t^2}\(-39+63\ln2-18\ln^22-(21-12\ln2)\ln\frac{n}{4c^3t}-2\ln^2\frac{n}{4c^3t}\)+
\nonumber \\
&&+\mathcal{O}(t^{4\sigma-3}\ln^4t)+\mathcal{O}(t^{\sigma-1}\ln^2t)+\mathcal{O}(t^{-\sigma}).
\end{eqnarray}
This gives us the statement of Corollary \ref{cor_1_q_el}.
\QED

In Sec.\ref{sect_mesoscopic} we  give an alternative proof of Corollary \ref{cor_1_q_el} which  has the advantage that it is also valid up to $\sigma\in[0,\frac{M}{M+1}).$
The number of the soliton to  which the point $(x,t)$ belongs at a particular time will be  expressed by the solution \eqref{gamma_eq_mes} whereas in this section this number was determined by $\frac{z}{2}$ (formula \ref{z_tBDelta}). In order to compare $z$ and $\gamma$ we need the following lemma:
\begin{lem}\label{lem_z_tBDelta_equation}
 The quantity $z$ in (\ref{z_tBDelta}) satisfies the following asymptotic bound:
$$8c^3vt+\(z+\frac12\)\(\ln\frac{z}{2t}-(\widehat Q+1)\)=\mathcal{O}(1),\quad t\to\infty$$
where  $\widehat Q$ satisfies 
$$\widehat Q+1=\ln\frac{B}{2\pi}+\frac{8c^3v\pi}{B}+\mathcal{O}\le(\frac{1}{t\eta}\ri).$$
\end{lem}
\noindent {\bf Proof.} 
 The proof is straightforward using (\ref{vineta}), (\ref{B}), (\ref{Deltapl12}).
We have $$8c^3vt+\(\frac{tB}{\pi}+\frac{\Delta}{\pi}+\frac12\)\(\ln\frac{B}{2\pi}+\ln\(1+\frac{\Delta}{tB}\)-(\widehat Q+1)\)=$$
$$=8c^3vt+\frac{tB}{\pi}\(\ln\frac{B}{2\pi}-(\widehat Q+1)\)+
\underbrace{\(\frac{\Delta}{\pi}+\frac12\)\(\ln\frac{B}{2\pi}+\ln\(1+\frac{\Delta}{tB}\)-(\widehat Q+1)\)}+\underbrace{\frac{tB}{\pi}\ln\(1+\frac{\Delta}{tB}\)}.$$
The underbraced terms are of the order $\mathcal{O}(1).$ This proves lemma \ref{lem_z_tBDelta_equation}.
For further use let us notice also that
$$\ln\frac{B}{2\pi}+\frac{8c^3v\pi}{B} = \(1+\ln(32c^3)\)+\(-1+\frac12\ln\frac{8}{\eta}\)\eta+$$
$$+\frac{1}{16}\(9-21\ln2+18\ln^22+(7-12\ln2)\ln\eta+2\ln^2\eta\)\eta^2+\mathcal{O}(\eta^3\ln^4\eta).$$
Substituting here $\eta$ in view of (\ref{aseq1}), we obtain 
$$ \ln\frac{B}{2\pi}+\frac{8c^3v\pi}{B} = \(1+\ln(32c^3)\) + \(-1+\frac32\ln2-\frac12\ln\frac{n}{4c^3t}\)\frac{n}{4c^3t}+\mathcal{O}(t^{2\sigma-2}).$$
\QED

\section{Logarithmic domain $D_{2a}$}\label{sect_as_sol}
This is the region where  $4c^2t-C\ln t<x<4c^2t,$ $C>0.$
In this domain we deal with a finite number, $n$, of asymptotic solitons and our goal is to prove Thm \ref{teorKK_refined}.
We start from the original Riemann-Hilbert problem \ref{RH_problem_1}  (see also \cite{KM}).
In this section we consider the regime $\xi=\frac{c^2}{3}-\frac{\rho}{12}\frac{\ln t}{t},$
for some fixed positive $\rho>0.$
We perform a  first transformation of the RH   problem which removes jump from the real line; fix a sufficiently small $r>0$ and introduce the lines $$L_1=\mathbb{R}+\i(c-r),\quad L_2=\mathbb{R}-\i(c-r),$$
with the natural orientation of $\mathbb R$. Denote by $\Omega_1$ the domain bounded by $\mathbb{R}$ and $L_1$,
the other domain bounded by $L_1$ is $D_3$, the domain bounded by $\mathbb{R}$ and $L_2$ denote by $\Omega_2,$
and the other domain bounded by $L_2$ denote by $\Omega_4.$
Introduce the following transformation of the RH problem:
\begin{equation}\label{M1}M^{(1)}:=M\begin{cases}
             \begin{pmatrix}
              1 & 0 \\ -r(k) \e^{2\i t\theta} & 1
             \end{pmatrix},\quad k\in\Omega_1,\\
         \begin{pmatrix}
              1 & \ol{r(\ol{k})} \e^{-2\i t\theta} \\ 0 & 1
             \end{pmatrix},\quad k\in\Omega_2,\\
          I, \quad k\in\Omega_3\cup\Omega_4.
            \end{cases}
\end{equation}
Then the jump $M_-^{(1)} = M_+^{(1)} J^{(1)}$ is
\begin{equation}\label{J1}J^{(1)} = \begin{cases} I, \quad k\in\mathbb{R}\cup(\i(c-r),-\i(c-r)),\\
          J, \quad k\in(\i c, \i (c-r))\cup (-\i (c-r),-\i c),\\
         \begin{pmatrix}
              1 & 0 \\ -r(k) \e^{2\i t\theta} & 1
             \end{pmatrix},\quad k\in L_1,\\
         \begin{pmatrix}
              1 & -\ol{r(\ol{k})} \e^{-2\i t\theta} \\ 0 & 1
             \end{pmatrix},\quad k\in L_2.
            \end{cases}
\end{equation}
We see that the jumps are exponentially close to the identity matrix everywhere except for the intervals $\pm(\i c, \i(c-r)).$
We then define 
a local change of spectral variable $k$ on the interval $\pm(\i c,\i (c-r))$ in terms of new variables $y, y_d$, respectively:
 \begin{equation}
 \label{y_yd}k=\i c-\i y,\qquad k=-\i c+\i y_d.
 \end{equation}
  Then we set 
$$2\i t \theta(k,\xi)=2c\rho\ln t-t\left[\left(16c^2+\frac{2\rho\ln
t}{t}\right)y - 24 c y^2+8 y^3\right]=:2c\rho\ln t-t z=:2c\rho\ln
t-\zeta,$$
$$-2\i t \theta(k,\xi)=2c\rho\ln t-t\left[\left( 
16c^2+\frac{2\rho\ln
t}{t}\right )y_d - 24 c y_d^2+8 y_d^3\right]=:2c\rho\ln t-t z_d=:2c\rho\ln
t-\zeta_d,$$
where we denoted
$$\frac{\zeta}{t}:= z:= \left[(16c^2+\frac{2\rho\ln t}{t})y -
24 c y^2+8 y^3\right];\qquad
\frac{\zeta_d}{t}:= z_d:= \left[(16c^2+\frac{2\rho\ln t}{t})y_d -
24 c y_d^2+8 y_d^3\right];$$
$$f(k)=\frac{2\sqrt{2}\ \i}{(h^*)^2\sqrt{c}}\sqrt{y}(1+\mathcal{O}(\sqrt{y}))=
\frac{2\sqrt{2}\ \i}{(h^*)^2\sqrt{c}}\sqrt{\frac{y}{z}}(1+\mathcal{O}(\sqrt{y}))\ \sqrt{z}=
:\i \phi(k,t)\sqrt{z},$$
$$-\ol{f(\ol{k})} = \frac{2\sqrt{2}\ \i}{(h^*)^2\sqrt{c}}\sqrt{y_d}(1+\mathcal{O}(\sqrt{y_d}))=\frac{2\sqrt{2}\ \i}{(h^*)^2\sqrt{c}}\sqrt{\frac{y_d}{z_d}}(1+\mathcal{O}(\sqrt{y_d}))\sqrt{z_d}=:\i\phi_d(k,t)\sqrt{z_d}.$$
Here $\phi(k,t)$, $\phi_d(k,t)$ are analytic in a
neighborhood of $k=\pm\i c,$ respectively, where they are separated both from $0$ and $\infty$ uniformly with respect to $t\geq 1.$
\begin{com}
 In the case of pure step initial function (\ref{pure_step_ic}) we have
$$\widetilde{f}(k)=\frac{2\i}{c}\sqrt{k^2+c^2}=\frac{2\i}{c}\sqrt{y(2c-y)},\qquad h^*=1$$
\end{com}

\noindent
In terms of the $z$, $z_d$ variables, the jump matrix $J^{(1)}$ on the intervals $\pm(\i c,\i (c-r))$  is expressed as  
\begin{equation}\label{jumps_J1}J^{(1)}=\begin{pmatrix}
1 & 0
\\
f(k)\e^{2\i t\theta(k,\xi)} & 1
\end{pmatrix}=\begin{pmatrix}
1 & 0
\\
\i\phi\ t^{2c\rho-\frac12}\sqrt{z t}\ \e^{-t z} & 1
\end{pmatrix}
=
\begin{pmatrix}
1 & 0
\\
\i\phi\ t^{2\gamma}\sqrt{\zeta}\e^{-\zeta} & 1
\end{pmatrix}, k\in(\i c,\i (c-r)),
\end{equation}
\begin{equation}\label{jumps_J1d}J^{(1)}=\begin{pmatrix}
1 & -\ol{f(\ol{k})}\e^{-2\i t\theta(k,\xi)} \\ 0 & 1
\end{pmatrix}=\begin{pmatrix}
1 & \i\phi_d\ t^{2c\rho-\frac12}\sqrt{z_d t}\ \e^{-t z_d} \\ 0 & 1
\end{pmatrix}
=
\begin{pmatrix}
1 & \i\phi_d\ t^{2\gamma}\sqrt{\zeta_d}\e^{-\zeta_d} \\ 0 & 1
\end{pmatrix}, k\in(-\i (c-r),-\i c),
\end{equation}
\begin{equation}\label{gamma}\gamma:=c\rho-\frac14.
\end{equation}

\subsection{Generalized Laguerre polynomials with index $\frac12.$}
Denote
$p_n(\zeta)=L^{(1/2)}_n(\zeta)=\frac{(-1)^n}{n!}\zeta^n+...,\quad
\pi_n(\zeta)=(-1)^nn!p_n(\zeta)=\zeta^n+...$
$$\int\limits_0^{+\infty}\zeta^{1/2}\e^{-\zeta}p_n(\zeta)p_m(\zeta)d\zeta=\frac{\Gamma(n+\frac32)}{n!}\delta_{m,n},
\int\limits_0^{+\infty}\zeta^{1/2}\e^{-\zeta}\pi_n(\zeta)\pi_m(\zeta)d\zeta=\Gamma(n+\frac32){n!}\delta_{m,n}.$$
The generalized Laguerre polynomials with index $\frac 1 2$ and degree $n$ solve a RHP of the form
\begin{eqnarray}
\nonumber
L_-(\zeta) &=& L_+(\zeta) J_L(\zeta), \ \  \ \zeta \in \mathbb R_+,
\\\nonumber
J_L(\zeta) &=&
\begin{pmatrix}
1 & 0
\\
-\sqrt{\zeta}\e^{-\zeta} & 1
\end{pmatrix},
\\
L(\zeta) &=& \left({\mathbf 1} + \mathcal O(\zeta^{-1})\right)  \zeta^{-n\sigma_3} ,\ \ \ \zeta \to \infty. \label{Las1}
\end{eqnarray}
and the solution is written as follows: for $n\geq 1$
\begin{equation}\label{LaguerreMatrixfiniten}L(\zeta)=\begin{pmatrix}\frac{-2\pi\i}{\Gamma(n+\frac12)\Gamma(n)}\displaystyle\frac{1}{2\pi\i}\int\limits_{0}^{+\infty}\frac{\sqrt{s}\,\e^{-s}\pi_{n-1}(s)\d
s}{s-\zeta} &
\frac{-2\pi\i}{\Gamma(n+\frac12)\Gamma(n)}\pi_{n-1}(\zeta)
\\\\
\displaystyle\frac{1}{2\pi\i}\int\limits_{0}^{+\infty}\frac{\sqrt{s}\,\e^{-s}\pi_n(s)\d
s}{s-\zeta}
 & \pi_n(\zeta)
\end{pmatrix},\end{equation}
 and for $n=0$
$$L(\zeta)=\begin{pmatrix}
            1 & 0 \\ \frac{1}{2\pi\i}\int\limits_{0}^{+\infty}\frac{\sqrt{s}\ \e^{s}\ \d s}{s-\zeta} & 1
           \end{pmatrix}.
$$

\noindent
Furthermore, the matrix function 
$$
L_d(\zeta)=\begin{pmatrix}0&1\\1&0\end{pmatrix}L(\zeta)\begin{pmatrix}0&1\\1&0\end{pmatrix}
$$
 solves a RHP of the form
\begin{eqnarray}\nonumber
L_{d,-}(\zeta) &=& L_{d,+}(\zeta) J_{L_d}(\zeta), \ \  \ \zeta \in (+\infty,0)\  (\textrm{the orientation is from } +\infty \textrm { to } 0),
\\\nonumber
J_{L_d}(\zeta) &=&
\begin{pmatrix}
1 & \sqrt{\zeta_d}\e^{-\zeta_d} \\ 0 & 1
\end{pmatrix},
\\
L_d(\zeta_d) &=& \left({\mathbf 1} + \mathcal O(\zeta_d^{-1})\right)  \zeta_d^{n\sigma_3} ,\ \ \ \zeta_n \to \infty. \label{Lasd1}
\end{eqnarray}
To show the relation with our RH problem for $M^{(1)}$, consider the functions $$L^{(1)}=\(-\i\phi t^{2\gamma}\)^{-\sigma_3/2}
L\(-\i\phi t^{2\gamma}\)^{\sigma_3/2},\qquad
L_d^{(1)}=\(\i\phi_d t^{2\gamma}\)^{\sigma_3/2}
L_d\(\i\phi_d t^{2\gamma}\)^{-\sigma_3/2};$$  they have
the following jumps on $\zeta\in(0,+\infty),$ $\zeta_d\in(+\infty,0),$ respectively (compare with (\ref{jumps_J1}), (\ref{jumps_J1d})):
$$(L_+^{(1)})^{-1}L_-^{(1)}=\begin{pmatrix}1&0\\ \i\phi t^{2\gamma}\sqrt{\zeta}\e^{-\zeta}&1\end{pmatrix},\qquad
(L_{d,+}^{(1)})^{-1}L_{d,-}^{(1)}=\begin{pmatrix}1&\i\phi_d t^{2\gamma}\sqrt{\zeta_d}\e^{-\zeta_d} \\ 0&1\end{pmatrix}.$$

\noindent Further, developing up to $\zeta^{-1}$, $\zeta_d^{-1}$ term in the asymptotics (\ref{Las1}), (\ref{Lasd1}) of $L$, $L_d$ as $\zeta,\zeta_d\to\infty,$ we obtain
\begin{equation}\label{Las2}L(\zeta)=\left[\begin{pmatrix}1+\frac{n^2+\frac{n}{2}}{\zeta} & \frac{-2\pi\i\ n}{\Gamma(n+\frac12)\, n!\ \zeta}
\\
\frac{-n!\ \Gamma(n+\frac32)}{2\pi\i\ \zeta} &
1-\frac{n^2+\frac{n}{2}}{\zeta}\end{pmatrix}+\mathrm{O}(\zeta^{-2})\right]\zeta^{-n\sigma_3},\quad\zeta\to\infty,
\end{equation}
\begin{equation}\label{Lasd2}L_d(\zeta)=\left[\begin{pmatrix}1-\frac{n^2+\frac{n}{2}}{\zeta_d} & \frac{-n!\ \Gamma(n+\frac32)}{2\pi\i\ \zeta_d}
\\
\frac{-2\pi\i\ n}{\Gamma(n+\frac12)\, n!\ \zeta_d} &
1+\frac{n^2+\frac{n}{2}}{\zeta_d}\end{pmatrix}+\mathrm{O}(\zeta_d^{-2})\right]\zeta_d^{n\sigma_3},\quad\zeta_d\to\infty.
\end{equation}
 The  formulas \eqref{Las2}, \eqref{Lasd2} include also the case $n=0.$

\subsection{First approximation of $M^{(1)}.$}
Let $\alpha\in \mathbb N$ be an integer  to be determined later. We look for  an approximation of $M^{(1)}$ of the form
\begin{equation}\label{Minf} M_{\infty}=\begin{cases}
\(\frac{k-\i c}{k+\i c}\)^{\alpha\sigma_3},\quad |k\mp\i c|>r, 
\\B L (-\i\phi)^{\sigma_3/2}t^{\gamma\sigma_3},\quad |k-\i c|<r, 
\\B_dL_d(\i\phi_d)^{-\sigma_3/2}t^{-\gamma\sigma_3},\quad
|k+\i c|<r,
\end{cases}
\end{equation}
In the process of the construction we will also determine the matrix-valued functions $B,$ $B_d$ analytic inside the disks $|k\mp\i c|<r$ , respectively. The driving logic is that of minimizing the distance of the error matrix $E=MM_{\infty}^{-1}$  from the identity matrix. To this end we inspect its jump
$J_E=(E_+)^{-1}E_-=M_{\infty,+}JM_{\infty,-}^{-1}.$

\noindent
On the interval $(\i c,\i(c-r))$ the jump is
$$J_E=BL_+(-\i\phi)^{\sigma_3/2}t^{\gamma\sigma_3}\begin{pmatrix}1 & 0 \\\i\phi t^{2\gamma}\sqrt{\zeta}\e^{-\zeta} & 1\end{pmatrix}
t^{-\gamma\sigma_3}(-\i\phi)^{-\sigma_3/2}L_-^{-1}B^{-1}=I.$$
Similarly, on $(-\i (c-r), -\i c)$ the jump $J_E=I.$
\noindent The jump $J_E$ on the disks 
$$\partial C=\left\{k:\ |k-\i c|=r\right\},\quad \partial C_d=\left\{k:\ |k+\i c|=r\right\}, \quad (\textrm {we take counterclockwise orientation })$$
 is
$$J_E=BL(-\i\phi)^{\sigma_3/2}t^{\gamma\sigma_3}\(\frac{k-\i c}{k+\i c}\)^{-\alpha\sigma_3},\ k\in\partial C,\qquad
J_E=B_dL_d(\i\phi_d)^{-\sigma_3/2}t^{-\gamma\sigma_3}\(\frac{k-\i c}{k+\i c}\)^{-\alpha\sigma_3},\ k\in\partial C_d.$$
To have $J_E$ close to $I,$ taking into account the asymptotics (\ref{Las1}), (\ref{Lasd1}) of $L$, $L_d$
on the circles (we have $\zeta=z t\to\infty$ as $t\to\infty$ when $k\in\partial C$), we take
\begin{equation}\label{B_Bd_finite_n}B=\(\frac{k-\i c}{k+\i c}\)^{\alpha\sigma_3}z^{n\sigma_3}(-\i\phi)^{-\sigma_3/2}t^{(n-\gamma)\sigma_3},\quad
B_d=\(\frac{k-\i c}{k+\i c}\)^{\alpha\sigma_3}z_d^{-n\sigma_3}(\i\phi_d)^{\sigma_3/2}t^{(\gamma-n)\sigma_3}.\end{equation}
In order not to have poles at $z=0$, $z_d=0,$ we henceforth choose
$\alpha=-n.$ Then
\begin{equation}\label{J_E_rough_pre1}
 J_E\hskip-1mm=\hskip-1mm\(\frac{k+\i c}{k-\i c}\cdot z\hskip-1mm\)^{\hskip-1mmn\sigma_3}\hskip-2mm\(\hskip-1mm-\i\phi\)^{\frac{-\sigma_3}{2}}t^{(n-\gamma)\sigma_3}\hskip-1mm
\begin{pmatrix}
 1+\mathcal{O}(\frac{1}{zt}) & \mathcal{O}(\frac{1}{zt}) \\
\mathcal{O}(\frac{1}{zt}) & 1+\mathcal{O}(\frac{1}{zt})
\end{pmatrix}\hskip-1mm(\hskip-1mm-\i\phi)^{\frac{\sigma_3}{2}}t^{(\gamma-n)\sigma_3}\hskip-1mm\(\frac{k-\i c}{k+\i c}\cdot\frac{1}{z}\)^{\hskip-1mmn\sigma_3}\hskip-1mm, k\in\partial C,
\end{equation}
\begin{equation}\label{J_E_rough_pre2}
 J_E\hskip-1mm=\hskip-1mm\(\frac{k+\i c}{k-\i c}\cdot \frac{1}{z_d}\hskip-1mm\)^{\hskip-1mmn\sigma_3}\hskip-2mm\(\i\phi_d\)^{\frac{\sigma_3}{2}}t^{(\gamma-n)\sigma_3}\hskip-1mm
\begin{pmatrix}
 1+\mathcal{O}(\frac{1}{z_dt}) & \mathcal{O}(\frac{1}{z_dt}) \\
\mathcal{O}(\frac{1}{z_dt}) & 1+\mathcal{O}(\frac{1}{z_dt})
\end{pmatrix}\hskip-1mm(\i\phi_d)^{\frac{-\sigma_3}{2}}t^{(n-\gamma)\sigma_3}\hskip-1mm\(\frac{k-\i c}{k+\i c}\cdot z_d\)^{\hskip-1mmn\sigma_3}\hskip-1mm, k\in\partial C_d,
\end{equation}
and hence
\begin{equation}\label{J_E_rough}
J_E=\begin{pmatrix}1+\mathrm{O}(t^{-1})&\mathrm{O}(t^{-2\gamma+2n-1})\\\mathrm{O}(t^{2\gamma-2n-1})&
1+\mathrm{O}(t^{-1})\end{pmatrix},k\in\partial C,\quad
J_E=\begin{pmatrix}1+\mathrm{O}(t^{-1})&\mathrm{O}(t^{2\gamma-2n-1})\\\mathrm{O}(t^{-2\gamma+2n-1})&
1+\mathrm{O}(t^{-1})\end{pmatrix},k\in\partial C_d.
\end{equation}
For every value of $\gamma$ which is not half-integer we can choose $n$ such that $\gamma-\frac12<n<\gamma+\frac12,$ and
both off-diagonal terms in the r.h.s. of the first expression in (\ref{J_E_rough}) will be vanishing. However, for a half-integer
$\gamma=m+\frac12$ we cannot make both off-diagonal terms in (\ref{J_E_rough}) small: indeed, if we choose $n=m,$ then the
$(1,2)$ entry is of the order $\mathcal{O}(t^{-2})$, but the $(2,1)$ entry is just $\mathcal{O}(1);$ vice versa, for
$n=m+1,$ then $(2,1)$ entry is $\mathcal{O}(t^{-2})$, but the  $(1,2)$ entry is just $\mathcal{O}(1).$
As we shall see below, this indicates the presence of asymptotic solitons, which correspond to half-integer $\gamma.$ Away from the asymptotic solitons,
the solution of MKdV equation is asymptotically vanishing.

To capture the asymptotic solitons, we are lead to make a further correction in the approximate solution; this is accomplished in the next section.
\subsubsection{Refined approximation of $M^{(1)}$}\label{sect_sub_ref1_second}
\noindent In order to have more freedom in choosing $n$ in (\ref{J_E_rough}), the idea is that of  ``removing'' the  $\frac{1}{\zeta}$ term in (2,1) entry in asymptotics (\ref{Las2}) of $L,$
and in (1,2) entry in asymptotics (\ref{Lasd2}) of $L_d$ at $\infty,$ so that they will start from $\zeta^{-2},$ $\zeta_d^{-2},$
$$\begin{pmatrix}1&0\\\frac{R_1}{\zeta}&1\end{pmatrix} L = \begin{pmatrix}1+\mathcal{O}(\frac{1}{\zeta}) & \mathcal{O}(\frac{1}{\zeta}) \\ \mathcal{O}(\frac{1}{\zeta^2}) & 1+\mathcal{O}(\frac{1}{\zeta}) \end{pmatrix},
\qquad
\begin{pmatrix}1&\frac{R_1}{\zeta_d}\\0&1\end{pmatrix} L_d = \begin{pmatrix}1+\mathcal{O}(\frac{1}{\zeta_d}) & \mathcal{O}(\frac{1}{\zeta_d^2}) \\ \mathcal{O}(\frac{1}{\zeta_d}) & 1+\mathcal{O}(\frac{1}{\zeta_d}) \end{pmatrix},
$$
where $$R_1=\frac{n!\Gamma(n+\frac32)}{2\pi\i}.$$
However, this will bring poles at $k=\pm\i c$ of the approximate solution and to compensate for this issue we  multiply all $M_{\infty}$ by an appropriate meromorphic matrix function from the left, which will remove these poles.
 In concrete, the above idea requires to define
 \begin{equation}
\label{G}G=I+\frac{A}{k-\i c}+\frac{\widetilde A}{k+\i c}
\end{equation}
\begin{equation}\label{Minf1} M^{(1)}_{\infty}=\begin{cases}
G\(\frac{k-\i c}{k+\i c}\)^{-n\sigma_3},\quad |k\mp\i c|>r,
\\GB\begin{pmatrix}1&0\\\frac{R_1}{\zeta}&1\end{pmatrix} L (-\i\phi)^{\sigma_3/2}t^{\gamma\sigma_3},\quad |k-\i c|<r,
\\
GB_d
\begin{pmatrix}
1&\frac{R_1}{\zeta_d}\\0&1
\end{pmatrix}L_d(\i\phi_d)^{-\sigma_3/2}t^{-\gamma\sigma_3},\quad
|k+\i c|<r,
\end{cases}
\end{equation}
where $B,$ $B_d$  are as in (\ref{B_Bd_finite_n}) with $\alpha=-n$.
The new error matrix $$E^{(1)}:=M^{(1)}\(M^{(1)}_{\infty}\)^{-1}$$ has the jump
$$J_{E^{(1)}}=(E^{(1)}_+)^{-1}E^{(1)}_-=M^{(1)}_{\infty,+}J^{(1)}\(M^{(1)}_{\infty,-}\)^{-1};$$
in the intervals $\pm(\i c,\i(c-r))$ the jump  is $J_{E^{(1)}}=I,$ and on the circles $k\in\partial C,$ $k\in\partial C_d$ it is of the form
\begin{equation}
 \label{J_E_refined_1}
J_{E^{(1)}}=G\begin{pmatrix}1+\mathrm{O}(t^{-1})&\mathrm{O}(t^{-2\gamma+2n-1})\\\mathrm{O}(t^{2\gamma-2n-2})&1+\mathrm{O}(t^{-1})\end{pmatrix}G^{-1}, k\in\partial C,
 \
J_{E^{(1)}}=G\begin{pmatrix}1+\mathrm{O}(t^{-1})&\mathrm{O}(t^{2\gamma-2n-2})\\\mathrm{O}(t^{-2\gamma+2n-1})&1+\mathrm{O}(t^{-1})\end{pmatrix}G^{-1},
k\in\partial C_d.
\end{equation}
We see that (\ref{J_E_refined_1}) provides us with better estimate than (\ref{J_E_rough}) provided that 
$G,$ $G^{-1}$ are uniformly bounded, as $t\to\infty$, on the circles $|k\mp\i c|=r.$ Now, for such 
$\gamma$ that $$\left\{\gamma\right\}\in[0,\frac12]$$ we choose 
$$n=\lfloor \gamma \rfloor,$$
where $\left\{\gamma\right\},$ $\lfloor \gamma \rfloor$ denote the
fractional part of $\gamma$ and the greatest integer not exceeding $\gamma,$ respectively.
Then $J_{E^{(1)}}$ in (\ref{J_E_refined_1}) admits the estimate $$J_{E^{(1)}}=I+\mathcal{O}(t^{-1}).$$
Other values of  $\gamma$, such that $$\left\{\gamma\right\}\in(\frac12,1)$$ are considered in the next section \ref{sect_sub_ref2_second}.

\begin{com}
 We could assign $n$ for wider range of $\gamma$ (actually for all $\gamma$), namely take $n-\frac14\leq\gamma<n+\frac34.$ Then $J_{E^{(1)}}$ in (\ref{J_E_refined_1}) would be of the order $$J_{E^{(1)}}=I+\mathcal{O}(t^{\frac{-1}{2}}).$$
This would give us a worse estimate $\mathcal{O}(t^{\frac{-1}{2}})$ instead of $\mathcal{O}(t^{-1}).$
\end{com}

We determine now the matrix  $G$ (\ref{G}) in such a way that $M^{(1)}_{\infty}$ is bounded (has no poles)
at $k=\pm\i c.$ Expanding the product, the terms responsible for poles at $k=\pm\i c$ in $M^{(1)}$ are
$$GB\begin{pmatrix}1&0\\\frac{R_1}{\zeta}&1\end{pmatrix}=\(I+\frac{A}{k-\i c}+\frac{\widetilde A}{k+\i c}\)\begin{pmatrix}1&0\\\frac{R_1}{z}z^{-2n}(\frac{k-\i c}{k+\i c})^{2n}t^{2\gamma-2n-1}(-\i\phi)&1\end{pmatrix}
B ,$$
$$GB_d\begin{pmatrix}
1&\frac{R_1}{\zeta_d}\\0&1
\end{pmatrix}=\(I+\frac{A}{k-\i c}+\frac{\widetilde A}{k+\i c}\)\begin{pmatrix}1&\frac{R_1}{z_d}z_d^{-2n}\(\frac{k-\i c}{k+\i c}\)^{-2n}(\i\phi_d)t^{2\gamma-2n-1}
\\0&1\end{pmatrix}B_{d}.$$

\noindent
We see that at most we can have the poles of the second order at
$z=0,$ $z_d=0.$ The requirement that the singular part vanishes yields a linear system for the matrices $A, \widetilde A:$ from the vanishing of the double-pole coefficient it is  seen that they must be of the form
$$A=\begin{pmatrix}a&0\\b&0\end{pmatrix},\qquad \widetilde A=\begin{pmatrix}0&\widetilde b\\0&\widetilde a\end{pmatrix}.$$
Writing down the conditions of vanishing of the residue we get the system of equations
\begin{equation}\label{eq_abtildeab}\begin{cases}
\frac{1}{-2\i c}\ \frac{ a\
\widehat{R}_{1,d}}{16c^2+\frac{2\rho\ln t}{t}}-\i
\widetilde b+\frac{\widehat{R}_{1,d}}{16c^2+\frac{2\rho\ln
t}{t}}=0,
\\
\frac{1}{2\i c}\
\frac{b\ \widehat{R}_{1,d}}{16c^2+\frac{2\rho\ln t}{t}}+\i
\widetilde a=0;
\end{cases}
\begin{cases}
\frac{1}{2\i c}\ \frac{ \widetilde a\
\widehat{R}_{1}}{16c^2+\frac{2\rho\ln t}{t}}+\i
b+\frac{\widehat{R}_{1}}{16c^2+\frac{2\rho\ln t}{t}}=0,
\\
\frac{1}{2\i c}\ \frac{\widetilde b\
\widehat{R}_{1}}{16c^2+\frac{2\rho\ln t}{t}}+\i
 a=0,
\end{cases}\end{equation}
which decomposes into 2 linear systems: one for $a, \widetilde b$, another for $ \widetilde a, b.$
\noindent Here $$\widehat R_1 = R_1 t^{2\gamma-2n-1}\lim\limits_{k\to\i c}  z^{-2n}\(\frac{k-\i c}{k+\i c}\)^{2n}(-\i\phi) =
\frac{-2\ n!\Gamma(n+\frac32) t^{2\gamma-2n-1}}{\pi(h^*)^2\ \(2c(16c^2+\frac{2\rho\ln t}{t})\)^{2n+\frac12}},$$
$$\widehat R_{1,d} = R_1 t^{2\gamma-2n-1}\lim\limits_{k\to-\i c}  z_d^{-2n}\(\frac{k-\i c}{k+\i c}\)^{-2n}(\i\phi_d)
=\frac{2\ n!\Gamma(n+\frac32) t^{2\gamma-2n-1}}{\pi(h^*)^2\ \(2c(16c^2+\frac{2\rho\ln t}{t})\)^{2n+\frac12}},$$
$$-\widehat R_1=\widehat{R}_{1,d}>0.$$
Solving system (\ref{eq_abtildeab}) for $a, b, \widetilde a, \widetilde b,$ we obtain
\begin{equation}\label{abtildeab_sol1}a=\frac{-2\i c\ \widehat R_1\
\widehat{R}_{1,d}}{4c^2\(16c^2+\frac{2\rho\ln
t}{t}\)^2-\widehat R_1\
\widehat{R}_{1,d}},
\qquad
 b=\frac{4\i c^2\(16c^2+\frac{2\rho\ln t}{t}\)\
\widehat{R}_1}{4c^2\(16c^2+\frac{2\rho\ln
t}{t}\)^2-\widehat R_1\
\widehat{R}_{1,d}},
\end{equation}
\begin{equation}\label{abtildeab_sol2}\widetilde a= \frac{2\i c\ \widehat R_1\
\widehat{R}_{1,d}}{4c^2\(16c^2+\frac{2\rho\ln
t}{t}\)^2-\widehat R_1\
\widehat{R}_{1,d}},
\qquad \widetilde b= \frac{-4\i c^2\(16c^2+\frac{2\rho\ln t}{t}\)\
\widehat{R}_{1,d}}{4c^2\(16c^2+\frac{2\rho\ln
t}{t}\)^2-\widehat R_1\
\widehat{R}_{1,d}}.
\end{equation}
We see, that $a,b,\widetilde a, \widetilde b$ are all bounded for
$t\to\infty$, hence, $G$ does not contribute to the error estimate (\ref{J_E_refined_1})
of $J_E.$
 Hence,
$$q^{(1)}_{\infty}(x,t):=\lim\limits_{k\to\infty}(M_{\infty})_{21}=\lim\limits_{k\to\infty}(M_{\infty})_{12}=2\i
b=2\i\widetilde b = \frac{2c}{\cosh\left[\ln\(\frac{\widehat R_{1,d}}{2c(16c^2+\frac{2\rho\ln t}{t})}\)\right]}=
$$
\begin{equation}\label{q_inf_1}=\frac{2c}{\cosh\left[\ln\(\frac{2\ n!\Gamma(n+\frac32) t^{2\gamma-2n-1}}{\pi(h^*)^2\ \(2c(16c^2+\frac{2\rho\ln t}{t})\)^{2n+\frac32}}\)\right]}=
\frac{2c}{\cosh\left[(2\gamma-2n-1)\ln t+\ln\(\frac{2\ n!\Gamma(n+\frac32)}{\pi(h^*)^2\ \(2c(16c^2+\frac{2\rho\ln t}{t})\)^{2n+\frac32}}\)\right]}
,\end{equation}
and
 the solution of the initial value problem $q(x,t)$
is approximated by $q^{(1)}_{\infty}$ with the accuracy
$\mathcal{O}(t^{-1}):$
$$q(x,t)=q^{(1)}_{\infty}(x,t)+\mathcal{O}(t^{-1/2}).$$

On the curve $\xi=\frac{c^2}{3}-\frac{\rho\ln t}{12 t}$ we have $v=1-\frac{3\xi}{c^2}=1-\frac{x}{4c^2t}=\frac{\rho\ln t}{4c^2t},$ and $$-2c(x-4c^2t)=8c^3vt=2c\rho\ln t=(2\gamma+\frac12)\ln t,$$
and substituting this into (\ref{q_inf_1}), we obtain
\begin{equation}\label{KK}q^{(1)}_{\infty}(x,t)=
\frac{2c}{\cosh\(2c(x-4c^2t)+(2n+\frac32)\ln
t-\ln\frac{4}{(h^*)^2}-\ln\frac{n!\Gamma(n+\frac32)}{2\pi}+(2n+\frac32)\ln[16c^3(2+v)]\)}=
\end{equation}
$$=\frac{2c}{\cosh\(2c(x-4c^2t)+(2n+\frac32)\ln
t-\ln\frac{2n!\Gamma(n+\frac32)}{\pi\ (h^*)^2}+(2n+\frac32)\ln(32c^3)\)}+\mathrm{O}(\frac{\ln
t}{t}),$$
This coincides with the formula by Khruslov and Kotlyarov \cite{KK}, Thm. \ref{teorKK} on p. \pageref{teorKK}.
\subsubsection{Second refined approximation of $\hskip-.5mmM^{\hskip-.5mm(1)\hskip-.5mm}\hskip-.5mm\hskip-.5mm$}
\label{sect_sub_ref2_second}
To deal with those $\gamma$ such that $\left\{\gamma\right\}\in(\frac12,1),$ we introduce another refined approximation of $M^{(1)}.$ Namely, by following a similar strategy as in the previous section we  now "remove" the $\frac{1}{\zeta}$ term in the $(1,2)$ entry in the asymptotics
(\ref{Las2}) of $L,$ and in the $(2,1)$ entry in asymptotics (\ref{Lasd2}) of $L_d$ at $\infty,$
$$\begin{pmatrix}
   1 & \frac{R_2}{\zeta} \\ 0 & 1
  \end{pmatrix}L =
\begin{pmatrix}
 1+\mathcal{O}(\frac{1}{\zeta}) & \mathcal{O}(\frac{1}{\zeta^2}) \\
\mathcal{O}(\frac{1}{\zeta}) & 1 + \mathcal{O}(\frac{1}{\zeta})
\end{pmatrix},
\qquad
\begin{pmatrix}
   1 & 0 \\ \frac{R_2}{\zeta_d} & 1
  \end{pmatrix}L_d =
\begin{pmatrix}
 1+\mathcal{O}(\frac{1}{\zeta_d}) & \mathcal{O}(\frac{1}{\zeta_d}) \\
\mathcal{O}(\frac{1}{\zeta_d}) & 1 + \mathcal{O}(\frac{1}{\zeta_d})
\end{pmatrix},
$$
where $$R_2=\frac{2\pi\i\ n}{n!\,\Gamma(n+\frac12)},\quad n\geq0,\qquad \textrm{ and } R_2=0\quad \textrm{ for }n=0.
$$
In keeping with the previous strategy, we introduce 
\begin{equation}
\label{G2}G_2=I+\frac{A_2}{k-\i c}+\frac{\widetilde A_2}{k+\i c}
\end{equation}
 and define the refined approximation of $M^{(1)}$ by
\begin{equation}\label{Minf2}  M^{(2)}_{\infty}=\begin{cases}
G_2\(\frac{k-\i c}{k+\i c}\)^{-n\sigma_3},\quad |k\mp\i c|>r,
\\G_2B\begin{pmatrix}1&\frac{R_2}{\zeta}\\0&1\end{pmatrix} L (-\i\phi)^{\sigma_3/2}t^{\gamma\sigma_3},\quad |k-\i c|<r,
\\
G_2B_d
\begin{pmatrix}
1&0\\\frac{R_2}{\zeta_d}&1
\end{pmatrix}L_d(\i\phi_d)^{-\sigma_3/2}t^{-\gamma\sigma_3},\quad
|k+\i c|<r,
\end{cases}
\end{equation}
with $B, B_d$ as in (\ref{B_Bd_finite_n}) with $\alpha=-n.$ The error matrix $$E^{(2)}:=M^{(1)}\(M^{(2)}_{\infty}\)^{-1}$$ has no discontinuity in the disks $|k\mp\i c|<r,$
and on the circles the jump is
\begin{equation}
 \label{J_E_refined_2}J_{E^{(2)}}=G\begin{pmatrix}1+\mathrm{O}(t^{-1})&\mathrm{O}(t^{-2\gamma+2n-2})\\\mathrm{O}(t^{2\gamma-2n-1})&1+\mathrm{O}(t^{-1})\end{pmatrix}G^{-1}, k\in\partial C,
\ \
J_{E^{(2)}}=G\begin{pmatrix}1+\mathrm{O}(t^{-1})&\mathrm{O}(t^{2\gamma-2n-1})\\\mathrm{O}(t^{-2\gamma+2n-2})&1+\mathrm{O}(t^{-1})\end{pmatrix}G^{-1},
k\in\partial C_d.
\end{equation}
We see that estimates in (\ref{J_E_refined_2}) are shifted with respect to estimates in (\ref{J_E_refined_1}), which allows to use both of them for different ranges of $\gamma.$
For $
 \left\{\gamma\right\}\in(\frac12,1)$  we take $$n=\lfloor \gamma\rfloor+1,
$$
 then $J_{E^{(2)}}$ in (\ref{J_E_refined_2}) is of the order
$$J_{E^{(2)}} = I+\mathcal{O}(t^{-1}).$$
The minimal possible value of $\gamma$ according to (\ref{gamma}), is $\frac{-1}{4},$ and in this case we take $n=0.$ Hence, we are able to handle all the cases $\gamma\geq\frac{-1}{4}$ with constructions in terms of Laguerre polynomials
with nonnegative index $n.$
The matrix  $G_2$ (\ref{G2}) is determined by the requirement that  $M^{(2)}$ is  regular at the points $k=\pm\i c$: similar arguments lead to 
$$A_2=\begin{pmatrix}
       0 &  b_2 \\ 0 & a_2
      \end{pmatrix},\quad \widetilde A_2=
\begin{pmatrix}
 \widetilde a_2 & 0\\ \widetilde b_2 & 0
\end{pmatrix},
$$
where $a_2,b_2,\widetilde a_2,\widetilde b_2$ satisfy the system (\ref{eq_abtildeab}) with $\widehat R_1$, $\widehat R_{1,d}$ replaced by $\widehat R_2$, $\widehat R_{2,d}$, respectively, and
$a,b,\widetilde a,\widetilde b$ replaced with $a_2,b_2,\widetilde a_2,\widetilde b_2,$ respectively. Hence, $a_2,b_2,\widetilde a_2,\widetilde b_2$ are defined by (\ref{abtildeab_sol1}), (\ref{abtildeab_sol2}), where we replace
$\widehat R_1,$ $\widehat R_{1,d}$ with $\widehat R_2,$ $\widehat R_{2,d}.$ Here
$$\widehat R_2=R_2t^{2n-2\gamma-1}\lim\limits_{k\to\i c}\(\frac{k-\i c}{k+\i c}\)^{-2n}z^{2n}(-\i\phi)^{-1}=\frac{-2\pi}{\Gamma(n)\Gamma(n+\frac12)}\frac{(h^*)^2}{4}\left[2c(16c^2+\frac{2\rho\ln t}{t})\right]^{2n+\frac12}t^{2n-2\gamma-1},$$
$$\widehat R_{2,d}=R_2t^{2n-2\gamma-1}\lim\limits_{k\to-\i c}\(\frac{k-\i c}{k+\i c}\)^{2n}z_d^{2n}(\i\phi_d)^{-1}=-\widehat R_2geq0.$$
Hence, again $G_2$ does not contribute into asymptotics (\ref{J_E_refined_2}), and \begin{equation}\label{q_inf_2}q^{(2)}_{\infty} = 2i\lim\limits_{k\to\infty}kM^{(2)}_{12}=2i\lim\limits_{k\to\infty}kM^{(2)}_{21}=2ib_2=2i\widetilde b_2=
\frac{2c}{\cosh\left[\ln\(\frac{2\Gamma(n)\Gamma(n+\frac12) t^{2\gamma-2n+1}}{\pi(h^*)^2\ \(2c(16c^2+\frac{2\rho\ln t}{t})\)^{2n-\frac12}}\)\right]}\end{equation}
for $n\geq 1,$ and $q_{\infty}^{(2)}=0$ for $n=0.$ Further,
$q(x,t)=q^{(2)}_{\infty}(x,t)+\mathcal{O}(t^{-1}).$
We see that (\ref{q_inf_2}) coincides with (\ref{q_inf_1}), if we replace in the latest $n$ with $n-1.$

To finish with the proof of Theorem \ref{teorKK_refined} it is enough to notice that all the estimates in subsections \ref{sect_sub_ref1_second}, \ref{sect_sub_ref2_second} are uniform with respect to finite shifts of
the parameter $\rho.$

\section{"Mesoscopic" regime}\label{sect_mesoscopic}
We extend the result of the previous section
to the case when the number of the soliton $n$ grows unboundedly but at a fixed rate in $t$.  Let
$(x,t)$ lie on the curve
\begin{equation}
\label{curve2}\frac{x}{12t}\equiv\xi=\frac{c^2}{3}-\frac{\beta
t^{\sigma}\ln t}{12t},\quad \sigma\in[0,1),\quad \beta>0.
\end{equation}
The results of  Section \ref{sect_as_sol}
suggests that $(x,t)$ is  constrained to an
asymptotic soliton with number $n$ of the order $\beta t^{\sigma}$
(indeed we will see further that it is of the order
$n\sim\frac{\beta t^{\sigma}}{1-\sigma}$), so that $(x,t)$ will
not be constrained to a one particular asymptotic soliton, but
will move move through the bulk of them.

It is not quite trivial to offer uniform estimates for the  behaviour with respect to $n$ in the constructions of Sec.  \ref{sect_as_sol}. For example, consider the construction of Sec.  \ref{sect_as_sol} using the 
matrix-valued function $L(\zeta)$ (\ref{LaguerreMatrixfiniten}). The
monic Laguerre polynomial $\pi_n(\zeta)$ there  has the following representation:
$$
\pi_n(\zeta)=\zeta^n\(1-\frac{(n+\frac12)n}{\zeta}+\frac{(n+\frac12)n(n-\frac12)(n-1)}{2!\zeta^2}+...\)=\zeta^n\(1-\frac{(n)_{(1)}}{\zeta}+\frac{(n)_{(2)}}{2!\zeta^2}+...\)=
$$
\begin{equation}\label{Lag_pol}=
\zeta^n\sum\limits_{j=0}^{n}\frac{(-1)^j(n)_{(j)}}{j!\zeta^j},
\end{equation}
where we denoted for brevity 
\begin{equation}
\label{red_d}(a)_{(b)}=(a+\frac12)a(a-\frac12)(a-1)...(a+\frac32-b)(a-b+1), \textrm{ for } b\geq 1, b\in\mathbb{N},\quad (a)_{(0)}=1.
\end{equation}

\noindent Its Cauchy transform 
\begin{equation}
\label{Cauchy_Transform}C_{n-1}(\zeta):=\frac{1}{2\pi\i}\int\limits_{0}^{+\infty}\frac{\pi_{n-1}(s)\sqrt{s}\e^{-s}\d s}{s-\zeta}
\end{equation}
admits an asymptotic expansion
\begin{eqnarray}
\frac{-\zeta^n}{\Gamma(n)\Gamma(n+\frac12)}\int\limits_{0}^{+\infty}\frac{\pi_{n-1}(s)\sqrt{s}\e^{-s}\d s}{s-\zeta}=1+\frac{n(n+\frac12)}{\zeta}+\frac{n(n+\frac12)(n+1)(n+\frac32)}{2!\zeta}+...=\nonumber \\
=1+\frac{(n)^{(1)}}{\zeta}+
\frac{(n)^{(2)}}{2!\zeta^2}+...=
\label{Cauchy_Transform_Asymp_Series}\sum\limits_{j=0}^{\infty}\frac{(n)^{(j)}}{j!\zeta^j},\end{eqnarray}
where we denoted for brevity 
\begin{equation}
\label{red_u}(a)^{(b)}=a(a+\frac12)(a+1)(a+\frac32)...(a+b-1)(a+b-\frac12), \textrm{ for } b\geq 1, b\in\mathbb{N},\quad (a)^{(0)}=1.
\end{equation}
While the $n$-dependence in $\pi_n(\zeta)$ can be controlled by making a change of variable $\zeta\mapsto n^2\zeta,$ this does not help in the asymptotic series for the Cauchy transform.
Instead,  the next lemma provides us with a uniform control in $n$ of the asymptotic expansion of  $L$. 
\begin{lem}
\label{lem_L}
Define the functions $h(\zeta), \delta(\zeta)$, analytic in  $\zeta\in\mathbb{C}\setminus[0,1]$
\begin{eqnarray}
 h(\zeta)&=&2\int\limits_{1}^{\zeta}\sqrt{\frac{s-1}{s}}\d s-2\zeta+\ln\zeta+\ln4\e\\
 &=&2\sqrt{\zeta(\zeta-1)}-2\zeta+\ln\(4\e\zeta(-1+2\zeta-2\sqrt{\zeta(\zeta-1)})\)
 \\
\delta(\zeta)&=&\(\zeta(-1+2\zeta-2\sqrt{\zeta(\zeta-1)})\)^{\frac14}.
\end{eqnarray}
Given the Laurent expansions of $h(\zeta)$ 
\begin{equation}
\label{h}h(\zeta)=\sum\limits_{j=1}^{\infty}\frac{2(2j-1)!}{(j+1)(j!)^2(4\zeta)^j}=:\sum\limits_{j=1}^{\infty}\frac{b_j}{\zeta^j};
\end{equation}
we denote by 
$$
 h_{K}(\zeta)=\sum\limits_{j=1}^{K-1}\frac{b_j}{\zeta^j}, \qquad r_{K}(\zeta)=\sum\limits_{j=K}^{\infty}\frac{b_j}{\zeta^j}.
$$
the partial sums and tails.
Define also the matrix--valued function 
$M_{mod}(\zeta)$ as follows: 
$$
M_{mod}(\zeta)=\begin{pmatrix}
                                               \frac12\(\gamma(\zeta)+\gamma^{-1}(\zeta)\) & \frac{i}{2}\(\gamma(\zeta)-\gamma^{-1}(\zeta)\)\\
                           \frac{-i}{2}\(\gamma(\zeta)-\gamma^{-1}(\zeta)\) & \frac12\(\gamma(\zeta)+\gamma^{-1}(\zeta)\)
                                              \end{pmatrix},\qquad \gamma(\zeta)=\sqrt[4]{\frac{\zeta-1}{\zeta}}.
$$
Finally, 
let $L$ be as in (\ref{LaguerreMatrixfiniten}) and consider the matrix-valued functions
\begin{equation}
\label{Q_from_L}\mathcal{Q}_n(\zeta) = \(\frac{n^{n+\frac14}}{\e^n}\)^{\sigma_3}L(4n\zeta)(4n\zeta)^{n\sigma_3}\e^{-nh(\zeta)\sigma_3}\(\frac{\e^n}{n^{n+\frac14}}\)^{\sigma_3},
\end{equation}
$$
\mathcal{E}_n(\zeta) = \(\frac{n^{n+\frac14}}{\e^n}\)^{\sigma_3}L(4n\zeta)(4n\zeta)^{n\sigma_3}\e^{-nh(\zeta)\sigma_3}\(\frac{\e^n}{n^{n+\frac14}}\)^{\sigma_3}
\(\sqrt{2}\delta(\zeta)\)^{-\sigma_3}\(M_{mod}(\zeta)\)^{-1},
$$
Let $U$ be any  open  set $U\supset[0,1]$ and fix $\mathbb N \ni M\geq 0$. 
The functions $\mathcal{Q}_n(\zeta)$ and $\mathcal{E}_n(\zeta)$ admit the expansion into series
$$
\mathcal{Q}_n(\zeta)=I+\sum\limits_{j=1}^{M-1}\frac{q_j(n)}{\zeta^j}+r_{\mathcal{Q},M}(\zeta;n),
$$
$$
\mathcal{E}_n(\zeta)=I+\sum\limits_{j=1}^{M-1}\frac{e_j(n)}{\zeta^j}+r_{\mathcal{E},M}(\zeta;n).
$$ 
The expansions are uniform in  $\zeta\in\mathbb{C}\setminus U$ in the sense that there is a constant $C_M<\infty$ such that  
 $$
 \sup\limits_{n\geq 1}|q_j(n)|<\infty, \qquad \sup\limits_{n\geq1}|r_{\mathcal{Q},M}(\zeta;n)|\leq\frac{C_M}{|\zeta|^M},\qquad \sup\limits_{n\geq1}|r_{\mathcal{E},M}(\zeta;n)|\leq\frac{C_M}{|\zeta|^M}.
 $$
Moreover,
$$\lim\limits_{n\to\infty}e_j(n)=0,\qquad \textrm{ and } \lim\limits_{n\to\infty}r_{\mathcal{E},M}(\zeta;n)=0 \textrm { for any fixed } \zeta, M.$$
\end{lem}
We prove Lemma \ref{lem_L} in Appendix \ref{sect_append}.
\begin{com}
As a byproduct of Lemma \ref{lem_L} we can control the $n$-behaviour of the asymptotic series (\ref{Cauchy_Transform_Asymp_Series}) for the Cauchy transform $C_{n-1}(\zeta)$(\ref{Cauchy_Transform}) of the 
Laguerre polynomials, namely 
(we use the notation (\ref{red_u}) below):
$$\e^{-nh(\zeta)}\frac{-(4n\zeta)^nC_{n-1}(4n\zeta)}{\Gamma(n)\Gamma(n+\frac12)}=\e^{-nh(\zeta)}\sum\limits_{j=0}^{\infty}\frac{(n)^{(j)}}{j!(4n\zeta)^n}
=
\e^{-nh(\zeta)}\sum\limits_{j=0}^{\infty}\frac{\Gamma(n+j)\Gamma(n+j+\frac12)}{j!\Gamma(n)\Gamma(n+\frac12)(4n\zeta)^n}
=$$ $$=1+\sum\limits_{j=1}^{M-1}\frac{s_j(n)}{\zeta^j}+r_M(\zeta;n),\qquad \textrm{ for any integer } M\geq 1
$$
and $$\sup\limits_{n\geq 1}|s_j(n)|\leq C_j,\quad \sup\limits_{n\geq 1}|r_M(s;n)|\leq\frac{C_M}{|\zeta|^M},\quad \textrm{ for any } \zeta\in\mathbb{C}\setminus U,$$
where $M\geq 1$ and $C_j$ are independent of $n,$
which we believe is also of independent interest.
\end{com}
\begin{com}
From the representation of Laguerre polynomials (\ref{Lag_pol}) and asymptotic series for its Cauchy transform (\ref{Cauchy_Transform_Asymp_Series}) we find an asymptotic series for $\mathcal{Q}_n(\zeta)$ (using the notations
(\ref{red_d}), (\ref{red_u})) $$\begin{pmatrix}\e^{-nh(\zeta)}\sum\limits_{j=0}^{\infty}\frac{(n)^{(j)}}{j!(4n\zeta)^j} & \(\frac{2\pi\ n^{2n-\frac12}}{\Gamma(n+\frac12)\Gamma(n)\e^{2n}}\)
\frac{(-i)\ \e^{nh(\zeta)}}{4\zeta}\sum\limits_{j=0}^{n-1}\frac{(-1)^j(n-1)_{(j)}}{j!(4n\zeta)^j}
\\
\(\frac{\Gamma(n+1)\Gamma(n+\frac32)\e^{2n}}{2\pi\ n^{2n+\frac32}}\)\frac{i\ \e^{-nh(\zeta)}}{4\zeta}\sum\limits_{j=0}^{\infty}\frac{(n+1)^{(j)}}{j!(4n\zeta)^j} &
\e^{nh(\zeta)}\sum\limits_{j=0}^{n}\frac{(-1)^j(n)_{(j)}}{j!(4n\zeta)^j}
\end{pmatrix}.
$$
In particular,
$$\mathcal{Q}_n(\zeta) = \begin{pmatrix}
                          1+\frac{1}{8\zeta} & \frac{2\pi\ n^{2n-\frac12}}{\Gamma(n+\frac12)\Gamma(n) \e^{2n}}\ \frac{-\i}{4\zeta} \\
\frac{\Gamma(n+1)\Gamma(n+\frac32)\e^{2n}}{2\pi\ n^{2n+\frac32}}\ \frac{\i}{4\zeta} & 1-\frac{1}{8\zeta}
                         \end{pmatrix}+\mathcal{O}(\zeta^{-2})
$$
and \begin{equation}\label{Q_ref1}\begin{pmatrix}
       1&0\\\frac{R_1}{\zeta} & 1
      \end{pmatrix}\mathcal{Q}_n(\zeta)=
\begin{pmatrix}
 1+\mathcal{O}(\frac{1}{\zeta}) & \mathcal{O}(\frac{1}{\zeta}) \\ \mathcal{O}(\frac{1}{\zeta^2}) & 1+\mathcal{O}(\frac{1}{\zeta})
\end{pmatrix},\end{equation}
\begin{equation}\label{Q_ref2}\begin{pmatrix}
       1&\frac{R_2}{\zeta} \\0 & 1
      \end{pmatrix}\mathcal{Q}_n(\zeta)=
\begin{pmatrix}
 1+\mathcal{O}(\frac{1}{\zeta^2}) & \mathcal{O}(\frac{1}{\zeta}) \\ \mathcal{O}(\frac{1}{\zeta}) & 1+\mathcal{O}(\frac{1}{\zeta})
\end{pmatrix},\end{equation}
with \begin{equation}\label{R1R2_mes}R_1=\frac{-\i}{4}\frac{\Gamma(n+1)\Gamma(n+\frac32)\e^{2n}}{2\pi\ n^{2n+\frac32}},\qquad R_2 = \frac{\i}{4}\frac{2\pi\ n^{2n-\frac12}}{\Gamma(n+\frac12)\Gamma(n)\e^{2n}},\end{equation}
and all $\mathcal{O}$ estimates are uniform with respect to $n\geq 1.$
\end{com}
\begin{cor}\label{cor_Y_L}
Consider the matrix--valued function
$$
Y(\zeta):=:\Lambda^{(n+\frac14)\sigma_3}L(\zeta\Lambda)\Lambda^{-\sigma_3/4}:=
\begin{pmatrix}
\frac{-\Lambda^{n+1/4}}{2\pi\i\
\Gamma(n+\frac12)\Gamma(n)}\int\limits_{0}^{+\infty}\frac{\pi_{n-1}(s\Lambda)\sqrt{s}\
\e^{-\Lambda s}ds}{s-\zeta} &\frac{-2\pi\i\
\Lambda^{n+\frac12}}{\Gamma(n+\frac12)\Gamma(n)}\
\pi_{n-1}(\zeta\Lambda)
\\ \frac{1}{2\pi\i\ \Lambda^n
}\int\limits_{0}^{+\infty}\frac{\pi_{n}(s\Lambda)\sqrt{s}\
\e^{-\Lambda s}ds}{s-\zeta} &
\frac{\pi_n(\zeta\Lambda)}{\Lambda^n}
\end{pmatrix}.
$$
The matrix $Y(\zeta)$ solves a RHP of the form
\begin{eqnarray}
\nonumber&&
 Y_-=Y_+\begin{pmatrix}1 & 0 \\ -\sqrt{\zeta}\e^{-\Lambda \zeta} & 1\end{pmatrix}, \zeta\in(0,+\infty),
\\\nonumber
&&Y(\zeta)=\(I+\mathrm{O}\(\frac{1}{\zeta}\)\)\zeta^{-n\sigma_3},
\end{eqnarray}
and it can be written as follows
\begin{equation}
\label{Y}Y(\zeta)=\(\frac{\Lambda^{n+\frac14}\e^{n}}{n^{n+\frac14}}\)^{\sigma_3}\mathcal{Q}_n\(\frac{\zeta\Lambda}{4n}\)\zeta^{-n\sigma_3}\e^{nh(\frac{\zeta\Lambda}{4n})\sigma_3}
\(\frac{n^{n+\frac14}}{\Lambda^{n+\frac14}\e^{n}}\)^{\sigma_3},
\end{equation}
where $\mathcal{Q}$ is defined in (\ref{Q_from_L}).
\end{cor}

Lemma \ref{lem_L} allows us to control the large  $n$-behaviour of $L$ in the approximate solutions $M_{\infty}$ (\ref{Minf}), $M_{\infty}^{(1)}$ (\ref{Minf1}), $M_{\infty}^{(2)}$ (\ref{Minf2}), regardless of whether $n\geq 1$ is bounded or
growing together with $t$.

But there is also another issue for growing $n,$ namely providing bounds for the  terms 
$$
\(\frac{k-\i c}{k+\i c}\)^{-n}z^{n},\qquad \(\frac{k+\i c}{k-\i c}\)^{n}z_d^{n}
$$
 on the circles $|k\mp\i c|=r$ in the estimates (\ref{J_E_rough_pre1}), (\ref{J_E_rough_pre2}).
While, for bounded $n$, these quantities are separated both from $0$ and $\infty$,  for growing $n$ this is no longer the case. We also need to control the term $e^{n h(\zeta)}$ in the  asymptotics (\ref{Q_from_L}), since $n h(\zeta)$ will be of the order
$\frac{n^2}{t}$ (as we will see shorthly), and for growing fast enough $n$ this term is not negligible. To overcome these difficulties, we  need to make a local conformal change of variables $z,$ $z_d,$ as described in
Lemma \ref{lem_conform_z} below.

Mimicking the approach of Sec. \ref{sect_as_sol}, we first transform  our original RH problem \ref{RH_problem_1} to a RH problem for  a new function $M^{(1)}$ (\ref{M1}) with jump $J^{(1)}$ (\ref{J1}).
The following estimates hold in the intervals
$k\in\pm(\i c,\i (c-r))$ (small enough parameter $r>0$ will still be fixed throughout this section)
$$
f(k)\e^{2it\theta}=\frac{2\sqrt{2}\ \i}{(h^{*})^2\sqrt{c}}\sqrt{y}\(1+\mathcal{O}(\sqrt{y})\)t^{2c\beta t^{\sigma}}\e^{-t z},
$$
$$
-\ol{f(\ol{k})}\e^{-2it\theta}=\frac{2\sqrt{2}\ \i}{(h^{*})^2\sqrt{c}}\sqrt{y_d}\(1+\mathcal{O}(\sqrt{y_d})\)t^{2c\beta t^{\sigma}}\e^{-t z_d},
$$
and
\begin{equation}
z=z(y;t)=(16c^2+\frac{2\beta t^{\sigma}\ln t}{t})y-24cy^2+8y^3,\qquad
z_d=z_d(y;t)=(16c^2+\frac{2\beta t^{\sigma}\ln t}{t})y_d-24cy^2_d+8y^3_d,
\label{zdef}
\end{equation} and $y,$ $y_d$ are as in (\ref{y_yd}).

Let $\widetilde z$, $\widetilde z_d$ be some local conformal transformations of variables $z,$ $z_d$, to  be specified later, and denote
\begin{equation}
\label{zeta_mesosc}\zeta := \frac{\widetilde z t}{\Lambda},\quad \zeta_d := \frac{\widetilde z_d t}{\Lambda},
\end{equation}
where $\Lambda$ is a large parameter  growing with $t$ as $\Lambda\asymp t^{\sigma}$. Then
$$f(k)\e^{2it\theta}=\frac{2\sqrt{2}\ \i}{(h^{*})^2\sqrt{c}}\sqrt{\frac{y}{\widetilde z}}\(1+\mathcal{O}(\sqrt{y})\)t^{2c\beta t^{\sigma}}\sqrt{\widetilde z}\e^{-t z}=:\i\phi(k,t)t^{2c\beta t^{\sigma}}\sqrt{\widetilde z}\e^{-t z},$$
 $$
 -\ol{f(\ol{k})}\e^{-2it\theta}=\frac{2\sqrt{2}\ \i}{(h^{*})^2\sqrt{c}}\sqrt{\frac{y_d}{\widetilde z_d}}\(1+\mathcal{O}(\sqrt{y_d})\)t^{2c\beta t^{\sigma}}\sqrt{\widetilde z_d}\e^{-t z_d}=:\i\phi_d(k,t)t^{2c\beta t^{\sigma}}
 \sqrt{\widetilde z_d}\e^{-t z_d},$$
The functions $\phi(k,t)$, $\phi_d(k,t)$, analytic near $k=\pm\i c$,  are uniformly separated from $0$ and $\infty$  with respect to $t\geq 1.$
Let us choose an integer $K\geq 0$ such that 
\begin{equation}
\label{K}K\geq \frac{2\sigma-1}{1-\sigma},\quad \sigma\leq\frac{K+1}{K+2}.
\end{equation} 
For $\sigma\in[0,\frac12]$ we take $K=0,$ for $\sigma\in(\frac12,\frac23]$ we take $K=1,$ etc.
\noindent The next lemma specifies the conformal maps $z\mapsto\widetilde z,$ $z_d\mapsto\widetilde z_d.$

\begin{lem}\label{lem_conform_z}
Denote
$$
v:=1-\frac{3\xi}{c^2}=\frac{\beta t^{\sigma}\ln t}{4c^2t},\quad
\delta = \frac{2n}{c^3t}.
$$
 For sufficiently small $v, \delta$
there exists a local conformal map $\widetilde{z}(y, \delta, v)$
such that
\begin{equation}
\label{z_conform} 
z(y) -\frac{2n}{t}\ln \frac{y}{2c-y}+\frac{2n}{t}\sum\limits_{j=1}^K\frac{A_j(\i c-\i y)}{(y(2c-y))^j}
=
\widetilde{z}-\frac{2n}{t}\ln\widetilde{z}+\frac{2n}{t}Q+\frac{2n}{t}\sum\limits_{j=1}^K\frac{b_j (4n)^j}{t^j\widetilde{z}^j}.
\end{equation}
($z(y) = z(y;t)$ as in \eqref{zdef}).
Furthermore, $\widetilde z$ and $Q$ 
admit expansions of the form
\begin{eqnarray}
\widetilde{z} &=& yc^2\left[16+8v-\frac{8+v}{2+v}\
\frac{n}{c^3t}+\mathcal{O}\(\frac{n^2}{t^2}\)\right]y+cy^2\(-24-\frac{24-4v+v^2}{4(2+v)^2}\ 
\frac{n}{c^3t}+\mathcal{O}\(\frac{n^2}{t^2}\)\)+\nonumber
\\
&&+y^3\(8+\mathcal{O}\(\frac{n}{t}\)\)+\mathcal{O}\(y^4nt^{-1}\),
\label{zy}
\end{eqnarray}
and
\begin{eqnarray}\nonumber
Q&=&\ln(16c^3(2+v))+\mathcal{O}\(\frac{n}{t}\)\qquad \textrm{ for } \quad K=0,\nonumber\\
Q&=&\ln(16c^3(2+v))-\frac{3(8+v)}{16(2+v)^2}\
\frac{n}{c^3t}+\mathcal{O}\(\frac{n^2}{t^2}\)\qquad \textrm{ for } \quad K=1,\nonumber\\
Q&=&\ln(16c^3(2+v))-\frac{3(8+v)}{16(2+v)^2}\
\frac{n}{c^3t}-\frac{400+48v+5v^2}{128(2+v)^4}\(\frac{n}{c^3t}\)^2+\mathcal{O}\(\frac{n^3}{t^3}\)\qquad \textrm{ for }\quad K=2.
\label{Q}
\end{eqnarray}
\end{lem}

Statements of the type of Lemma \ref{lem_conform_z} fall within the general class of  ``normal form'' of singularities. A completely similar lemma can be found in \cite{Bertola Buckingma Lee Pierce} (Prop. 2.2) with the only difference that in our case $z(y;t)  = \mathcal O(y)$ uniformly as $t\to \infty$ while in the case of   \cite{Bertola Buckingma Lee Pierce} we have $\mathcal O(y^2)$ and hence the normal form starts with $\widetilde z^2$ in the right side of \eqref{z_conform}. We omit the detailed proof because it is not substantially different from the one referred to above.

Let $\widetilde z_d$ be the conformal map of $z_d$ determined by \eqref{z_conform}, where we substituted $y$ with $y_d$, $z$ with $z_d.$
Then we have
\begin{equation}
\label{conform_z}\e^{-t z}=\e^{-t\widetilde z}\e^{-2nQ}\ \cdot \ \frac{\widetilde z^{2n}(k+\i c)^{2n}}{(k-\i c)^{2n}}\ \cdot \  \e^{-2n\sum\limits_{j=1}^K\frac{b_j(4n)^j}{t^j\widetilde z^j}}\ \cdot \  \e^{2n\sum\limits_{j=1}^K\frac{A_j k}{(k^2+c^2)^j}},\end{equation}
and
\begin{equation}
\label{conform_zd}\e^{-t z_d}=\e^{-t\widetilde z_d}\e^{-2nQ}\ \cdot \ \frac{\widetilde z_d^{2n}(k-\i c)^{2n}}{(k+\i c)^{2n}}\ \cdot \  \e^{-2n\sum\limits_{j=1}^K\frac{b_j(4n)^j}{t^j\widetilde z_d^j}}\
\cdot \  \e^{-2n\sum\limits_{j=1}^K\frac{A_j k}{(k^2+c^2)^j}}.
\end{equation}


\noindent Now we are ready to construct two approximations $M^{(j)}_{\infty}, j=1,2,$ of the function $M^{(1)}$ (analogues of (\ref{Minf1}), (\ref{Minf2}) in section \ref{sect_as_sol}):
\begin{equation}\label{M_inf_mes}\hskip-0cm M^{(j)}_{\infty}=
\begin{cases}
G_{j}\(\frac{k-\i c}{k+\i
c}\)^{-n\sigma_3}\exp{\(n\sum\limits_{j=1}^K\frac{A_j
k}{(k^2+c^2)^j}\sigma_3\)},\quad |k\mp\i c|>r,
\\
\\
G_jB\Delta_{u,j}
Y(\zeta)
(-\i\phi)^{\frac{\sigma_3}{2}}\(t^{c\beta t^{\sigma}}\sqrt[4]{\frac{\Lambda}{t}}\ \e^{-nQ}\)^{\sigma_3}
\(\frac{\widetilde z\ (k+\i c)}{k-\i
c}\)^{n\sigma_3}
\times
\\
\hfill\times\exp\(-n\sum\limits_{j=1}^K\frac{b_j(4n)^j}{\Lambda^j\zeta^j}\sigma_3\)\exp\(n\sum\limits_{j=1}^K\frac{A_j
k}{(k^2+c^2)^j}\sigma_3\), \quad  |k-\i c|<r.
\\
\\
G_jB_d\Delta_{d,j}
Y_d(\zeta_d)
(\i\phi_d)^{\frac{-\sigma_3}{2}}\(t^{c\beta t^{\sigma}}\sqrt[4]{\frac{\Lambda}{t}}\ \e^{-nQ}\)^{-\sigma_3}\(\frac{k+\i c}{\widetilde z_d(k-\i c)}\)^{n\sigma_3}
\times
\\
\hfill\times\exp\(n\sum\limits_{j=1}^K\frac{b_j(4n)^j}{\Lambda^j\ \zeta_d^j}\sigma_3\)
\exp\(n\sum\limits_{j=1}^K\frac{A_j
k}{(k^2+c^2)^j}\sigma_3\), \quad|k+\i c|<r.
\end{cases}\end{equation}
Here $b_j$ are as in (\ref{h}), function $Y(\zeta)$ as in (\ref{Y}), and $$Y_d(\zeta_d):=\sigma_1Y(\zeta_d)\sigma_1,\qquad \mathcal{Q}_{d,n}(\zeta_d)=\sigma_1\mathcal{Q}_n(\zeta_d)\sigma_1,\quad \sigma_1=\begin{pmatrix}0&1\\1&0\end{pmatrix}.$$
Formulas (\ref{conform_z}), (\ref{conform_zd}) ensure that $M^{(j)}_{\infty},$ $j=1,2,$ have the same jump on the intervals $\pm(\i c,\i (c-r))$ as $M^{(1)}$ has.
For the same reasons as in section \ref{sect_as_sol}, i.e. in order to minimize jump on the circles $\partial C,$ $\partial C_d$ of the error matrix $E^{(j)}:=M^{(1)}\(M_{\infty}^{(j)}\)^{-1}, j=1,2,$ we take
$$B = \(\frac{t^{n+\frac14}\ \e^{nQ}}{\Lambda^{n+\frac14}\ t^{c\beta t^{\sigma}}}\)^{\sigma_3}(-\i\phi)^{-\sigma_3/2},\quad B_d = \(\frac{\Lambda^{n+\frac14}\ t^{c\beta t^{\sigma}}}{t^{n+\frac14}\ \e^{nQ}}\)^{\sigma_3}(\i\phi_d)^{\sigma_3/2}.$$
Then the jump matrix for $E^{(j)}_-=E^{(j)}_+J_{E^{(j)}},$ $j=1,2,$ on $k\in\partial C,$ $k\in\partial C_d,$ respectively, is (see (\ref{Q_from_L}))
\begin{equation}\label{J_E_mesosc_1}J_{E^{(j)}} = G_j\(\frac{\e^{n(Q+1)}\ t^{n+\frac14}}{n^{n+\frac14}\ t^{c\beta t^{\sigma}}}\)^{\sigma_3}\Delta_{u,j}\mathcal{Q}_n(\frac{\zeta\Lambda}{4n})
(-\i\phi)^{\frac{\sigma_3}{2}}
\e^{n(h(\frac{\zeta\Lambda}{4n})-\sum\limits_{j=1}^K\frac{b_j4^jn^j}{\Lambda^j\zeta^j})\sigma_3}
\(\frac{n^{n+\frac14}\ t^{c\beta t^{\sigma}}}{\e^{n(Q+1)}\ t^{n+\frac14}}\)^{\sigma_3}G_j^{-1}.\end{equation}
\begin{equation}\label{J_E_mesosc_2}J_{E^{(j)}} = G_j\hskip-1mm\(\hskip-1mm\frac{n^{n+\frac14}\ t^{c\beta t^{\sigma}}}{\e^{n(Q+1)}\ t^{n+\frac14}}\hskip-1mm\)^{\hskip-1.5mm\sigma_3}\hskip-1mm\Delta_{d,j}\mathcal{Q}_{d,n}(\frac{\zeta_d\Lambda}{4n})
(\i\phi_d)^{\frac{-\sigma_3}{2}}
\e^{-n(h(\frac{\zeta\Lambda}{4n})-\sum\limits_{j=1}^K\frac{b_j4^jn^j}{\Lambda^j\zeta^j})\sigma_3}
\(\frac{\e^{n(Q+1)}\ t^{n+\frac14}}{n^{n+\frac14}\ t^{c\beta t^{\sigma}}}\)^{\sigma_3}G_j^{-1}.\end{equation}
Under the assumption that  $n\asymp t^{\sigma}$,  we have that the middle exponential term in (\ref{J_E_mesosc_1}), ((\ref{J_E_mesosc_2})) is of the order
$$
\e^{n\left(
h(\frac{\zeta\Lambda}{4n})-\sum\limits_{j=1}^K\frac{b_j4^jn^j}{\Lambda^j\zeta^j}\right)\sigma_3}=I+\mathcal{O}(t^{K(\sigma-1)+2\sigma-1}),
$$
due to the choice of $K$ (\ref{K}), and the definition of $\zeta$ (\ref{zeta_mesosc}). Hence,
$$
J_{E^{(1)}} = G_1\begin{pmatrix}
                  1+\mathcal{O}\(\frac{\Lambda}{\widetilde z t}\) & \mathcal{O}\(\frac{\Lambda}{\widetilde z t}\)\frac{1}{\omega}
\\
\mathcal{O}\(\frac{\Lambda^2}{\widetilde z^2 t^2}\)\omega & 1+\mathcal{O}\(\frac{\Lambda}{\widetilde z t}\)
                 \end{pmatrix}\Omega G_1^{-1}, \partial C,\quad
J_{E^{(1)}} = G_1\begin{pmatrix}
                  1+\mathcal{O}\(\frac{\Lambda}{\widetilde z t}\) & \mathcal{O}\(\frac{\Lambda^2}{\widetilde z^2 t^2}\) \omega
\\
\mathcal{O}\(\frac{\Lambda}{\widetilde z t}\)\frac{1}{\omega} & 1+\mathcal{O}\(\frac{\Lambda}{\widetilde z t}\)
                 \end{pmatrix}\Omega G_1^{-1}, \partial C_d,
$$
$$J_{E^{(2)}} = G_1\begin{pmatrix}
                  1+\mathcal{O}\(\frac{\Lambda}{t}\) & \mathcal{O}\(\frac{\Lambda^2}{t^2}\)\frac{1}{\omega}
\\
\mathcal{O}\(\frac{\Lambda}{t}\)\omega & 1+\mathcal{O}\(\frac{\Lambda}{t}\)
                 \end{pmatrix}\Omega G_1^{-1}, \partial C,\quad
J_{E^{(2)}} = G_1\begin{pmatrix}
                  1+\mathcal{O}\(\frac{\Lambda}{t}\) & \mathcal{O}\(\frac{\Lambda}{t}\) \omega
\\
\mathcal{O}\(\frac{\Lambda^2}{t^2}\)\frac{1}{\omega} & 1+\mathcal{O}\(\frac{\Lambda}{t}\)
                 \end{pmatrix}\Omega G_1^{-1}, \partial C_d,
$$
where we denoted for brevity
\begin{equation}
\omega = \frac{t^{2c\beta t^{\sigma}}\ n^{2n+\frac12}}{\e^{2n(Q+1)}\ t^{2n+\frac12}},\quad \textrm { and } \quad \Omega = I+\mathcal{O}\(t^{K(\sigma-1)+2\sigma-1}\)
.
\label{defomega}
\end{equation} We need to choose such an $n$ that makes $\omega$ as close to
1 as possible. Denote by $\gamma$ exact solution of equation
\begin{equation}\label{gamma_eq_mes}\omega|_{n\mapsto\gamma}=1,\quad \textrm{ i.e. }\quad
\frac{t^{2c\beta t^{\sigma}}\
\gamma^{2\gamma+\frac12}}{\e^{2\gamma(Q(\gamma)+1)}\
t^{2\gamma+\frac12}} = 1.\end{equation}
Then
$$
\gamma\sim\frac{\beta t^{\sigma}}{1-\sigma}\quad \textrm{ and }
\quad \omega\asymp t^{2(1-\sigma)(\gamma-n)}.
$$ 
For the same reasons as the triangular factors in (\ref{Minf1}),
(\ref{Minf2}) in the previous section \ref{sect_as_sol} we introduce the
triangular matrices $\Delta_{u,j},$ $\Delta_{d,j},$ $j=1,2,$:
$$\Delta_{u,1}= \begin{pmatrix}
 1&0\\
 \frac{R_1 4n}{\Lambda \zeta}\ \frac{n^{2n+\frac12}}{\Lambda^{2n+\frac12}\ \e^{2n}}&1
 \end{pmatrix},
\quad
\Delta_{d,1}= \begin{pmatrix}
 1& \frac{R_1 4n}{\Lambda \zeta_d}\ \frac{n^{2n+\frac12}}{\Lambda^{2n+\frac12}\ \e^{2n}}
\\0&1
 \end{pmatrix},$$
$$
\Delta_{u,2}=\begin{pmatrix}1 & \frac{R_24n}{\zeta\Lambda}\ \frac{\Lambda^{2n+\frac12}\e^{2n}}{n^{2n+\frac12}} \\ 0&1\end{pmatrix},\quad
\Delta_{d,2}=\begin{pmatrix}1 & 0 \\ \frac{R_24n}{\zeta_d\Lambda}\ \frac{\Lambda^{2n+\frac12}\e^{2n}}{n^{2n+\frac12}} &1\end{pmatrix}.
$$
where,  due to
(\ref{Q_ref1}), (\ref{Q_ref2}), we have set $R_1,$ $R_2$ as in
\ref{R1R2_mes}.

Similarly to Sec. \ref{sect_as_sol}, the matrices  $M^{(j)}_{\infty}, j=1,2,$ are regular at $k=\pm\i c$ provided that $G_j, j=1,2,$ are of the form
$$
G_1=\begin{pmatrix}1+\frac{a_1}{k-\i c} & \frac{\widetilde b_1}{k+\i c} \\ \frac{b_1}{k-\i c} & 1+\frac{\widetilde a_1}{k+\i c}\end{pmatrix},\quad
G_2=\begin{pmatrix} 1+\frac{\widetilde a_2}{k+\i c} & \frac{b_2}{k-\i c} \\ \frac{\widetilde b_2}{k+\i c} & 1+\frac{a_2}{k-\i c}\end{pmatrix},
$$
and $a_j,b_j,\widetilde{a}_j,\widetilde b_j, j=1,2,$ are determined by the regularity condition at $k=\pm\i c$ of the expressions
$$G_1\begin{pmatrix}
      1 & 0\\\frac{-\widehat R_1}{\widetilde z } & 1
     \end{pmatrix},\quad G_1\begin{pmatrix} 1 & \frac{\widehat R_1}{\widetilde z_d} \\ 0 & 1 \end{pmatrix},\quad
G_2\begin{pmatrix}
    1 & \frac{-\widehat R_2}{\widetilde z} \\ 0 & 1
   \end{pmatrix},\quad
G_2\begin{pmatrix}
    1 & 0 \\ \frac{\widehat R_2}{\widetilde z_d} & 1
   \end{pmatrix},
$$
where  $$\widehat R_1 :=
\frac{\Gamma(n+1)\Gamma(n+\frac32)t^{2c\beta
t^{\sigma}}\lim\limits_{k\to\i c}\phi(k)}{2\pi\
t^{2n+\frac32}\e^{2nQ}},\quad \widehat R_2 := \frac{2\pi\
t^{2n-\frac12}\ \e^{2nQ}}{\Gamma(n+\frac12)\Gamma(n)\ t^{2c\beta
t^{\sigma}}\lim\limits_{k\to\i c}\phi(k)}.$$

Solving  the analogues of  (\ref{eq_abtildeab}), we obtain analogs of (\ref{abtildeab_sol1}), (\ref{abtildeab_sol2})
(we denote by $\widetilde z_y$ the derivative of $\widetilde z$ \eqref{zy} with respect to $y$ at the point $y=0$):
\begin{equation}
\label{abtildeab_sol1_mes}a_1=\frac{2\i c\ \widehat R_1^2}{4c^2\(\widetilde z_y\)^2+\widehat R_1^2},
\qquad
 b_1=\frac{-4\i c^2\ \widetilde z_y\
\widehat{R}_1}{4c^2\(\widetilde z_y\)^2+\widehat R_1^2},
\end{equation}
\begin{equation}\label{abtildeab_sol2_mes}\widetilde a_1= \frac{-2\i c\ \widehat R_1^2}{4c^2\(\widetilde z_y\)^2+\widehat R_1^2},
\qquad \widetilde b_1= \frac{-4\i c^2\ \widetilde z_y\
\widehat{R}_{1}}{4c^2\(\widetilde z_y\)^2+\widehat
R_1^2}.
\end{equation}

Finally,  for 
$\left\{\gamma\right\}\in[0,\frac12]$ we take the first approximation
$M^{(1)}_{\infty}$ with index $n=\lfloor \gamma\rfloor ,$ and for  $\left\{\gamma\right\}\in(\frac12,1)$ the
second approximation $M^{(2)}_{\infty}$ with index $n=\lfloor \gamma \rfloor+1$.  We then obtain the following asymptotics for $q(x,t):$ let $(x,t)$ be
on curve (\ref{curve2}), and let $\gamma$ be solution of equation
(\ref{gamma_eq_mes}). Take $n=\lfloor \gamma \rfloor.$ Then
\begin{eqnarray}
q(x,t)=\frac{2c}{\cosh\ln\left[2c(x-4c^2t)+(2n+\frac32)\ln t+2nQ+\frac32\ln(2c\widetilde z_y)-\ln\frac{\Gamma(n+1)\Gamma(n+\frac32)}{2\pi}
-\ln\frac{4}{(h^*)^2}\right]}
+
\nonumber\\ 
\label{q_mesoscopic}\hfill+\mathcal{O}(t^{K(\sigma-1)+2\sigma-1}).
\end{eqnarray}

 \begin{com} Multiplication by
$e^{n\sum\limits_{j=1}^K\frac{A_j k}{(k^2+c^2)^j}\sigma_3}$ in (\ref{M_inf_mes}) does not affect the error estimates for the 
 jumps outside of the disks $|k\mp\i c|<r.$ Indeed, $A_j\asymp\frac{n^j}{t^j}\asymp t^{j(\sigma-1)},$ hence the $j^{th}$ term gives the contribution of the order
$\exp\(t^{\sigma(j+1)-j}\),$ and since the error matrix $J^{(1)}$ outside of the disks is of the order $J^{(1)}=I+\mathcal{O}(\e^{-b t})$ for some positive $b>0,$ the jump for the error matrix 
$$E^{(j)}=M^{(1)}\(M_{\infty}^{(j)}\)^{-1}, j=1,2$$ is of the order $$J_{E^{(j)}}-I=\mathcal{O}\(\exp\(-b t + t^{\sigma(j+1)-j}\)\)=
\mathcal{O}\(\exp\left[-b t(1  + \mathcal{O}(t^{-(1-\sigma)(j+1))}\right]\),$$ which is still of the order $\mathcal{O}(\e^{-\widetilde b t})$ with some other constant $\widetilde b>0.$
\end{com}
\par\vskip 14pt
Let us now focus on the case $\sigma\in[0,\frac12).$ Lemma \ref{lem_z_tBDelta_equation} allows us to conclude that the quantity $\omega$ in 
\eqref{defomega} will be of the order $1+\mathcal{O}(t^{2\sigma-1}\ln t)$ provided that
we substitute $\gamma$ with $\frac{z}{2}$ in \eqref{gamma_eq_mes}.
This gives us the statement of Theorem \ref{thrm_as_sol_mes}. 

Beyond the regime $\sigma\in[0,\frac12),$ we obtain 
\vskip0cm

\begin{teor}\label{teor_mesoscopic}
 Let $(x,t)$ be on a curve $$\frac{x}{12t}\equiv\xi=\frac{c^2}{3}-\frac{\beta t^{\sigma}\ln t}{4c^2 t}.$$
Let $K\geq 0$ be an integer satisfying (\ref{K}), i.e.,  $K\geq\frac{2\sigma-1}{1-\sigma}.$
Let $Q,$ $\widetilde z_y$ be as in (\ref{Q}), (\ref{zy}) and $\gamma$ be solution of (\ref{gamma_eq_mes}). Let $n$ be an integer such that $n\leq \gamma\leq n+1.$

Then the solution of the initial value problem (\ref{mkdv}), (\ref{ic}) is given by formula (\ref{q_mesoscopic}).
In particular, for $\sigma\leq\frac12,$ $K=0,$ $n\asymp t^{\sigma}$ we have 
$$
q=\frac{2c}{\cosh\(2c(x-4c^2t)+(2n+\frac32)\ln t+(2n+\frac32)\ln(32c^3)-\ln\frac{\Gamma(n+1)\Gamma(n+\frac32)}{2\pi}-\ln\frac{4}{(h^*)^2}\)}+\mathcal{O}(t^{2\sigma-1}\ln t).
$$
\end{teor}

\begin{com}
 To check the consistency of Theorem \ref{teor_mesoscopic} with Theorem \ref{thrm_Elliptic_est} and Corollary \ref{cor_1_q_el}, one need to match the  phases and the center lines of the peaks, in the spirit of the end of 
Sec. \ref{sect_ell}.  We verified this for $K=0.$ While this can be done for several first $K=1,2,..$ it is unclear how to deal with a general integer $K.$ The reason is that $\gamma$ and 
$\frac{z}{2}$, which determine the number of soliton to which point $(x,t)$ is constrained, are defined in quite a different implicit manner. In any 
case, this would  not bring a new result since theorems \ref{thrm_Elliptic_est}, 
\ref{teorKK_refined}, \ref{thrm_as_sol_mes} and Corollary \ref{cor_1_q_el} together give a complete description of asymptotics of $q(x,t)$ in the transition zone $4c^2t-\varepsilon t\leq x\leq4c^2t.$
\end{com}
\appendix
\section{Asymptotics of modified Laguerre polinomials}
\label{sect_append} In this section we prove Lemma \ref{lem_L} and hence Corollary \ref{cor_Y_L}. Large $n\to\infty$ asymptotics of the RHP as in Corollary \ref{cor_Y_L} with $\Lambda = n$ was studied in
\cite{VanLessen}, but that  treatment is not suitable for our present purposes.

\begin{figure}[t]
\begin{center}
\includegraphics[width=40mm]{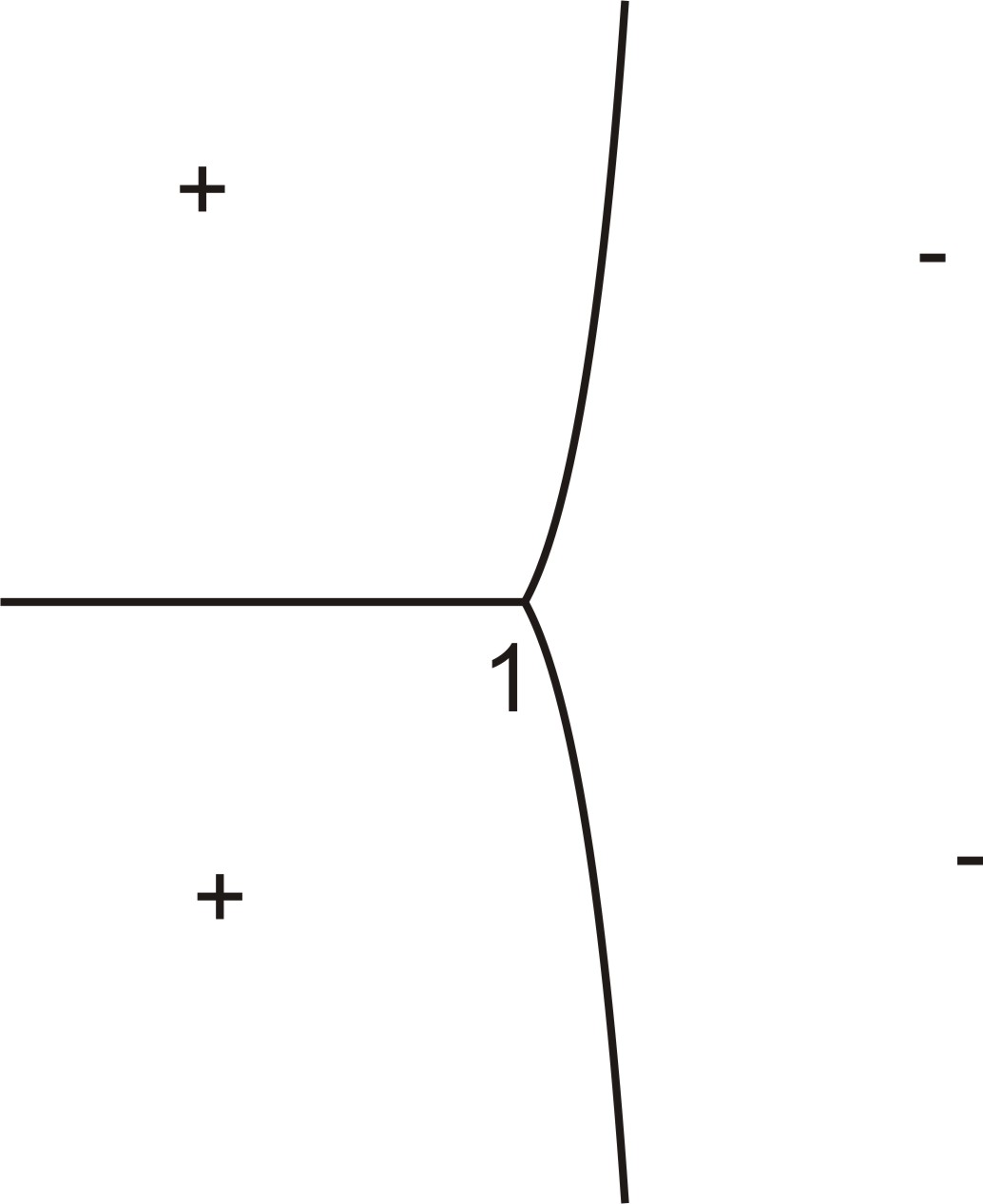}
\end{center}
\caption{Distribution table of signs for $\mathrm{Re}\ \psi(\lambda).$} \label{Signature_table_Repsi}
\end{figure}

In order to study large $n\to\infty$ asymptotics, first we introduce a new matrix-valued function
$$W(\zeta)=\e^{n\ell \sigma_3}Y(\zeta)\e^{n(g(\zeta)-\ell)\sigma_3},$$
where $g(\zeta)$ is the function analytic in $\mathbb{C}\setminus(-\infty,a]$ defined by:
$$g(\zeta)=\frac{2}{\pi a}\int\limits_{0}^a\ln(\zeta-s)\sqrt{\frac{a-s}{s}}\d s=\frac{-2}{a}\sqrt{\zeta(\zeta-a)} - \ln a +
\ln\(2\zeta-a+2\sqrt{\zeta(\zeta-a)}\)+\frac{2\zeta}{a}+\ell,$$
A simple computation shows that $g(\zeta)$ has the asymptotic expansion
\begin{equation}
g(\zeta)=\ln \zeta-\frac{a}{4\zeta}-\frac{a^2}{16\zeta^2}-\frac{5a^3}{192\zeta^3}+\cdots=\ln \zeta-\sum\limits_{j=1}^{\infty}\frac{2\ (2n-1)!}{(n+1)\ n!^2}\frac{a^n}{4^n \zeta^n}.\label{g_asymp}
\end{equation}
The constant $a, \ell$ are given by 
$$a=\frac{4n}{\Lambda},\quad \ell :=\ln\frac{a}{4\e},\quad .$$
At $\zeta=\infty$ we have
$W(\zeta)=\(I+\mathcal{O}(\zeta^{-1})\), $ and $W$ satisfies the boundary value condition
$$W_+=W_-\begin{pmatrix}\e^{n(g_+-g_-)}&0\\\sqrt{\zeta}\ \e^{n(g_-+g_+-2l-\frac{\Lambda \zeta}{n})}&\e^{-n(g_+-g_-)}\end{pmatrix},\zeta\in(0,+\infty),\qquad W_+=W_-\e^{n(g_+-g_-)\sigma_3},\zeta\in(-\infty,0).$$
Further, we introduce the "effective" potential
$$\varphi(\zeta):=-\frac{\Lambda}{2n}\zeta+g(\zeta)-\ell,$$
so that  the jump condition  can be written in the following form:
$$W_+=W_-\begin{pmatrix}\e^{n(\varphi_+-\varphi_-)}&0\\\sqrt{\zeta}\ \e^{n(\varphi_-+\varphi_+)}&\e^{-n(\varphi_+-\varphi_-)}\end{pmatrix}.$$
%
%
%
%
It is convenient now to scale the interval $(0,a)$ into $(0,1),$ i.e.
$$W^{(1)}\(\lambda\):=W\(\zeta\),\qquad \lambda:=\frac{\zeta}{a}=\frac{\zeta\ \Lambda}{4n}.$$
Denote $$\psi\(\lambda\):= \varphi(\zeta)=-2\int\limits_{1}^{\lambda}\sqrt{\frac{s-1}{s}}\ \d s=-2\sqrt{\lambda(\lambda-1)}+\ln\(-1+2\lambda+2\sqrt{\lambda(\lambda-1)}\)=$$
$$=-2\lambda+\ln\lambda+\ln4e-h(\lambda)=-2\lambda+\ln\lambda+\ln4e+\mathcal{O}(\lambda^{-1}),$$
where $h(.)$ is as in (\ref{h}).
The scaled effective
potential $\psi(\lambda),$ which is analytic in $\lambda\in\mathbb{C}\setminus(-\infty,1],$ has the following properties on the
interval $(-\infty,1]:$
\begin{equation}\label{psi_prop}\psi_+-\psi_-=2\pi\i,\quad \lambda\in(-\infty,0),\qquad\qquad \psi_-+\psi_+ = 0,\ \ \ \lambda\in(0,1).\end{equation}
Due to the above properties (\ref{psi_prop}), function
$$W^{(2)}(\lambda) = a^{\frac{\sigma_3}{4}}W^{(1)}(\lambda)a^{\frac{-\sigma_3}{4}}$$
does not have jump along $\lambda\in(-\infty,0),$ and solves a RHP of the form
\begin{eqnarray}
\nonumber&&
 W^{(2)}_+=W^{(2)}_-\begin{pmatrix}\e^{n(\psi_+-\psi_-)} & 0 \\ \sqrt{\lambda} & \e^{-n(\psi_+-\psi_-)}\end{pmatrix}, \lambda\in(0,1),
\qquad W^{(2)}_+=W^{(2)}_-\begin{pmatrix}1 & 0 \\ \sqrt{\lambda}\ \e^{2n\psi} & 1\end{pmatrix}, \lambda\in(1,\infty),
\\\nonumber
&&W^{(2)}(\lambda)=I+\mathrm{O}\(\frac{1}{\lambda}\).
\end{eqnarray}
Tracking back the relation between $W^{(2)}(\lambda)$ and $W^{(1)}(\lambda),$ $W(\zeta)$, $Y(\zeta),$ $L(\zeta),$ we see that
$$W^{(2)}(\lambda) = \(\frac{\sqrt{2}\ n^{n+\frac14}}{\e^n}\)^{\sigma_3}L(4n\lambda)(4n\lambda)^{n\sigma_3}\e^{-nh(\lambda)\sigma_3}\(\frac{\e^n}{\sqrt{2}\ n^{n+\frac14}}\)^{\sigma_3},$$
Important role now is played by the signature table of
$\Re\psi.$ It is shown in the figure \ref{Signature_table_Repsi}. 

Now we introduce the $\delta-$ transformation, which removes the
term $\sqrt{\lambda}$ from the jump matrix. It must solve the following
scalar conjugation problem:
$$\delta_+\delta_-=\sqrt{\lambda},\ \lambda\in(0,1).$$
The solution is given by the formula
$$\delta(\lambda)=\exp\left\{\frac{R(\lambda)}{2\pi\i}\int\limits_{0}^{1}\frac{\ln\sqrt{s}\ \d s}{(s-\zeta)\ R_+(s)}\right\},\quad R(\lambda)=\sqrt{\lambda(\lambda-1)},$$
$$\delta(\lambda)=\(\frac{\lambda}{2\lambda-1+2\sqrt{\lambda(\lambda-1)}}\)^{1/4},\quad \delta(\infty) = \frac{1}{\sqrt{2}}.$$
With a series of transformations, we now show how to  remove the oscillating jump from the interval $(0,1).$ First of them is
$$W^{(3)}(\lambda):=\delta^{\sigma_3}(\infty)W^{(2)}(\lambda)\delta^{-\sigma_3}(\lambda).$$
The jump matrix $W^{(3)}_+=W^{(3)}_+J^{(3)}$ is
$$J^{(3)}=\begin{pmatrix}\frac{\delta_-}{\delta_+}\e^{n(\psi_+-\psi_-)}&0\\
\e^{n(\psi_-+\psi_+)}&\frac{\delta_+}{\delta_-}\e^{-n(\psi_+-\psi_-)}\end{pmatrix},\
\lambda\in(0,1), \qquad
J^{(3)}=\begin{pmatrix}1&0\\\frac{\sqrt{\lambda}}{\delta^2}\
\e^{2n\psi}&1\end{pmatrix},\ \lambda\in(1,+\infty).
$$
Now we need to factorize the jump matrix on the interval $[0,1]$
according to the signature table, namely
$$\begin{pmatrix}\frac{\delta_-}{\delta_+}\e^{n(\psi_+-\psi_-)}&0\\
\e^{n(\psi_-+\psi_+)}&\frac{\delta_+}{\delta_-}\e^{-n(\psi_+-\psi_-)}\end{pmatrix}
=
\begin{pmatrix} 1 & \frac{\delta_-^2\
\e^{-2n\psi_-}}{\sqrt{\lambda}}
\\0 & 1
\end{pmatrix}
\begin{pmatrix} 0 & -\e^{-n(\psi_-+\psi_+)}
\\\e^{n(\psi_-+\psi_+)} & 0
\end{pmatrix}
\begin{pmatrix} 1 & \frac{\delta_+^2\
\e^{-2n\psi_+}}{\sqrt{\lambda}}
\\0 & 1
\end{pmatrix}.
$$
The next step is to "open lenses", we transform the RH
problem, "moving" the left and the right factor in the above
formula to the lower and upper half-planes, respectively. For a
new RH problem $W^{(4)}$ the jumps $W^{(4)}_+=W^{(4)}_-J^{(4)}$
are $$J^{(4)} = \begin{cases}\begin{pmatrix} 1 & \frac{\delta^2\
\e^{-2n\psi}}{\sqrt{\lambda}}
\\0 & 1
\end{pmatrix}, \lambda\in l_u (\textrm{ the upper lense }),
\qquad
\begin{pmatrix} 1 & \frac{\delta^2\
\e^{-2n\psi}}{\sqrt{\lambda}}
\\0 & 1
\end{pmatrix}, \lambda\in l_d (\textrm{ the lower lense }),
\\
\begin{pmatrix} 0 & -1
\\1 & 0
\end{pmatrix}, \lambda\in (0,1),
\qquad
\begin{pmatrix}1&0\\\frac{\sqrt{\lambda}}{\delta^2}\
\e^{2n\psi}&1\end{pmatrix},\ \lambda\in(1,+\infty).
\end{cases}$$
The RHP for $W^{(4)}$ gives us the first statement of Lemma \ref{lem_L} concerning function 
$$\mathcal{Q}_n(\lambda):=W^{(4)}(\lambda)\(\delta(\lambda)\delta^{-1}(\infty)\)^{\sigma_3}.$$
Indeed, the fact that $\mathcal{Q}_n(\lambda)$ decomposees into an asymptotic series follows, for example, from an explicit representation (\ref{LaguerreMatrixfiniten}) of function $L(.).$
Then RHP for $W^{(4)}$ gives us uniformity with respect to $n\geq 1$ of all the ingredients in this asymptotic series. Further, since for large $n\to\infty$ the jump matrix is concentrated
mainly on the interval $(0,1),$ the solution to the RH problem for
$W^{(4)}$ is close everywhere except for a neighborhood of the
points $0,$ $1$ to the solution 
$$\hskip-0cmM_{mod}=\begin{pmatrix}a(\lambda)&\i b(\lambda)\\-\i b(\lambda)&a(\lambda)\end{pmatrix},\quad a(\lambda)=\frac12\(\gamma(\lambda)+\gamma^{-1}(\lambda)\),\ b(\lambda)=\frac12\(\gamma(\lambda)-\gamma^{-1}(\lambda)\),\
\gamma(\lambda)=\sqrt[4]{\frac{\lambda-1}{\lambda}}$$
to the model problem with jumps
only on $(0,1).$ 
This gives us the second statement of Lemma \ref{lem_L} concerning function
$$\mathcal{E}_n(\lambda) := W^{(4)}(\lambda)M_{mod}^{-1}(\lambda).$$

However, rigorous proof of the fact that ingredients in asymptotic series for $\mathcal{E}_n$ tend to $0$ as $n\to\infty$ involves construction of local parametrices around the
points $0,1,$ since jump matrix for $W^{(4)}$ is not uniformly close to $I$ in vicinities of these points. The details can be found for example in \cite{VanLessen}.

\subsection{Airy Parametrix}
For the benefit of the reader we recall the form of the so--called ``Airy'' parametrix, which is used to provide local solutions of a large class of Riemann Hilbert problems 
\cite{DKMVZ}. This is defined by
\begin{equation}\label{Psi_Ai}\Psi_{Ai}(\zeta) = \left\{\begin{array}{l}
\begin{pmatrix}v_1&v_0\\v'_1&v'_0\end{pmatrix},\arg\zeta\in(0,2\pi/3),
\quad
\begin{pmatrix}v_{-1}&v_0\\v'_{-1}&v'_0\end{pmatrix},\arg\zeta\in(0,-2\pi/3),
\\
\begin{pmatrix}v_1&-\i v_{-1}\\v'_1&-\i v'_{-1}\end{pmatrix},\arg\zeta\in(2\pi/3,\pi),
\quad
\begin{pmatrix}v_{-1}&\i v_{1}\\v'_{-1}&\i v'_{1}\end{pmatrix},\arg\zeta\in(-\pi,-2\pi/3),
\end{array}\right\}
\times \e^{\pi\i\sigma_3/4}.\end{equation} 
where we have used the notation
 \begin{equation}
 \label{v01m1}v_0(\zeta)=\sqrt{2\pi}Ai(\zeta),\quad
v_1(\zeta)=\sqrt{2\pi}\e^{-\pi\i/6}Ai(\zeta\e^{-2\pi\i/3}),\quad
v_{-1}(\zeta)=\sqrt{2\pi}\e^{\pi\i/6}Ai(\zeta\e^{2\pi\i/3}).
\end{equation} We
have $$v_0-\i v_1+\i v_{-1}\equiv0.$$

This function $\Psi_{Ai}(\zeta)$
has the following jumps $\Psi_{Ai,+}=\Psi_{Ai,-}J_{\Psi_{Ai}}$:
$$J_{\Psi_{Ai}}=\begin{pmatrix}1&0\\1&1\end{pmatrix},\arg\zeta=0,\ \begin{pmatrix}1&1\\0&1\end{pmatrix},\arg\zeta=\frac{2\pi}{3},
\
\begin{pmatrix}1&1\\0&1\end{pmatrix},\arg\zeta=\frac{-2\pi}{3},\
\begin{pmatrix}0&-1\\1&0\end{pmatrix},\arg\zeta=\pi.$$
and the following behaviour at infinity:
$$\Psi_{Ai}(\zeta)=\zeta^{-\sigma_3/4}\frac{1}{\sqrt{2}}\begin{pmatrix}1&1\\1&-1\end{pmatrix}\(I+
\begin{pmatrix}
\frac{-1}{48}&\frac{-1}{8}\\
\frac{1}{8} & \frac{1}{48}
\end{pmatrix}\zeta^{-3/2}+\mathcal{O}(\zeta^{-3})\)\e^{\frac23\zeta^{3/2}\sigma_3}\e^{\pi\i\sigma_3/4}.$$


\begin{thebibliography}{99 }
\bibitem{DKMVZ} P. Deift, T. Kriecherbauer, K. T-R McLaughlin, S. Venakides, X. Zhou. \textit{Uniform asymptotics for polynomials
orthogonal with respect to varying exponential weights and
applications to universality questions in random matrix theory}.
Communications on Pure and Applied Mathematics. Volume 52, Issue
11, pages 1335--1425, November 1999.

\bibitem{Bertola Buckingma Lee Pierce} M.~Bertola, R.~Buckingham,
S.~Lee, V.Pierce. Spectra of random Hermitian matrices with a
small-rank external source: supercritical and subcritical regimes,
2010.

\bibitem{Bertola Tovbis}M.~Bertola, A.~Tovbis. Universality for
the focusing Schroedinger equation at the gradient catastrophe
point: Rational breathers and poles of the tritronquee solution to
Painleve I, 2012.

\bibitem{Claeys Grava 2010} T. Claeys,  T. Grava. Solitonic asymptotics for the Korteweg-de Vries equation in the small dispersion limit. SIAM J. Math. Anal. 42 (2010), no. 5, 2132-2154

\bibitem{CK} A. Cohen, T. Kappeler. Scattering and inverse scattering for steplike potentials in the Schrödinger equation. Indiana Univ. Math. J. 34 (1985), no. 1, 121-180.

\bibitem{GP} A. V. Gurevich, L. P. Pitaevskii.   Decay of Initial Discontinuity in the Korteweg-de Vries Equation
\emph{JETP Letters} \textbf{17/5} 193 (1975).

\bibitem{BikN1} R. F. Bikbaev, V. Yu. Novokshenov.  The Korteveg-de Vries Equation with Finite Gap Boundary Conditions and Self-Similar Solutions of Whitham Equations \emph{ Proc. III International Workshop "Nonlinear and Turbulent Processes in Physics"} Kiev \textbf{1} 32-35 (1988).

\bibitem{BikN2} R. F. Bikbaev, V. Yu. Novokshenov. Existence and uniqueness of the solution of the Whitham equation (Russian) \emph{ Asymptotic methods for solving problems in mathematical physics  Akad. Nauk SSSR Ural. Otdel. Bashkir. Nauchn. Tsentr Ufa} 81-95 (1989).

\bibitem{Bikb1} R.F. Bikbaev. Structure of a shock wave in the theory of the Korteweg-de Vries equation.  \emph{Phys. Lett. A} \textbf{141/5-6} 289-293 (1989).

\bibitem{Bikb2} R.F. Bikbaev, R. A.   Sharipov. The asymptotic behaviour, as $t\to\infty$, of the solution of the Cauchy problem for
the Korteweg-de Vries equation in a class of potentials with
finite-gap behaviour as $x\to\pm\infty$.  \emph{ Teoret. Mat.
Fiz.}  \textbf{78/3} 345-356  translation in \emph{Theoret. and
Math. Phys.} \textbf{78/3} 244-252 (1989).

\bibitem{Bikb3} R. F. Bikbaev. The Korteweg-de Vries equation with finite-gap boundary conditions and Whitham deformations
of Riemann surfaces (Russian) \emph{ Funktsional. Anal. i Prilozhen.}
\textbf{23/4} 1-10  translation in \emph{Funct. Anal. Appl.}
  \textbf{23/4} 257-266 (1990)

\bibitem{Bikb4} R. F. Bikbaev. The influence of viscosity on the structure of shock waves in the MKdV model (Russian)
\emph{Zap. Nauchn. Sem. S.-Peterburg. Otdel. Mat. Inst. Steklov. (POMI)  Voprosy Kvant. Teor. Polya Statist. Fiz.} \textbf{11} 37-42
184 translation in \emph{J. Math. Sci.}  \textbf{77}  (1995) \textbf{2} 3042-3045

\bibitem{Bikb5} R. F. Bikbaev. Complex Whitham deformations in problems with "integrable instability" (Russian)
\emph{Teoret. Mat. Fiz.}  \textbf{104/3} 393-419  translation in
\emph{Theoret. and Math. Phys.} (1996) \textbf{104/3} 1078-1097

\bibitem{Bikb6} R. F. Bikbaev. Modulational instability stabilization via complex Whitham deformations: nonlinear Schrodinger equation \emph{
Zap. Nauchn. Sem. S.-Peterburg. Otdel. Mat. Inst. Steklov. (POMI)}
\textbf{215}  \emph{Differentsialnaya Geom. Gruppy Li i Mekh. } \textbf{14}
65-76 310 translation in \emph{J. Math. Sci. (New York)}  \textbf{85}  (1997)
\textbf{1} 1596-1604

\bibitem{BK07} A. Boutet de Monvel, V.P.  Kotlyarov. Focusing non\-linear Schrodinger equation on the quarter plane with time-periodic boundary condition: a Riemann-Hilbert approach \emph{J. Inst. Math. Jussieu} \textbf{6/4} 579-611 (2007)

\bibitem{BIK07} A. Boutet de Monvel, A. R.  Its, V. P.  Kotlyarov. Long-time asymptotics for the focusing NLS equation with time-periodic
boundary condition. \emph{C. R. Math. Acad. Sci. Paris.} \textbf{
345/11} 615-620 (2007)

\bibitem{BIK09}  A. Boutet de Monvel, A. R. Its, V. P.  Kotlyarov.  Long -time asymptotics for the focusing NLS equation with time -- periodic
boundary condition on the half line. \emph{Comm. Math. Phys.}  \textbf{290/2} 479-522 (2009).

\bibitem{BKS11}  A. Boutet de Monvel, V. P.  Kotlyarov, D. G. Shepelsky.
 Focusing NLS equation: long-time dynamics of step-like initial
data.  \emph{International Mathematics Research Notices} \textbf{7} 1613-1653 (2011).

\bibitem{BV} R. Buckingham, S.  Venakides. Long-time asymptotics of the non-linear Schrodinger equation shock problem. \emph{Comm. Pure Appl. Math.} \textbf{60/9} 1349-1414 (2007).

\bibitem{C&K} R. M. Corless, G. H.  Gonnet,D. E. G.  Hare, D. J.  Jeffrey, D. E. Knuth. On the Lambert W Function \emph{ Advances in Computational Mathematics}
\textbf{5} 329-359 (1996).

\bibitem{DT} P. Deift, E.  Trubowitz. Inverse scattering on the line. Comm. Pure Appl. Math. 32, no. 2, 121-251 (1979).

\bibitem{DZ93} P. Deift, X.  Zhou. A steepest descent method for oscillatory Riemann -- Hilbert problems. Asymptotics for the MKdV equation \emph{
 Annals of Mathematics} \textbf{137/2} 295-368 (1993).

\bibitem{EGKT} I. Egorova, Z. Gladka, V.  Kotlyarov, G. Teschl.  Long-Time Asymptotics for the Korteweg-de Vries Equation with Steplike Initial
Data. \emph{Nonlinearity } \textbf{26/7} 1839-1864 (2012).

\bibitem{Jakovleva} A. I. Jakovleva. Master thesis "Application of inverse scattering transform method to a Cauchy problem for the modified Korteweg-de Vries equation", Kharkov, 1980. [Russian]

\bibitem{Kh1} E. Ya. Khruslov. Splitting of an initial step-like perturbation for the KdV equation \emph{ Letters to JETP} \textbf{21/4} 469-472 (1975).

\bibitem{Kh2} E. Ya. Khruslov. Asymptotics of the solution of the Cauchy prob\-lem for the Korteweg de Vries equation with initial data of step type. \emph{Matem. Sbornik (New Series)} \textbf{99(141):2} 261-281 (1976).

\bibitem{KK} E. Ya.  Khruslov, V. P.  Kotlyarov. Asymptotic
solitons of the modified Korteweg-de Vries equation \emph{  Inverse
problems} \textbf{5/6} 1075-1088 (1989).

\bibitem{KK2} E. Ya. Khruslov, V. P. Kotlyarov. Soliton asymptotics of nondecreasing solutions of nonlinear completely integrable evolution equations \emph{    Spectral operator theory and
related topics} Adv. Soviet Math. \textbf{19} Amer. Math. Soc.
Providence, RI 129-180 (1994).

\bibitem{KK3} E. Ya. Khruslov, V. P.  Kotlyarov.  Generation of asymptotic solitons in an integrable model of stimulated Raman scattering by periodic boundary data \emph{Mat. Fiz.Anal. Geom.}  \textbf{10/3} 366-384 (2003).

\bibitem{KM} V. Kotlyarov, M.  Alexander.
Riemann-Hilbert problem to the modified Korteveg de Vries equation:
Long-time dynamics of the steplike initial data
\emph{Journal of Mathematical Physics} \textbf{51} 093506 (2010).

\bibitem{KM2}  V. Kotlyarov, A.  Minakov.
Step-Initial Function to the MKdV Equation:
Hyper-Elliptic Long-Time Asymptotics of the Solution \emph{ Journal of Mathematical Physics, Analysis, Geometry} \textbf{ 8/1} 37-61 (2011).

\bibitem{KM2015} V. Kotlyarov, A.  Minakov. Modulated elliptic wave and asymptotic solitons in a shock problem to the modified Kortweg--de Vries equation. J. Phys. A 48 (2015), no. 30, 305201, 35 pp.

\bibitem{Lavrentiev Sabat} M. A. Lavrent'ev, B. V.  Sabat.  Metody teorii funkcii kompleksnogo peremennogo. (Russian) Methods of the theory of functions of a complex variable \emph{ Gosudarstv. Izdat. Tehn.-Teor. Lit. Moscow-Leningrad} (1951).

\bibitem{Leach_Needham}  J. A. Leach. An initial-value problem for the modified Korteweg-de Vries equation.
IMA J. Appl. Math. 78 (2013), no. 6, 1196-1213.

\bibitem{Marchant}  T. R. Marchant. Undular bores and the initial-boundary value problem for the modified Korteweg-de Vries equation. Wave Motion 45 (2008), no. 4, 540-555.

\bibitem{Mar}  V. A. Marchenko. {\it Operatory Shturma-Liuvillya i ikh prilozheniya.} (Russian) [Sturm-Liouville operators and their applications] Izdat. "Naukova Dumka'', Kiev, 1977. 331 pp.

\bibitem{M1} A. Minakov. Long-time behaviour of the solution to the mKdV
equation with step-like initial data \emph{ J. Phys. A: Math. Theor.} \textbf{44} 085206 (2011).

\bibitem{M2} A. Minakov.
Asymptotics of Rarefaction Wave Solution
to the mKdV Equation \emph{Journal of Mathematical Physics, Analysis, Geometry} \textbf{ 7/1} 59-86
(2011).

\bibitem{MK} E. A. Moskovchenko, V. P. Kotlyarov. A new Riemann -- Hilbert problem in a model of stimulated Raman scattering \emph{
J.Phys.A.: Math. Gen.} \textbf{39} 014591 (2006).

\bibitem{Mos} E. A. Moskovchenko. Simple periodic boundary data and Rie\-mann -- Hilbert problem for integrable model of the stimulated Raman scattering
    \emph{Journal of mathematical physics,analysis, geometry} \textbf{5/1} 82-103 (2009).

\bibitem{MK10} E. A. Moskovchenko, V. P.  Kotlyarov.  Periodic boundary data for an integrable model of stimulated Raman scattering: long-time asymptotic behavior
    \emph{Journal of Physics A: Mathematical and Theoretical}\textbf{ 43/5} 055205

\bibitem{M_disser} A. Minakov, PhD thesis. "Riemann-- Hilbert problems and the modified Korteweg de Vries equation: asymptotic analysis of solutions with step-like initial data", 2013.

\bibitem{Novik} V. Yu. Novokshenov. Time asymptotics for soliton equations in problems with step initial conditions (Russian) \emph{Sovrem. Mat. Prilozh., Asimptot. Metody Funkts. Anal.} \textbf{ 5} 138-168 translation in \emph{J. Math. Sci. (N. Y.)} \textbf{ 125} \textbf{5} 717-749  (2005)

\bibitem{Novokshenov80}  V. Yu. Novoksenov.  Asymptotic behavior as $t\to\infty$ of the solution of the Cauchy problem for a nonlinear
Schrodinger equation (Russian) \emph{ Dokl. Akad. Nauk SSSR}\textbf{ 251/4} 799-802 (1980).

\bibitem{Novokshenov82}  V. Yu. Novokshenov. Asymptotic Formulae for the Solutions of the System of Nonlinear Schrodinger
Equations \emph{ Uspekhi Matem. Nauk} \textbf{37/2} 215-216 (1982).

\bibitem{Novokshenov85} V. Yu. Novokshenov.  Asymptotics as $t\to\infty$ of the Solution to a Two-Dimentional Generalisation of the Toda Lattice
\emph{Doklady AN SSSR} \textbf{265/6} 1320-1324 translation in \emph{Soviet Math. Dokl}  \textbf{26/1} 264-268 (1982)

\bibitem{Shabat} A. B. Shabat. An inverse scattering problem. (Russian) Differentsialnye Uravneniya 15 (1979), no. 10, 1824--1834 (1918).

\bibitem{VanLessen} M.~Vanlessen. Strong Asymptotics of
Laguerre-Type Orthogonal Polynomials and Applications in Random
Matrix Theory, Constr.Approx, 25: 125-175 (2007).

\bibitem{Wadati}  M. Wadati. The modified Korteweg-de Vries equation. J. Phys. Soc. Japan 34 (1973), 1289-1296.

\bibitem{Zakharov-Shabat} V. E. Zaharov, A. B. Shabat. A plan for integrating the nonlinear equations of mathematical physics by the method of the inverse scattering problem. I. 
(Russian) Funkcional. Anal. i Prilozen. 8 (1974), no. 3, 43-��53.

\bibitem{Zhou} X. Zhou. The Riemann-Hilbert problem and inverse scattering. SIAM J. Math. Anal., 1989, Vol. 20, 966--986.

\end{thebibliography}
\end{document}